\documentclass[iop]{emulateapj-rtx4}
\usepackage[pass]{geometry}

\usepackage[linkcolor=blue,urlcolor=blue,citecolor=black,colorlinks=true,letterpaper]{hyperref}
\usepackage{natbib, multirow, rotating}

\shorttitle{YSOVAR: IRAS 20050+2720}
\shortauthors{K.\ Poppenhaeger for the YSOVAR team}

\begin{document}

\title{YSOVAR:\\ mid-infrared variability of young stellar objects and their disks\\ in the cluster IRAS 20050+2720}

\author{K. Poppenhaeger\altaffilmark{1,2}}
\affil{\altaffilmark{1}Harvard-Smithsonian Center for Astrophysics, 60 Garden Street, Cambridge, MA 02138, USA\\
\altaffilmark{2}NASA Sagan Fellow}

\and

\author{A.M. Cody}
\affil{NASA Ames Research Center, Moffett Field, CA 94035, USA}

\and

\author{K.R. Covey}
\affil{Western Washington University, 516 High Street, Bellingham, WA 98225, USA}

\and

\author{H.M. G\"unther}
\affil{Massachusetts Institute of Technology, Kavli Institute for Astrophysics and Space Research, 77 Massachusetts Avenue,  Cambridge, MA 02139, USA}

\and

\author{L.A. Hillenbrand}
\affil{Department of Astronomy, California Institute of Technology, Pasadena, CA 91125, USA}

\and

\author{P. Plavchan}
\affil{Missouri State University, 901 S.\ National Ave., Springfield, MO 65897, USA}

\and

\author{L.M. Rebull}
\affil{Spitzer Science Center/Caltech, 1200 E. California Blvd., Pasadena, CA 91125, USA}

\and

\author{J.R. Stauffer}
\affil{Spitzer Science Center/Caltech, 1200 E. California Blvd., Pasadena, CA 91125, USA}

\and

\author{S.J. Wolk}
\affil{Harvard-Smithsonian Center for Astrophysics, 60 Garden Street, Cambridge, MA 02138, USA}

\and

\author{C. Espaillat}
\affil{Boston University, 725 Commonwealth Avenue, Boston, MA 02215, USA}

\and

\author{J. Forbrich}
\affil{University of Vienna, Department of Astrophysics, T\"urkenschanzstr.\ 17, 1180, Vienna, Austria}

\and

\author{R.A. Gutermuth}
\affil{Dept. of Astronomy, University of Massachusetts, Amherst, MA 01003, USA}

\and

\author{J.L. Hora}
\affil{Harvard-Smithsonian Center for Astrophysics, 60 Garden Street, Cambridge, MA 02138, USA}

\and

\author{M. Morales-Calder{\'o}n}
\affil{Centro de Astrobiolog{\'i}a (INTA-CSIC), ESAC Campus, P.O. Box 78, E-28691 Villanueva de la Canada, Spain}

\and

\author{Inseok Song}
\affil{Physics and Astronomy Department, University of Georgia, Athens, GA 30602-2451, USA}

\submitted{}
\received{June 08 2015}
\accepted{July 13 2015}

\begin{abstract}
We present a time-variability study of young stellar objects in the cluster 
IRAS 20050+2720, performed at 3.6 and 4.5 $\mu$m with the \textit{Spitzer} Space Telescope; 
this study is part of the Young Stellar Object VARiability project (YSOVAR). 
We have collected light curves for 181 cluster members over 40 days. We find a high 
variability fraction among embedded cluster members of ca.\ 70\%, 
whereas young stars without a 
detectable disk display variability less often (in ca.\ 50\% of the cases) 
and with lower amplitudes. We detect 
periodic variability for 33 sources with periods primarily in the range 
of 2-6 days. Practically all embedded periodic sources display additional variability on 
top of their periodicity. Furthermore, we analyze the slopes of the tracks that our 
sources span in the  color-magnitude diagram (CMD). We find that sources with long 
variability time scales tend to display CMD slopes that are at least partially 
influenced by accretion processes, while sources with short variability time scales 
tend to display extinction-dominated slopes. We find a tentative trend of X-ray 
detected cluster members to vary on longer time scales than the X-ray undetected 
members.

\end{abstract}

\section{Introduction}
\label{sect:introduction}

Protoplanetary disks are the birthplaces of exoplanets. As a star forms from a contracting cloud of gas and dust, the rotational collapse causes parts of the material to form a protoplanetary disk around it. The disk material close to the young stellar object (YSO) accretes onto the YSO's surface along the stellar magnetic field lines. Due to the stellar radiation, the temperature of the disk decreases radially from the inner parts of the disk to the outer parts. There has been observational progress in studying the formation of exoplanets \textit{in situ} through interferometric observations with high spatial resolution, for example with \textit{ALMA} \citep[e.g.][]{deGregorio-Monsalvo2013, vanderMarel2013, Casassus2013, Pineda2014}, \textit{CARMA} \citep[e.g.][]{Eisner2008, Isella2009, Enoch2009, Isella2010}, \textit{SMA} \citep[e.g.][]{Andrews2007, Jorgensen2007, Andrews2009, Brown2009, Andrews2011}, or \textit{PdBI} \citep[e.g.][]{Pietu2006, Hughes2009}. Such observations have been successful in resolving the protoplanetary disk to distances below $100$ AU from the central star; a recent example is the successful imaging of ring structures in the protoplanetary disk of the young star HL~Tau \citep{ALMA2015arXivHLTau}. However, it is very challenging to spatially resolve the processes happening in the disk within a few AU of the host star. Our prime observational tool in the study of the inner and mid disk is therefore the variation in brightness of the star-disk system caused by the disk or an interaction between the star and the disk, such as accretion of disk material onto the star.

%\pagebreak

\begin{figure*}[ht!]
\includegraphics[width=0.99\textwidth]{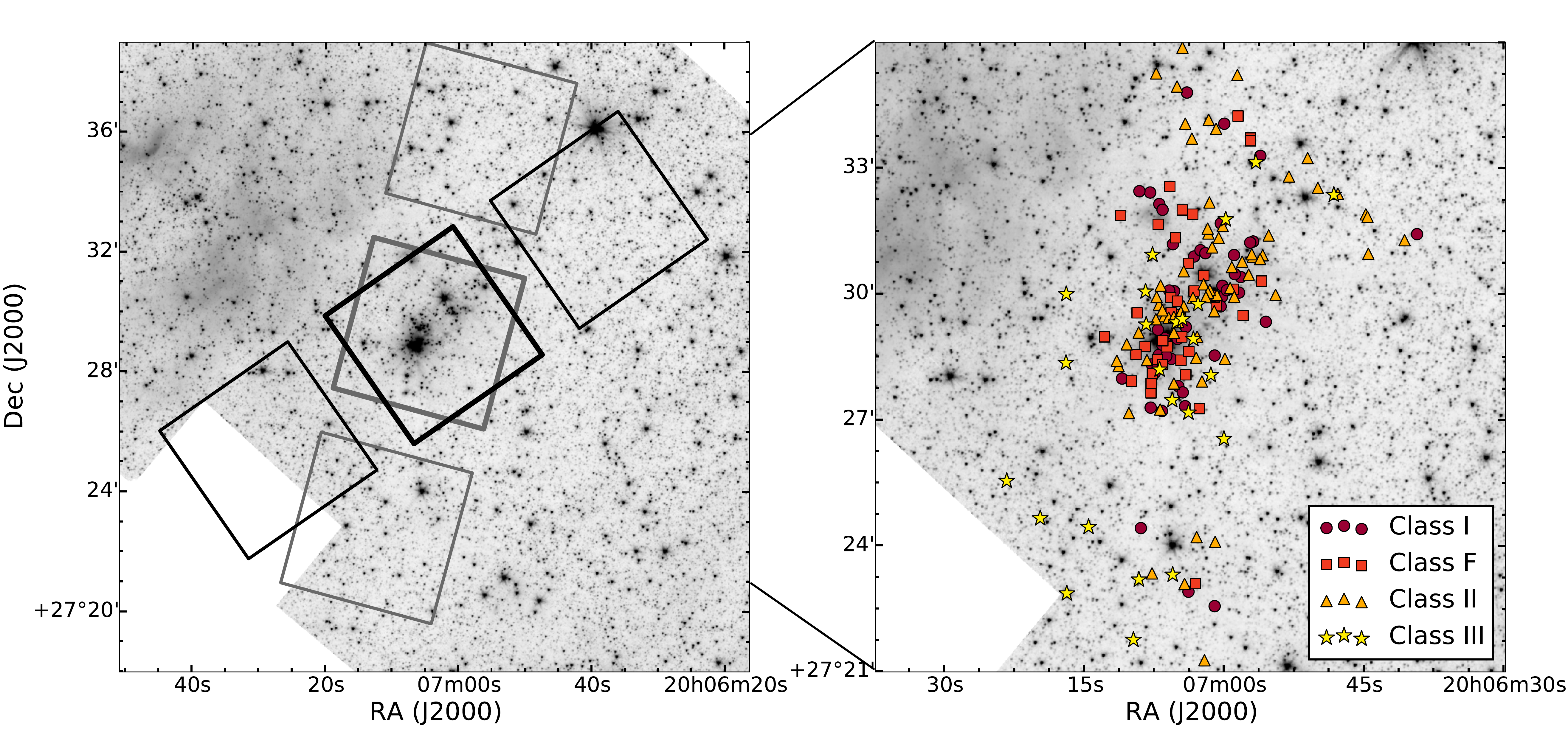}
\caption{\textit{Left:} Fields of view of the Spitzer observations collected for IRAS 20050+2720, overlaid on a \textit{Spitzer} $3.6\,\mu\mathrm{m}$ cryogenic-era image of the region. The field of view gradually rotated over the two months of observations, with sources inside the grey regions being observed at the start of the observing program and sources inside the black regions observed at the end. The middle field received coverage in both the [3.6] and [4.5] bands, while the upper fields were observed only in [3.6] and the lower fields in [4.5]. \textit{Right:} Zoomed-in image of the central field with the identified cluster members indicated.} 
\label{skyview_fov}
\end{figure*}

Given the temperature profile of a protoplanetary disk, we can observe processes at the inner rim of the disk in the near-infrared ($J$, $H$, and $K$ bands) since the inner edge of the disk is determined by the dust sublimation temperature of $\sim 1500$\,K\footnote{The dust sublimation temperature depends on the grain species and the local density \citep{Pollack1994}; 1500\,K is a typically assumed value for protoplanetary disks \citep{Dullemond2007PPV}, with some observations matching sublimation temperatures of 1000-1500\,K \citep{Monnier2005}, while some models use slightly higher temperatures of 1800-2000\,K \citep{dAlessio1998}.}. This temperature implies that the blackbody radiation from the inner part of the disk peaks around 1--2 $\mu$m. Observations in the mid-infrared from $\sim 3$--$20\,\mu$m provide observational access to parts of the disk which display surface temperatures of a few hundred Kelvin, i.e.\ parts of the disk with semimajor axes around 0.5 AU. 

Variability of YSOs has been apparent since the earliest observations; initially, the main focus was the stellar photosphere studied through optical observations \citep{Joy1942, Rydgren1976, Bouvier1986, Vrba1986, Herbst1994}. To study the inner rim of the disk, monitoring in the near-infrared (mostly $JHK$) bands has been used \cite[for example]{Skrutskie1996, Carpenter2001, Makidon2004}. These studies showed that a very large fraction ($\ge 90\%$) of the YSOs are variable in the near-infrared, for example as determined for Orion, the Chamaeleon I Molecular Cloud, and Cyg OB7 \citep{Carpenter2001, Carpenter2002, Rice2012}. Near-infrared variability was found to occur on multiple time scales \citep{Cohen2004, Grankin2007, AlvesdeOliveira2008}, and the fraction of YSOs detected to be variable was reported to grow when multi-year baselines are taken into account \citep{Scholz2012}.

While optical and near-infrared variability have been well studied for YSOs, it was initially unclear how parts of the disk farther out from the YSO behave, as they are thought to be more dynamically stable. Early mid-infrared observations suggested variability (see \citealt{Rebull2011proc} for a review). Long-term monitoring in the mid-IR revealed that YSOs in the cluster IC1396A fall into two classes of periodic and aperiodic variability in the mid-IR \citep{Morales-Calderon2009}, and new AA Tau-like objects and eclipsing binaries were found with those mid-IR data in the ONC \citep{Morales-Calderon2011, Morales-Calderon2012}. A large-scale observational effort to study these parts of disks was made by the Young Stellar Object VARiability (YSOVAR) project \citep{Rebull2014}. In the scope of the project, 12 young stellar clusters were observed in the mid-infrared with the \textit{Spitzer} Space Telescope \citep{Werner2004}. The aim of the project is to compare variability properties and associated disk processes over a wide range of cluster ages.
A recent comprehensive study by \cite{Cody2014} combined \textit{Spitzer} data with simultaneous monitoring with \textit{CoRoT} in the optical. They presented a detailed analysis of different light curve morphologies for YSOs in the cluster NGC 2264, for example ``dippers'' and ``bursters'', the former showing sudden downward dips in the light curves, interpreted as changes in extinction, the latter showing upward spikes interpreted as accretion events \citep{Stauffer2014}. \cite{Guenther2014} have presented an analysis of YSOVAR data for the cluster Lynds 1688 showing that variability amplitudes are larger for the most embedded objects, and Wolk et al.\ (submitted) have shown for the YSOVAR cluster GDD 12-15 that YSOs with X-ray detections show variability on longer time scales than those with no X-ray detection. \cite{Flaherty2013} found a correlation of infrared variability fraction with X-ray luminosity for class II sources in the cluster IC 348, possibly due to accretion-induced hot spots changing the dust sublimation radius of the inner disk.

This present work discusses data collected and results derived for one of the YSOVAR clusters, IRAS~20050+2720. In this paper, we will analyze the variability of the IRAS~20050+2720 members in the mid-infrared. Section~\ref{iras} discusses the cluster IRAS~20050+2720, section~\ref{observations} describes the collected observations, section~\ref{classification} demonstrates how we classified the detected sources and the types of variability, section~\ref{results} describes our results and discusses them in the context of disk processes, and section~\ref{conclusion} summarizes our findings.

% The data analysis and results for three other clusters of the YSOVAR project have been published recently: NGC 2264 \citep{Cody2014}, Lynds 1688 \citep{Guenther2014}, and GGD12-15 (Wolk et al. submitted). 

\section{The young cluster IRAS 20050+2720}
\label{sect:theyoungclusteriras200502720}

\label{iras}
IRAS 20050+2720 is a young stellar cluster in which no massive stars have been detected (see \citealt{Guenther2012} and their Fig.~7). This means that the evolution of its young stellar objects and their disks is not significantly altered by ultraviolet (UV) irradiation; such a UV influence on disks has been predicted \citep{Johnstone1998} and observed for other clusters that do host massive stars \citep{Guarcello2007, Balog2007, WrightNick2012, Guarcello2013}.

IRAS~20050+2720 is located in the Cygnus rift at a distance of 700\,pc from the Sun \citep{Wilking1989}. Objects in the Cygnus region can have substantial errors in their kinematically derived distance estimates if their galactic longitudes are close to $90^\circ$, because the local Galactic arm, the Perseus arm, and the outer parts of the Galaxy are lined up in that direction \citep{Schneider2006}. However, IRAS~20050+2720 is located at a galactic longitude of ca.\ $66^\circ$ where the Cygnus X region and the Cygnus rift can be distinguished from each other, and therefore its distance estimate is deemed to be reliable \citep{Beltran2008}. The central region of IRAS~20050+2720 has been found to display several radio lobes, which are likely due to jets from protostellar objects \citep{Bachiller1995, Codella1999, Beltran2008}. 

IRAS~20050+2720 consists of two cluster cores, the main cluster is located to the west (cluster core W), and a smaller core to the east (cluster core E) has been identified by \cite{Guenther2012}. Our observations target the main cluster core W, and all references to IRAC 20050+2720 in the remainder of the paper refer to cluster core W.

A first classification of likely cluster members has been performed in the infrared by \cite{Chen1997}. They identified ca.\ 100 sources with IR excesses. \cite{Gutermuth2009} revisited this cluster and used \textit{Spitzer} cryo-era observations to identify a total of 177 young stellar objects. \cite{Guenther2012} have performed additional optical and X-ray observations of the cluster and have identified ca.\ 300 cluster members, among them ca.\ 50 sources which are young stars without visible disks (weak-lined T Tauri stars). Those objects could only be classified and distinguished from older foreground stars through X-ray observations, because young stars are brighter in X-rays than older stars\footnote{The X-ray to bolometric luminosity ratio of young stars is of the order of $10^{-3}$--$10^{-4}$, while it is typically smaller than $10^{-6}$ for older stars.}.

\begin{figure}[ht!]
\centering\includegraphics[width=0.5\textwidth]{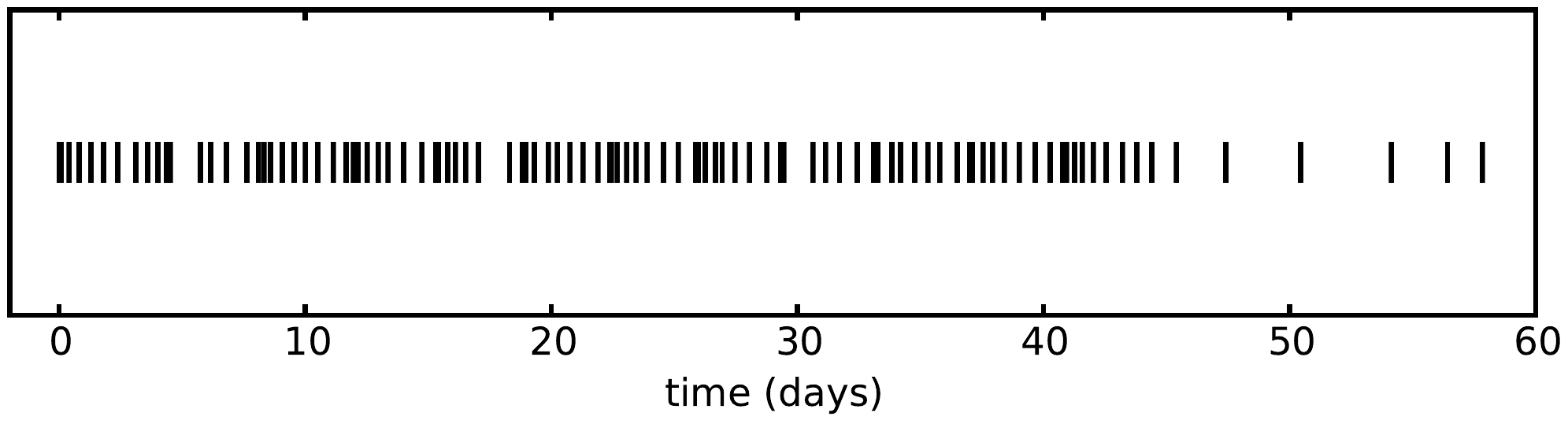}
\caption{Time sampling of the \textit{Spitzer} observations of IRAS~20050+2720, displayed for the full observational window of ca.\ 60 days. The time steps range from $\sim 4$ hours to $\sim 18$ hours for most of the light curve, with some sparser time sampling every 2-3 days at the very end. The time lag between the [3.6] and [4.5] band is small in comparison ($\sim 2$ minutes).} 
\label{cadence}
\end{figure}

\begin{table*}[ht!]
\caption{List of data sets used in this work.}
\begin{center}
\begin{tabular}{l c c c}
\hline \hline
Telescope	& Band	&	Epoch	& Reference \\ 
\hline
{\it Spitzer IRAC} cold mission& 3.6/4.5/8.0/24 $\mu$m	& 2007	& \cite{Gutermuth2009}\\
{\it Spitzer IRAC} warm mission& 3.6/4.5 $\mu$m	& 2010	& this work\\
2MASS		& J H K$_s$			& 1997	& \cite{2MASS}\\
PAIRITEL	& J H K$_s$			& 2010-2012	& this work \\
FLWO/KeplerCam	& U B V R I			& 2009	& \cite{Guenther2012}\\
IPHAS		& r i H$\alpha$		& 2003-2005	& \cite{Gonzales-Solares2008}\\
{\it Chandra ACIS-I}	& X-ray (0.25-12\,keV)	& 2006, 2007	& \cite{Guenther2012}\\

\hline
\end{tabular}
\end{center}
\label{aux_data}
\end{table*}

\section{Observations}
\label{sect:observations}

\label{observations}
We obtained mid-infrared light curves of objects in IRAS~20050+2720 during the {\it Spitzer} Warm Mission \citep{Storrie-Lombardi2010}; these observations are our main focus in this work, and we describe those data in detail in section \ref{spitzer_obs}. An in-depth presentation of the data reduction is given in \cite{Rebull2014}; for the convenience of the reader, we summarize the most important points in section \ref{spitzer_obs}. We supplement these \textit{Spitzer} observations with the auxiliary observational data listed in Table~\ref{aux_data}. Data from \textit{Chandra} was used in our source classification scheme and data from \textit{PAIRITEL} was specifically reduced for this work; we describe the reduction of those data sets in sections \ref{chandra_obs} and \ref{pairitel_obs}.

\subsection{Spitzer data}
\label{sect:spitzerdata}

\label{spitzer_obs}
The central region of IRAS~20050+2720 was observed with {\it Spitzer} from 2010-06-12 to 2010-08-10, using the 3.6$\mu$m and 4.5$\mu$m channel (hereafter [3.6] and [4.5]) of the IRAC camera \citep{Fazio2004}. Spitzer is able to observe simultaneously in [3.6] and [4.5], but the fields of view are adjacent and do not overlap. To obtain near-simultaneous light curves in both bands, the telescope first centers the [4.5] channel on the object, then the [3.6] channel, which results in a central area of the cluster which receives coverage in both [3.6] and [4.5], as well as two secondary areas which receive coverage in either [3.6] or [4.5]. We show a schematic representation of the fields of view in Figure~\ref{skyview_fov}.

The cadence of the observations was chosen to reduce aliasing when searching for periodic variability. We show a schematic representation of the observational cadence in Figure~\ref{cadence}. It consists of a total of ca.\ 100 epochs with repeating sequences of increasing time steps inbetween them. Specifically, the time increments relative to a preceding epoch are roughly 4, 6, 8, 10, 12, 14, 16, and 18 hours. For the central field of view, which received coverage in both [3.6] and [4.5], the [3.6] pointings follow the [4.5] pointings with a time lag of $\sim 2$ minutes. Each individual observation was performed as an IRAC mapping mode Astronomical Observation Request using the high-dynamic-range mode (HDR), which consists of a single 0.4\,s and 10.4\,s exposure.

\subsubsection{Spitzer data processing}
\label{sect:spitzerdataprocessing}

The photometry of the warm-era {\it Spitzer} observations was performed with the code {\it Cluster Grinder} \citep{Gutermuth2009} written in Interactive Data Language (IDL). The starting point was the \textit{Spitzer Science Center}-released basic calibrated data (BCD); the individual frames were combined into mosaics after processing for cosmic rays and bright source artifacts. Since the observations were performed in High Dynamic Range (HDR) mode, the two HDR exposures were combined into a single epoch by appropriately scaling the short-frame values for bright sources and replacing compromised pixels in the long frames. 

For each epoch and channel we performed a point source detection and aperture photometry, using a source extraction radius of 2.4$^{\prime\prime}$ (2 pixels) and an annulus from 2.4$^{\prime\prime}$ to 7.2$^{\prime\prime}$ (6 pixels) for background estimation. Source matching of epochs and the two channels was performed by position with a required cross-match radius of $\le 1^{\prime\prime}$. The final position was taken to be the mean of the individual positions. A comparison with the 2MASS catalogue \citep{2MASS} shows uncertainties of $<200$\,mas after a single global WCS cross-match to the catalogue. A detailed discussion of the accuracy of the photometry and the associated noise floor is given in \cite{Rebull2014}, sections 2.5 and 2.6. The data used in this article will be delivered to the NASA/IPAC Infrared Science Archive (IRSA); it was retrieved from the YSOVAR database on 2014-09-10. 

For the subsequent analysis of the light curves, we developed a set of Python routines named pYSOVAR \citep{pysovar} which are publicly available on github\footnote{see \url{https://github.com/YSOVAR}}. These routines calculate a set of statistical properties such as the mean, median, standard deviation, maximum and minimum of the light curves, as well as more sophisticated quantities like autocorrelation time scales and a fit to the color-magnitude diagram with data uncertainties on both axes.

\begin{table*}
\begin{center}
\caption{\label{tab:tab2} Source designations, flux densities and lightcurve properties. This table is published in its entirety in the electronic verion of the journal. Here the table columns are described as a guide to form and content.}
\begin{tabular}{r l c c l}
\hline\hline
ID & Name & Unit & Channel & Comment \\
\hline
1 & RA & deg & -- & J2000.0 Right ascension \\
2 & DEC & deg & -- & J2000.0 Declination \\
3 & IAU\_NAME & None & -- & IAU designation within the YSOVAR program \\
4 & simbad\_MAIN\_ID & None & -- & Main identifier for an object \\
5 & SEDclass & None & -- & IR class according to SED slope \\
6 & StandardSet & None & -- & Source in YSOVAR standard set (cluster member with light curve)? \\
7 & Member\_without\_LC & None & -- & Is source cluster member without light curve? \\
8 & n\_36 & ct & $3.6\;\mu$m & Number of datapoints \\
9 & n\_45 & ct & $4.5\;\mu$m & Number of datapoints \\
10 & median\_36 & mag & $3.6\;\mu$m & median magnitude \\
11 & median\_45 & mag & $4.5\;\mu$m & median magnitude \\
12 & mean\_36 & mag & $3.6\;\mu$m & mean magnitude \\
13 & mean\_45 & mag & $4.5\;\mu$m & mean magnitude \\
14 & min\_36 & mag & $3.6\;\mu$m & minimum magnitude in lightcurve \\
15 & min\_45 & mag & $4.5\;\mu$m & minimum magnitude in lightcurve \\
16 & max\_36 & mag & $3.6\;\mu$m & maximum magnitude in lightcurve \\
17 & max\_45 & mag & $4.5\;\mu$m & maximum magnitude in lightcurve \\
18 & delta\_36 & mag & $3.6\;\mu$m & width of distribution from 10\% to 90\% \\
19 & delta\_45 & mag & $4.5\;\mu$m & width of distribution from 10\% to 90\% \\
20 & LC\_ok & None & -- & Variability caused by light curve artifacts? (0: Yes, 1: No) \\
21 & stetson\_36\_45 & None & $3.6\;\mu$m, $4.5\;\mu$m & Stetson index for a two-band lightcurve. \\
22 & redchi2tomean\_36 & None & $3.6\;\mu$m & reduced $\chi^2$ to mean \\
23 & redchi2tomean\_45 & None & $4.5\;\mu$m & reduced $\chi^2$ to mean \\
24 & period & d & -- & Adopted period of lightcurve \\
25 & FAP & None & -- & false alarm probability of adopted period \\
26 & coherence\_time\_36 & d & $3.6\;\mu$m & decay time of autocorrelation function \\
27 & coherence\_time\_45 & d & $4.5\;\mu$m & decay time of autocorrelation function \\
28 & cmd\_angle\_360 & degrees & 3.6 $\mu$m, 4.5 $\mu$m & fitted CMD slope angle in degrees \\
29 & cmd\_angle\_error\_360 & degrees & 3.6 $\mu$m, 4.5 $\mu$m & 1\,$\sigma$ error of fitted CMD slope angle \\
30 & redchi2\_phased\_36 & None & $3.6\;\mu$m & residual scatter in phase-folded fitted light curves \\
31 & redchi2\_phased\_45 & None & $4.5\;\mu$m & residual scatter in phase-folded fitted light curves \\
32 & median\_J & mag & $J$ & median PAIRITEL J band magnitude \\
33 & median\_H & mag & $H$ & median PAIRITEL H band magnitude \\
34 & median\_K & mag & $K_S$ & median PAIRITEL $K_S$ band magnitude \\
35 & Umag & mag & $U$ & G\"unther et al.\ (2012) FLWO U band magnitude \\
36 & e\_Umag & mag & $U$ & Statistical error in Umag \\
37 & Bmag & mag & $B$ & G\"unther et al.\ (2012) FLWO B band magnitude \\
38 & e\_Bmag & mag & $B$ & Statistical error in Bmag \\
39 & Vmag & mag & $V$ & G\"unther et al.\ (2012) FLWO V band magnitude \\
40 & e\_Vmag & mag & $V$ & Statistical error in Vmag \\
41 & Rmag & mag & $R$ & G\"unther et al.\ (2012) FLWO R band magnitude \\
42 & e\_Rmag & mag & $R$ & Statistical error in Rmag \\
43 & Imag & mag & $I$ & G\"unther et al.\ (2012) FLWO I band magnitude \\
44 & e\_Imag & mag & $I$ & Statistical error in Imag \\
45 & Jmag & mag & $J$ & 2MASS J band magnitude \\
46 & e\_Jmag & mag & $J$ & Statistical error in Jmag \\
47 & Hmag & mag & $H$ & 2MASS H band magnitude \\
48 & e\_Hmag & mag & $H$ & Statistical error in Hmag \\
49 & Kmag & mag & $K_S$ & 2MASS $K_S$ band magnitude \\
50 & e\_Kmag & mag & $K_S$ & Statistical error in Kmag \\
51 & 3.6mag & mag & $3.6\;\mu$m & cryogenic Spitzer/IRAC 3.6 $\mu\mathrm{m}$ band magnitude \\
52 & e\_3.6mag & mag & $3.6\;\mu$m & Statistical error in 3.6mag \\
53 & 4.5mag & mag & $4.5\;\mu$m & cryogenic Spitzer/IRAC 4.5 $\mu\mathrm{m}$ band magnitude \\
54 & e\_4.5mag & mag & $4.5\;\mu$m & Statistical error in 4.5mag \\
55 & 5.8mag & mag & $5.8\;\mu$m & cryogenic Spitzer/IRAC 5.8 $\mu\mathrm{m}$ band magnitude \\
56 & e\_5.8mag & mag & $5.8\;\mu$m & Statistical error in 5.8mag \\
57 & 8.0mag & mag & $8.0\;\mu$m & cryogenic Spitzer/IRAC 8.0 $\mu\mathrm{m}$ band magnitude \\
58 & e\_8.0mag & mag & $8.0\;\mu$m & Statistical error in 8.0mag \\
59 & 24mag & mag & $24\;\mu$m & cryogenic Spitzer/MIPS 24 $\mu\mathrm{m}$ band magnitude \\
60 & e\_24mag & mag & $24\;\mu$m & Statistical error in 24mag \\
61 & FX & 1e-7 ct cm$^{-2}$ s$^{-1}$ & X-ray & G\"unther et al.\ (2012) X-ray flux count rate \\
\hline
\end{tabular}
\end{center}
\end{table*}

\subsection{Chandra data}
\label{sect:chandradata}

\label{chandra_obs}
IRAS~20050+2720 was observed with the {\it Chandra} X-ray Observatory in three observations using the ACIS-I camera in VFAINT mode (ObsIDs 6438, 7254, 8492, PI Wolk). The exposure times and observation epochs were 21\,ks (2006 Jan 07), 23\,ks (2006 Dec 10), and 51\,ks (2007 Jan 29), with a total exposure time of 93.95\,ks, i.e. $\sim 26$ hours. The data reduction is described in detail in \cite{Guenther2012}; in short, the three exposures were merged, an automatic source detection was performed, and sources detected at a significance over $2\sigma$ were recorded. For sources with more than 20 X-ray counts a spectral fit was performed, using an optically thin thermal plasma model with a single temperature component with variable temperatue, absorbing column $N_H$, and emission measure. This work was performed in the context of a single-epoch IR study of IRAS~20050+2720 \citep{Guenther2012}, and we use their derived X-ray properties here.

\subsection{PAIRITEL data}
\label{sect:pairiteldata}

\label{pairitel_obs}
\begin{figure}[ht!]
\includegraphics[width=0.45\textwidth]{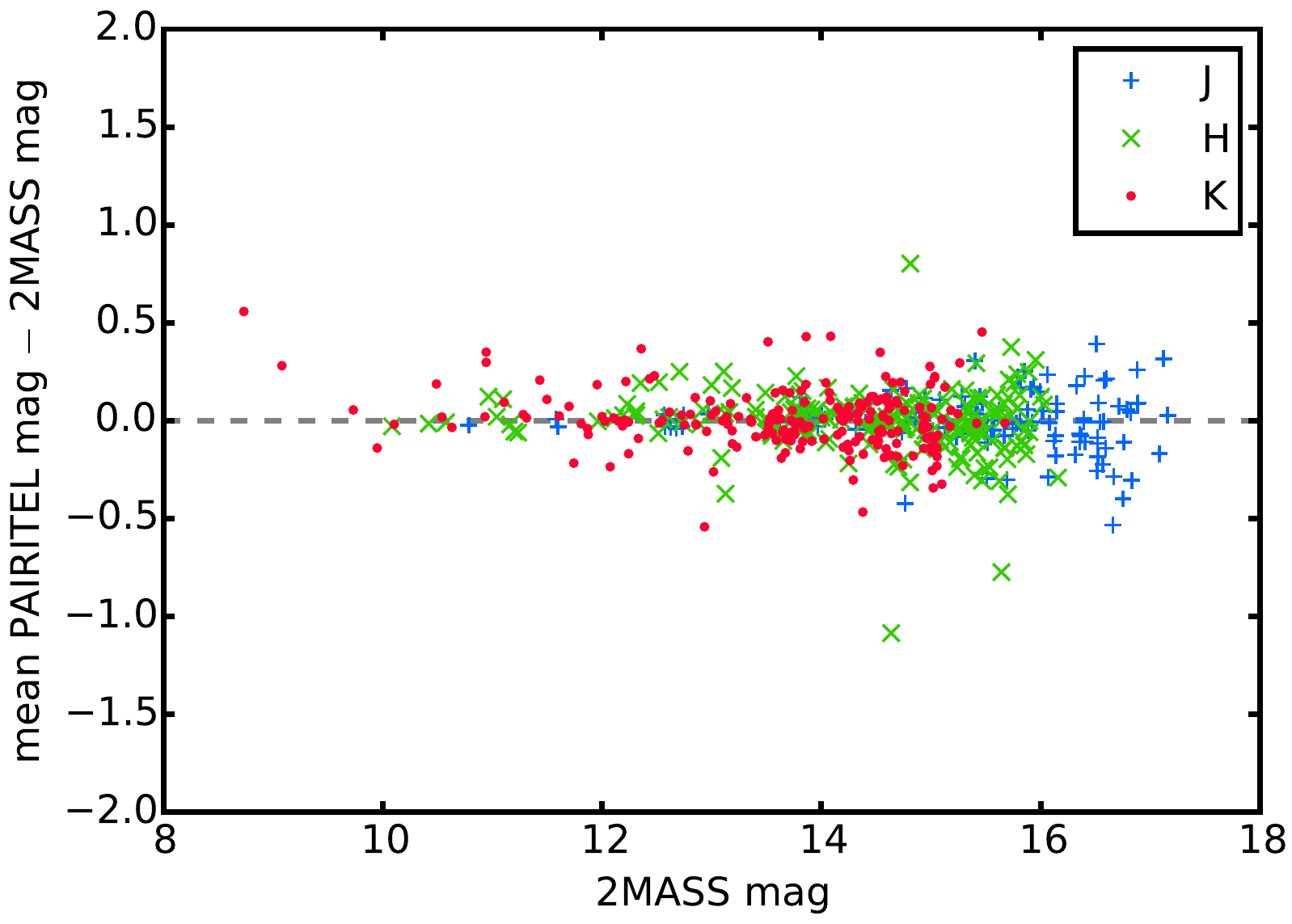}
\caption{Sources in the field of view of the \textit{Spitzer} observations of IRAS 20050+2720 
for which PAIRITEL observations in $J\, H\, K_S$ exist. The measured median PAIRITEL 
magnitudes for each source agree reasonably well with the magnitudes from the 2MASS catalog, with a typical standard deviation in the magnitude differences of ca.\ 0.15 mag.}
\label{fig:Pairitel}
\end{figure}

PAIRITEL is a 1.3m robotic telescope operated by the Smithsonian Astrophysical Observatory on Mount Hopkins, Arizona \citep{Bloom2006}; it was operational from 2004 to 2013. It was equipped with the $JHK_s$ camera previously used for the 2MASS project. Photometric observations of IRAS~20050+2720 were obtained during 31 nights between June 19 2010 and July 10 2012. We used Pyraf, a Python interface to IRAF (Image Reduction and Analysis Facility, \citealt{IRAF}), to perform the data reduction. The photometry was performed with the point-spread function photometry tools of IRAF's DAOPHOT package \citep{Stetson1987}. The observational frames were flat fielded and bias-corrected. We then used \verb+msccmatch+ to match the world coordinate system of the PAIRITEL observations to the 2MASS catalog. We detected the source positions using \verb+daofind+ and performed the psf photometry using tasks in the DAOPHOT package. The extracted instrumental magnitudes needed to be corrected for nightly extinction. We therefore matched the detected sources of each night to the 2MASS catalogue, using a $1^{''}$ matching radius, and performed a linear fit of the instrumental magnitudes to the 2MASS catalog magnitudes. This procedure yielded calibrated $JHK_s$ light curves for 341 sources; among the subset of sources that we will identify as cluster member in section~\ref{membership}, we have collected 77 $JHK_s$ light curves. The error bars of those photometric data points are often of a similar magnitude as the apparent variability, which is why we restrict ourselves in this work to only reporting the median $JHK_s$ source magnitudes measured with PAIRITEL as listed in our data table \ref{tab:tab2}. The median magnitudes of the light curves agree well with the magnitudes in the 2MASS catalog (see Fig.~\ref{fig:Pairitel}), with a standard deviation of 0.14/0.16/0.15~mag in the $J$/$H$/$K_s$ bands, respectively.

\section{Data classification}
\label{sect:dataclassification}

\label{classification}

\subsection{Cluster membership criteria}
\label{sect:clustermembershipcriteria}

\label{membership}

In order to compare the properties of YSOs in the different young clusters of the YSOVAR project, we have selected a uniform system to identify likely members of these clusters. This system is based on an IR selection to identify YSOs with disks, as well as an additional X-ray selection to identify young disk-free stars. 

We follow the IR selection by \cite{Gutermuth2009}, who used cryogenic-era \textit{Spitzer} data of the YSOVAR clusters to perform a multi-color selection of candidate members. This is thought to be a statistically well-defined sample with low contamination rates, as discussed in \cite{Gutermuth2008} and \cite{Gutermuth2009}. We adopt all sources classified by \cite{Gutermuth2009} to have an IR excess (spectral energy distribution (SED) class I or II) as members.

Furthermore, we use X-ray observations of IRAS~20050+2720 to identify young diskless stars. We require an X-ray detection of at least $2\sigma$ significance and a positional match to a source identified in the cryo-era \textit{Spitzer} catalog, as well as a star-like SED shape (see \ref{SEDclass}). The sources which fulfill these requirements are added to our list of members.

In the scope of the YSOVAR project we refer to a thus constructed list of members as the \textit{standard set of members}. These are the sources for which we analyze mid-IR light curves in detail. Light curves of sources that are fainter than 16th magnitude in [3.6] or [4.5] are generally dominated by noise; they are usually prevented from entering the standard set because their spectral energy distributions are not well enough measured to have been identified as a YSO by \cite{Gutermuth2009}. Since we are also mainly interested in time-variable processes here, we restrict ourselves to sources for which we actually have a light curve with at least 5 epochs in the post-cryogenic \textit{Spitzer} data in either the [3.6] or [4.5] band. It is possible for a source to be a cluster member according to our definition, but have a light curve with too few data points to allow a useful analysis.

  For our cluster IRAS\,20050+2720, we have 187 sources that fulfill 
  the membership criteria, and 181 of those have at least five data 
  points in either their [3.6] or [4.5] band light curve.
  We will refer to those 181 sources as the standard set of members 
  throughout this paper. In our data table (Tab.~\ref{tab:tab2}) they 
  are marked with the flag \mbox{``StandardSet''}. To identify the 6 sources
  that fulfill the membership criteria, but do not have light curves, we 
  also list a flag ``Member\_without\_LC'' in Tab.~\ref{tab:tab2}.
  We summarize the numbers of cluster members in Table~\ref{membertable}.

  \begin{table}
  \caption{Numbers of identified cluster members in IRAS 20050+2720.}
  \label{membertable}
  \begin{tabular}{p{7cm} c}
  \hline \hline
  Type of group        & number \\ \hline
  members & 187 \vspace{0.2cm}\\ 
  ``Standard Set of members'' (members with $\ge5$ post-cryogenic light curve epochs) & 181 \vspace{0.2cm}\\
  \hspace*{0.3cm}Standard Set of members with 2-band light curves & 138 \vspace{0.2cm}\\
  \hspace*{0.3cm}Standard Set of members with [3.6] light curve only & 25 \vspace{0.2cm}\\
  \hspace*{0.3cm}Standard Set of members with [4.5] light curve only & 18 \vspace{0.2cm}\\
  additional X-ray identified candidate YSOs from G\"unther et al. (2012) & 18 \vspace{0.2cm}\vspace{-0.2cm}\\
  \hline
  \end{tabular}
  \end{table}

Recent publications on other YSOVAR clusters \citep{Rebull2014, Guenther2014} also analyze the statistical properties of all sources with light curves with at least five data points, regardless of their cluster membership status. This group of sources is referred to as the \textit{standard set for statistics}. In the present work, however, we restrict ourselves to the member sources, because especially in the cluster IRAS~20050+2720, the majority of the fainter sources are badly affected by image artifacts (see section \ref{artefacts}). For this reason, we defer a discussion of variables beyond the standard set of members to a later paper in the YSOVAR series that will compare variability properties across several clusters.

  Apart from the standard set, we will also take a separate look at 18 additional 
  sources identified by \cite{Guenther2012}. Those sources have not been detected 
  in enough infrared bands to have a well-constrained SED, but they have been detected 
  in X-rays and are all located close to the center of the cluster.
  \cite{Guenther2012} call these objects ``X-ray identified YSOs'' (``XYSOs''), and 
  we follow their nomenclature.
  Sources at the center of clusters are usually subject to strong extinction caused by the surrounding gas and dust and
  are therefore hard to detect in optical or infrared bands; however, X-rays can typically 
  penetrate much larger column densities of absorbing gas. It is therefore not unusual to
  detect such sources at the cluster cores mainly in X-rays or in radio bands
  \citep{Hughes2001, Comito2007, Schneider2009, Pravdo2009}.
  We have collected warm-era \textit{Spitzer} light curves 
  for 10 of these sources in IRAS~20050+2720, which we will discuss in section \ref{xysos}.

\subsection{SED classification}
\label{sect:sedclassification}

\label{SEDclass}

During the different evolutionary stages of a young stellar object, its spectral energy distribution undergoes characteristic changes. Very young objects are still embedded in the envelope of gas and dust from which they are forming, and their SEDs peak at long wavelengths $>10\,\mu$m (see for example \citealt{Lada1984}, \citealt{Adams1987}, \citealt{Greene1996}). Objects with thick circumstellar disks will have an SED which is dominated by the flux at long wavelengths, while objects with thinner and dissolving disks will be dominated by the flux from the central object, but display NIR contributions from the disk emission. Once the disk is mostly dissipated, the SED is determined by the photosphere of the star.

  \begin{table}
  \caption{Assigned SED classes of IRAS 20050+2720 members with light curves (down); 
  previously assigned classes from \cite{Gutermuth2008} listed across.}
  \label{slopes}
  \begin{tabular}{c  c | cccccc | c}
  & & \multicolumn{5}{r}{Gutermuth} & \\
  & Class        & I & I* & II & II* & III & unclassified & Sum \\ \hline
  \multirow{6}{*}{\begin{sideways} YSOVAR \end{sideways} } 
  & I            & 35 & 6 & 6 & 0 & 0 & 0 & 47 \\
  & F            & 13 & 2 & 25 & 0 & 0 & 0 & 40 \\
  & II           & 1 & 0 & 65 & 4 & 0 & 0 & 70 \\
  & III          & 0 & 0 & 1 & 0 & 17 & 6 & 24 \\
  & unclassified & 0 & 0 & 0 & 0 & 0 & 0 & 0 \\ \hline
  & Sum   & 49 & 8 & 97 & 4 & 17 & 6 & 181 \  \end{tabular}
  \end{table}

We classify these SED slopes in a similar manner to \cite{Wilking2001}, using the slope of the SED in the near- and mid-IR. Specifically, we use the data points from 2 to 24 $\mu$m as given in \cite{Gutermuth2008}, add the means of the collected post-cryogenic \textit{Spitzer} light curves at 3.6\,$\mu$m and 4.5\,$\mu$m as additional data points (i.e.\ in addition to the cryogenic-era 3.6\,$\mu$m and 4.5\,$\mu$m data points), and calculate the slope $\alpha = d\log \lambda F_\lambda / d\log \lambda$. The inclusion of a second epoch of [3.6] and [4.5] measurements reflects that our sources are expected to be variable; however, if only the cryogenic [3.6] and [4.5] data points are used, SED slopes generally do not change strongly enough to change the assigned class of a source. We refer to slopes with $\alpha > 0.3$ as Class I, $0.3 > \alpha > -0.3$ as Flat (F), $-0.3 > \alpha > -1.6$ as Class II, and $\alpha < -1.6$ as Class III. Details on this procedure are given in \cite{Rebull2014}, Appendix B. As discussed there, fitting the observed SEDs (versus fitting de-reddened SEDs) is sufficient for our dataset, because we only fit SED data points redward of $K_S$. Substantial extinction of $A_V \sim 40$ would be needed to have a source be misclassified in our scenario. 

Indeed, the overlap of our classification with the de-reddened classification scheme of \cite{Gutermuth2008} is satisfactory. We show a correspondence table between the two classification schemes for the members of IRAS~20050+2720 with at least 5 data points in their post-cryogenic \textit{Spitzer} light curves in Table~\ref{slopes}.

\subsection{Variability classification}
\label{sect:variabilityclassification}

\begin{figure}[ht!]
\includegraphics[width=0.45\textwidth]{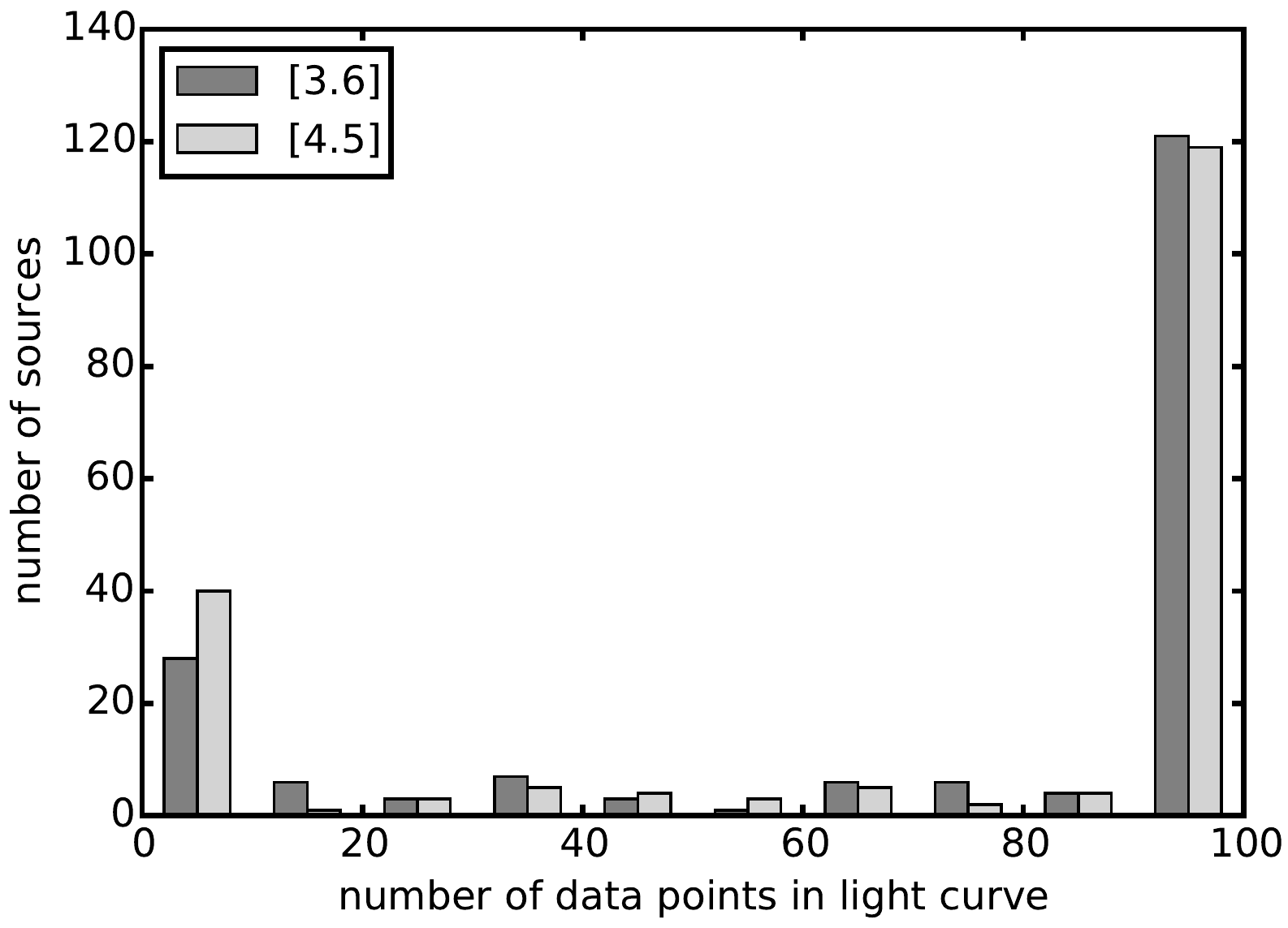}
\caption{Number of sources in the standard set of members that contain a certain number of data points in the [3.6] and [4.5] light curves. Typically these light curves contain 100 data points each.}
\label{fig:LCpoints}
\end{figure}

We investigate various types of time-variability in the \textit{Spitzer} [3.6] and [4.5] light curves of our sources. Typically, the standard set members have around 100 data points in their [3.6] and [4.5] light curves each; see Fig.~\ref{fig:LCpoints} for an overview. Sources that are positioned at the very edge of the \textit{Spitzer} field of view tend to have $\le 10$ light curve epochs. As listed in Table~\ref{membertable}, 138 of the standard set members have light curves in both the [3.6] and [4.5] band, since the cluster core where most of the members are located was targeted by both IRAC channels. 25 standard set members only have light curves in the [3.6] band, 18 only in the [4.5] band.

\begin{figure*}[ht!]
\includegraphics[width=0.32\textwidth]{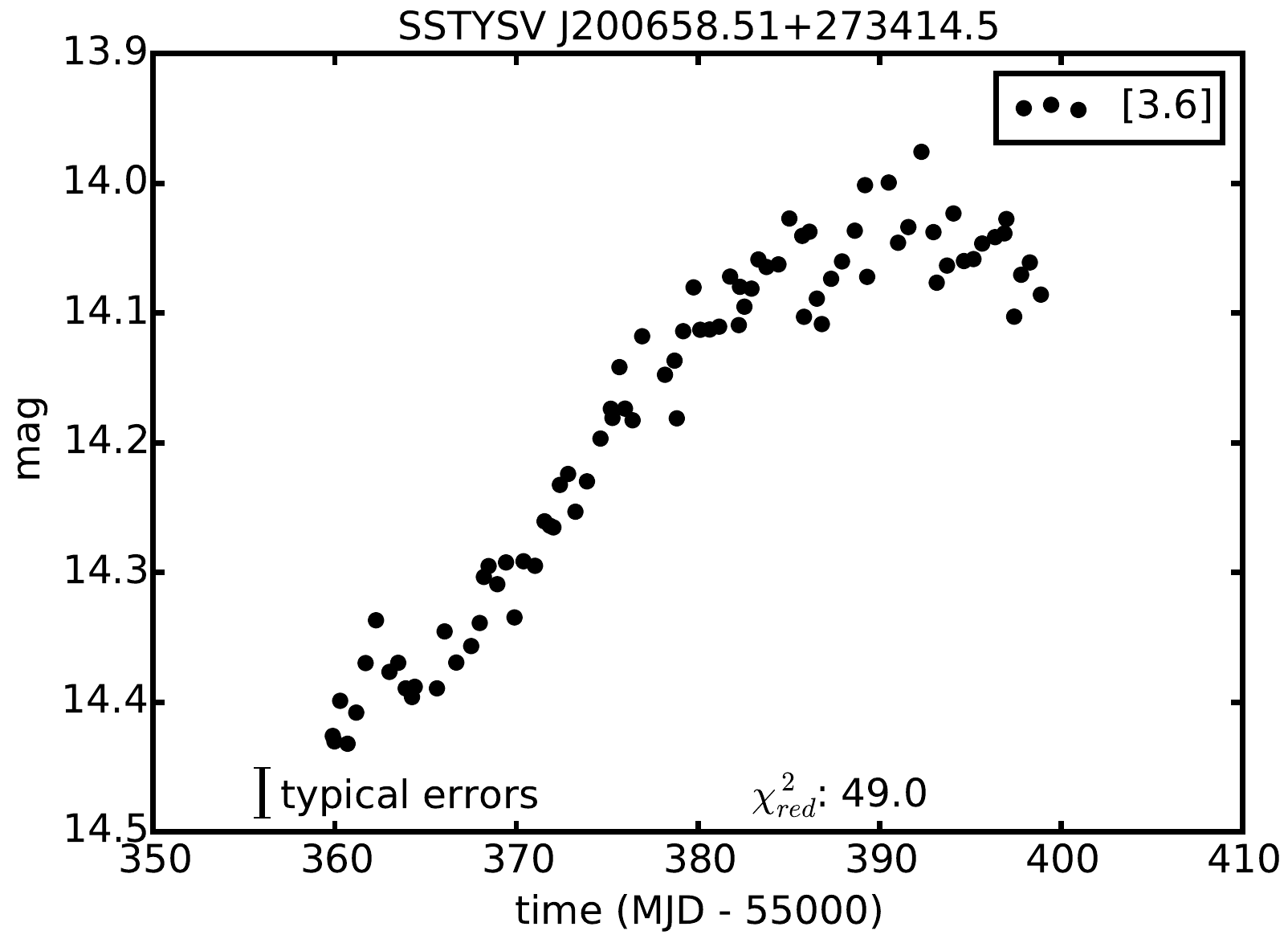}
\includegraphics[width=0.32\textwidth]{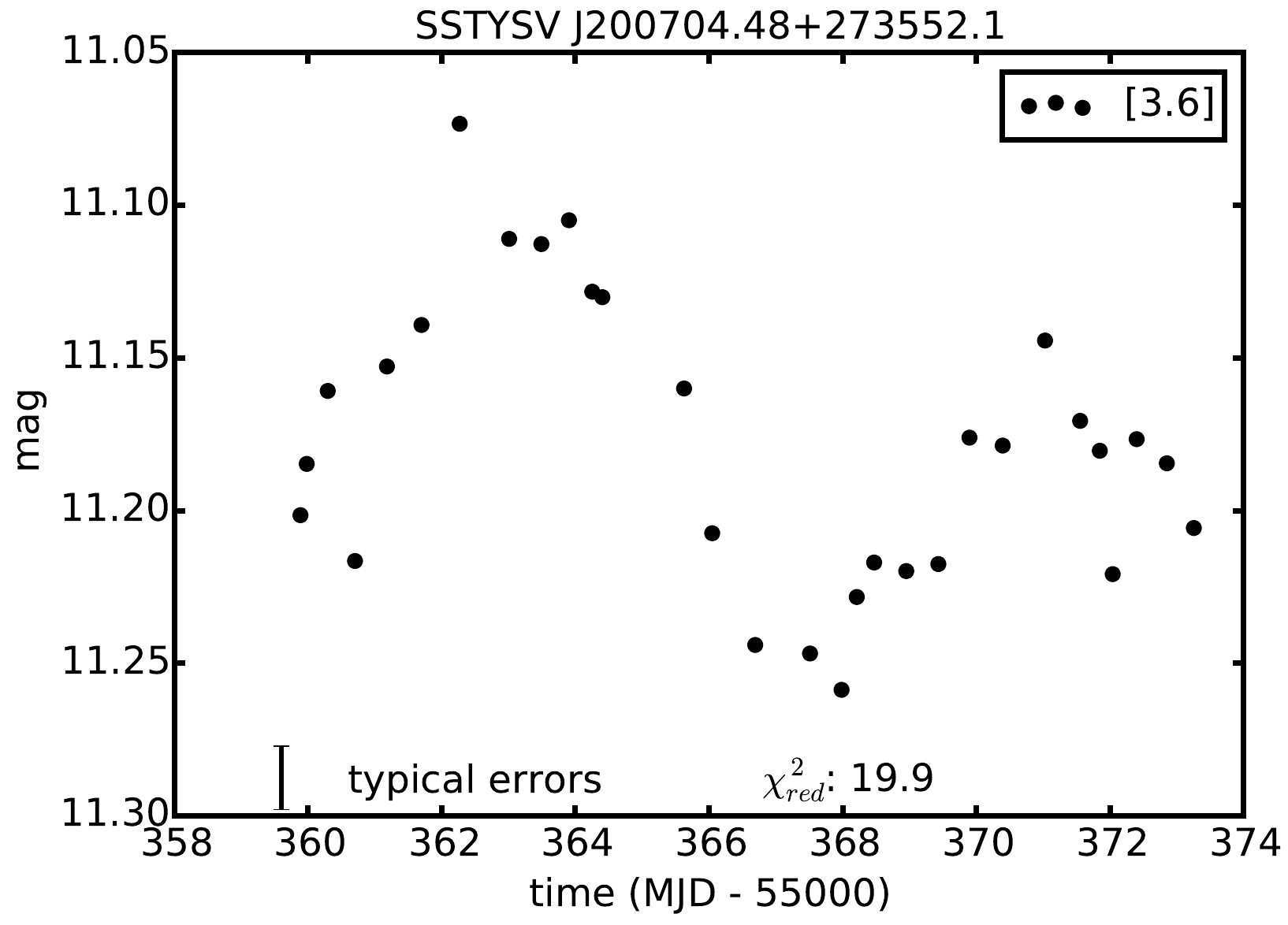}
\includegraphics[width=0.32\textwidth]{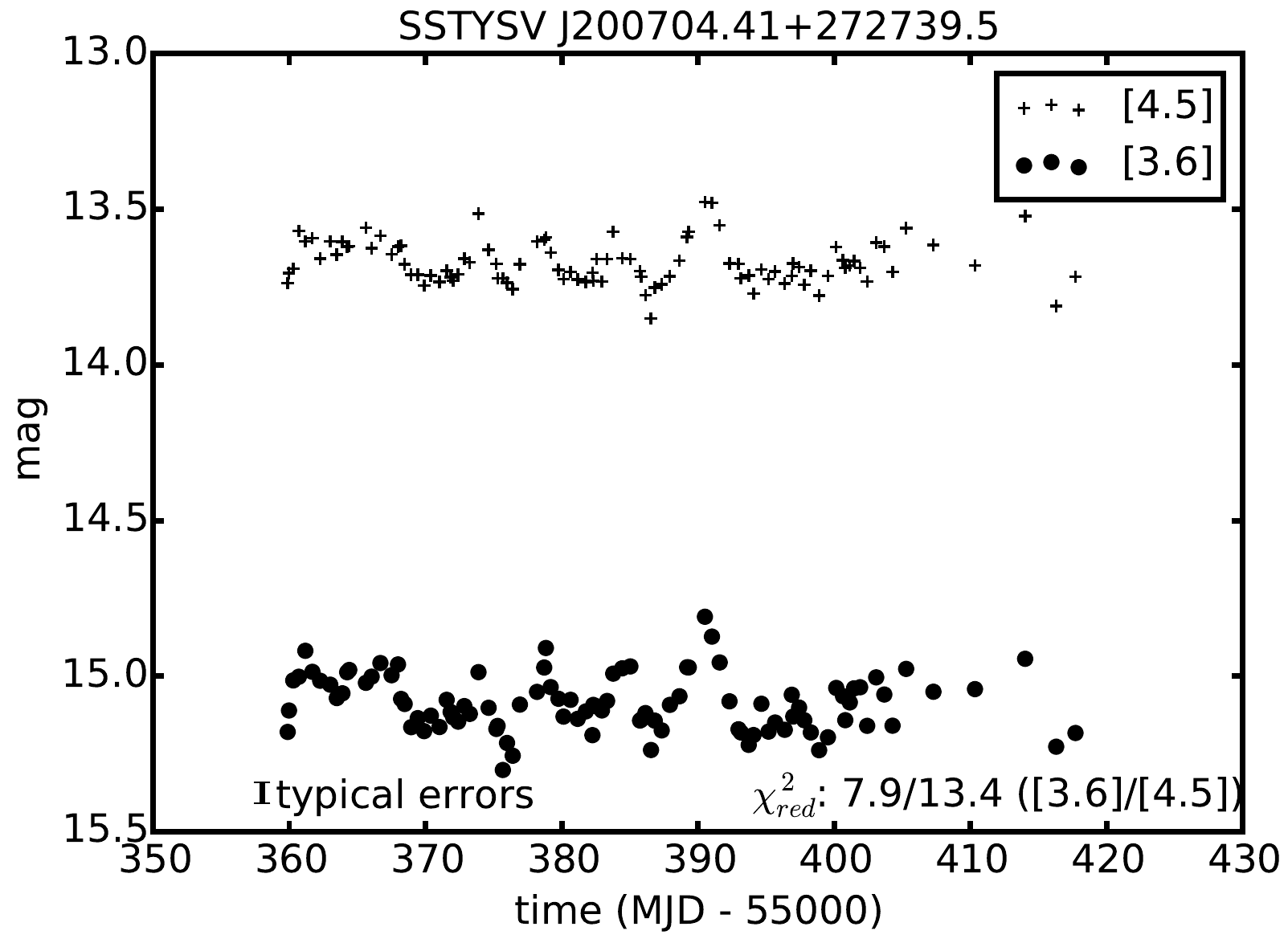}
\caption{Examples of sources displaying $\chi^2$ variability. Typical errors on the light curve data points and reduced $\chi^2$ with respect to the mean of the light curve given at the bottom of each plot.}
\label{var_chisq}
\end{figure*}

\begin{figure*}[ht!]
\includegraphics[width=0.32\textwidth]{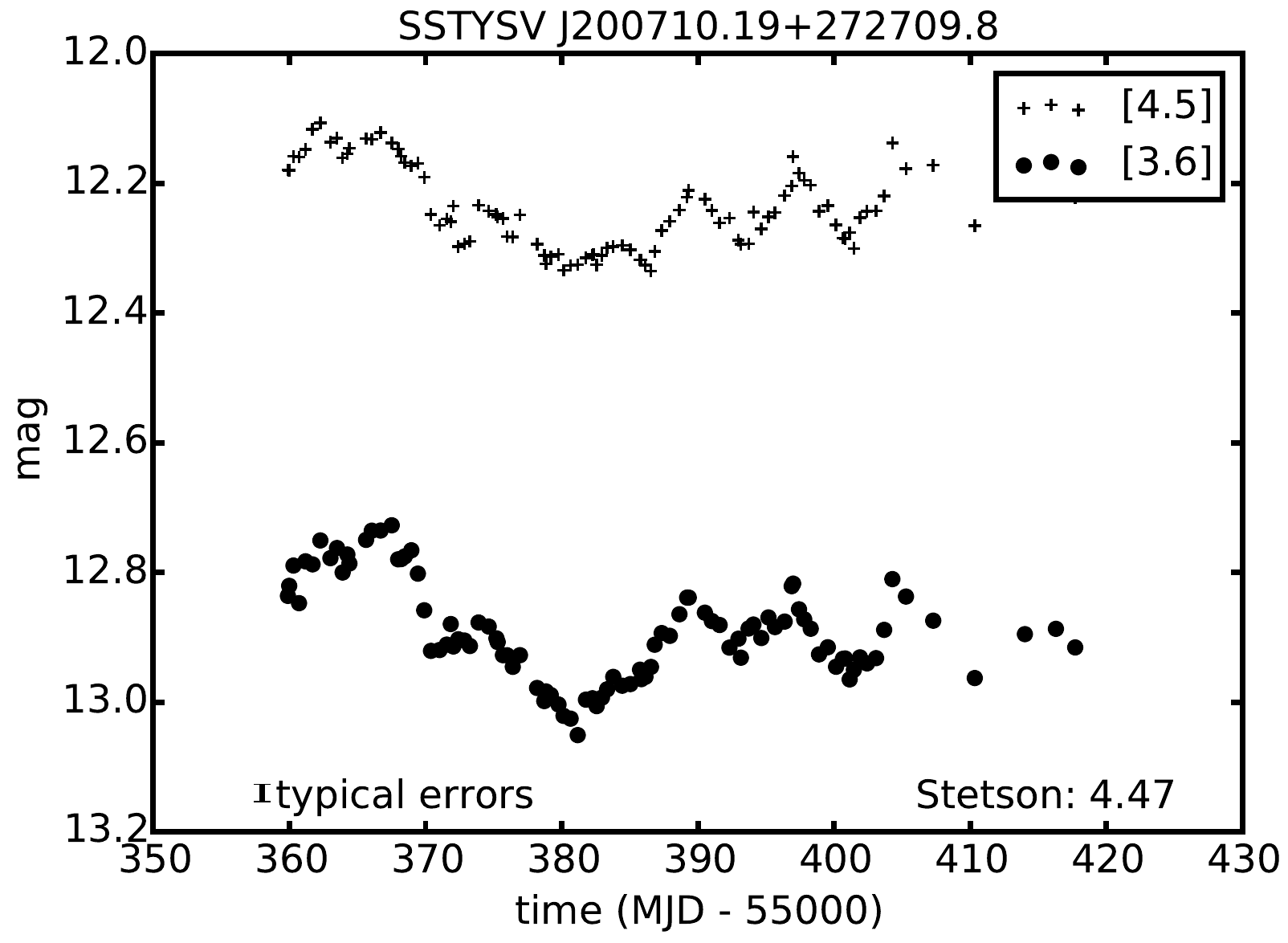}
\includegraphics[width=0.32\textwidth]{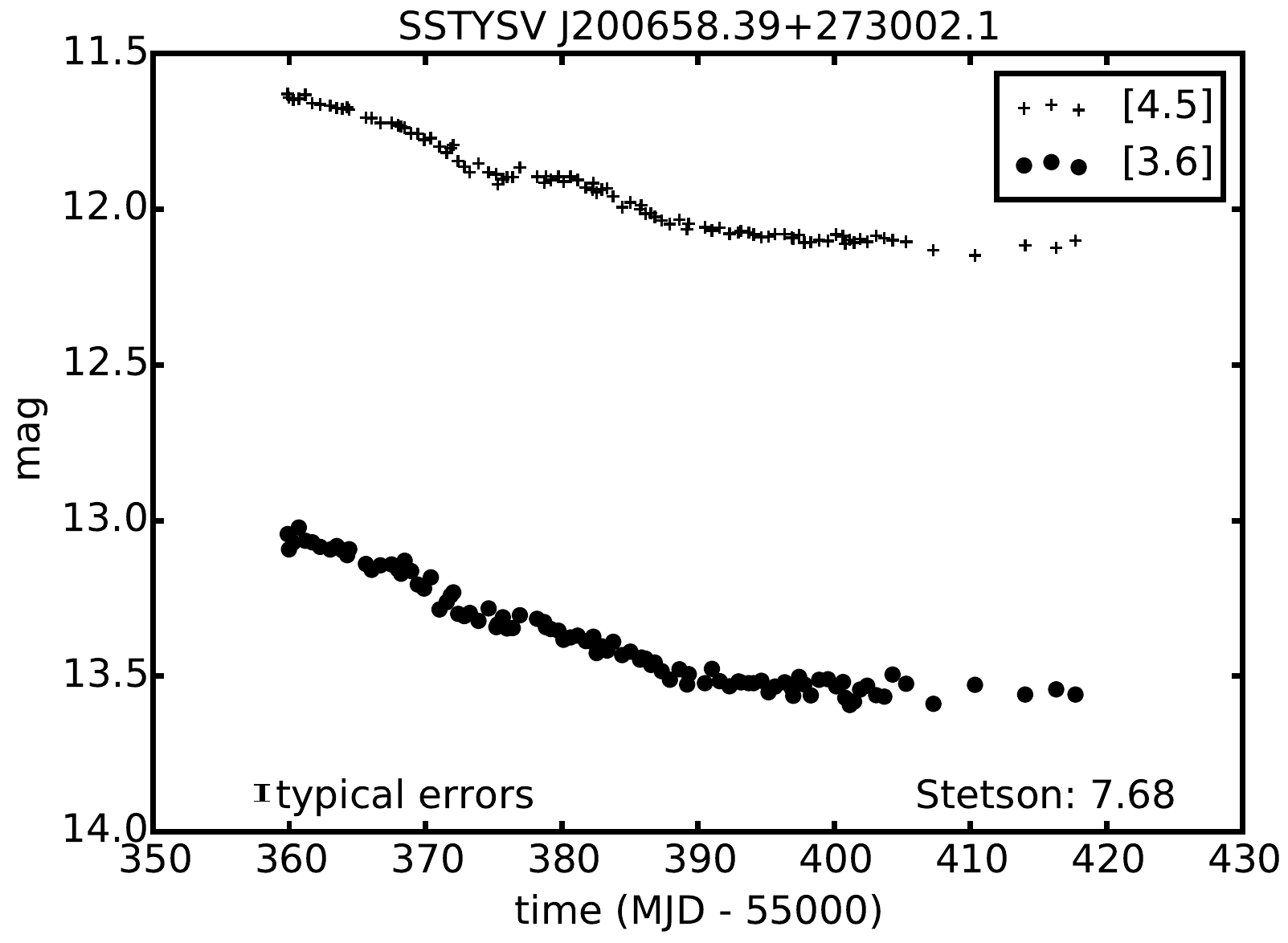}
\includegraphics[width=0.32\textwidth]{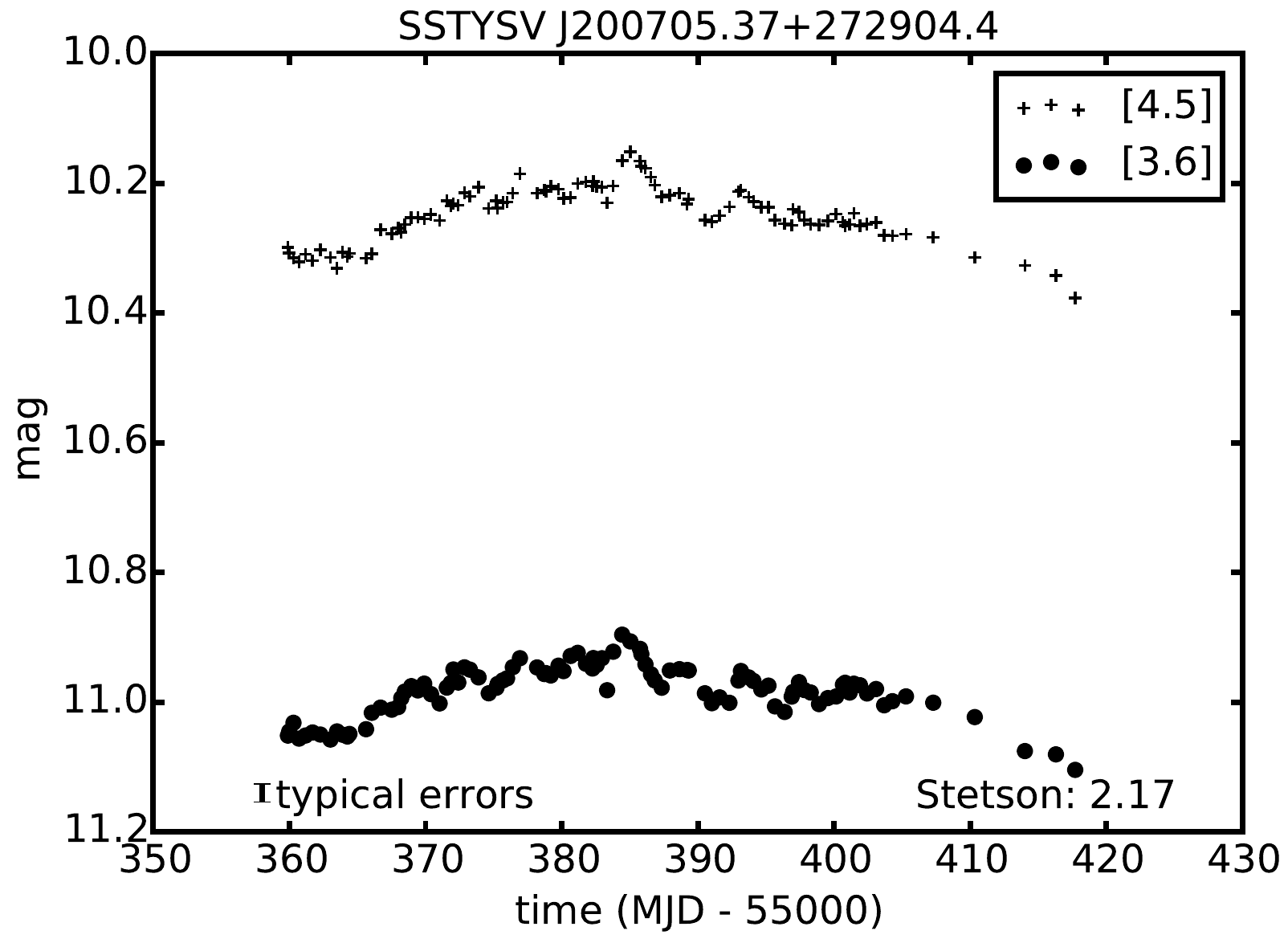}
\caption{Examples of sources displaying correlated variability. Typical errors on the light curve data points and Stetson index given at the bottom of each plot.}
\label{var_stetson}
\end{figure*}
\begin{figure*}[ht!]
\includegraphics[width=0.32\textwidth]{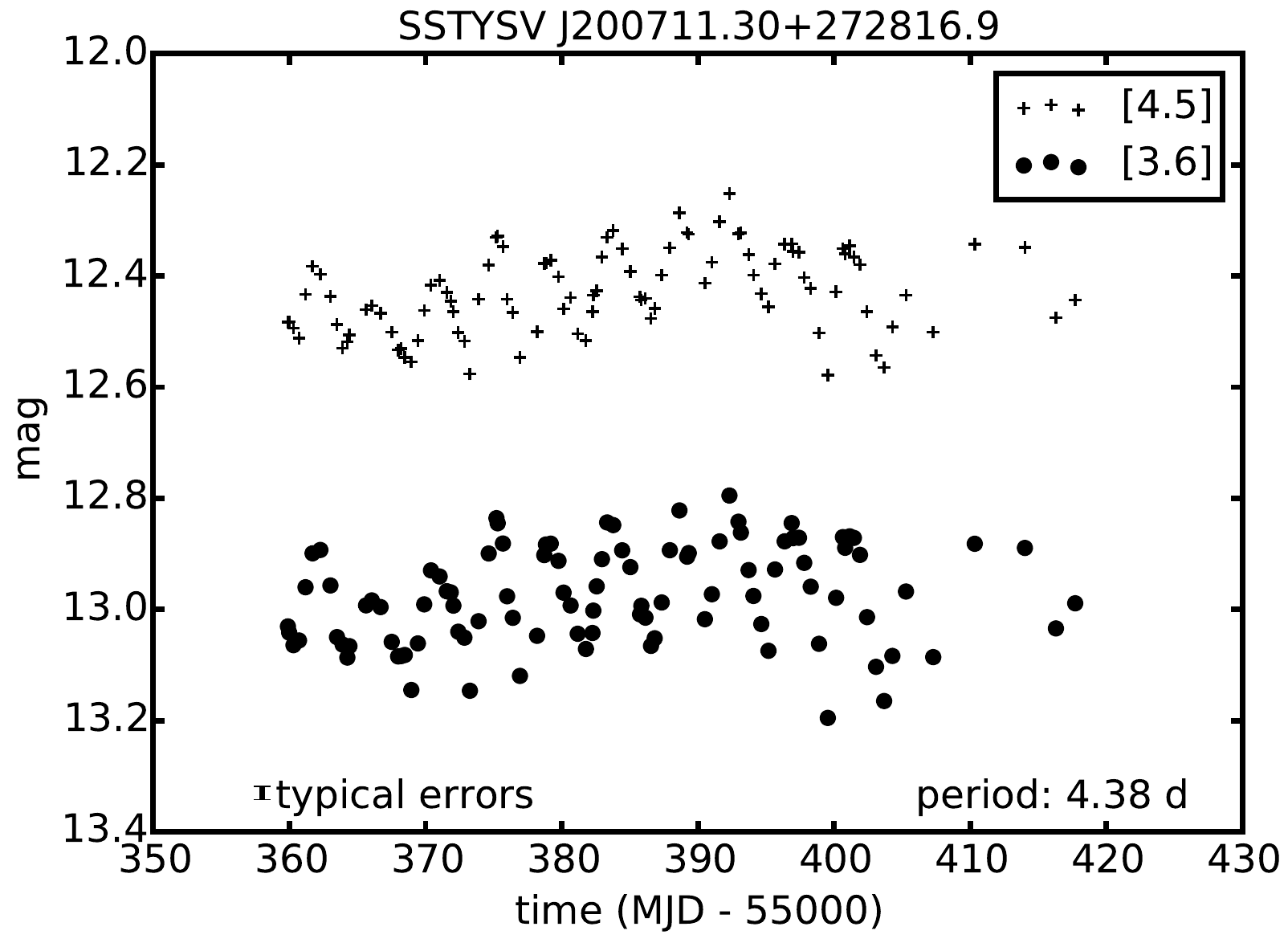}
\includegraphics[width=0.32\textwidth]{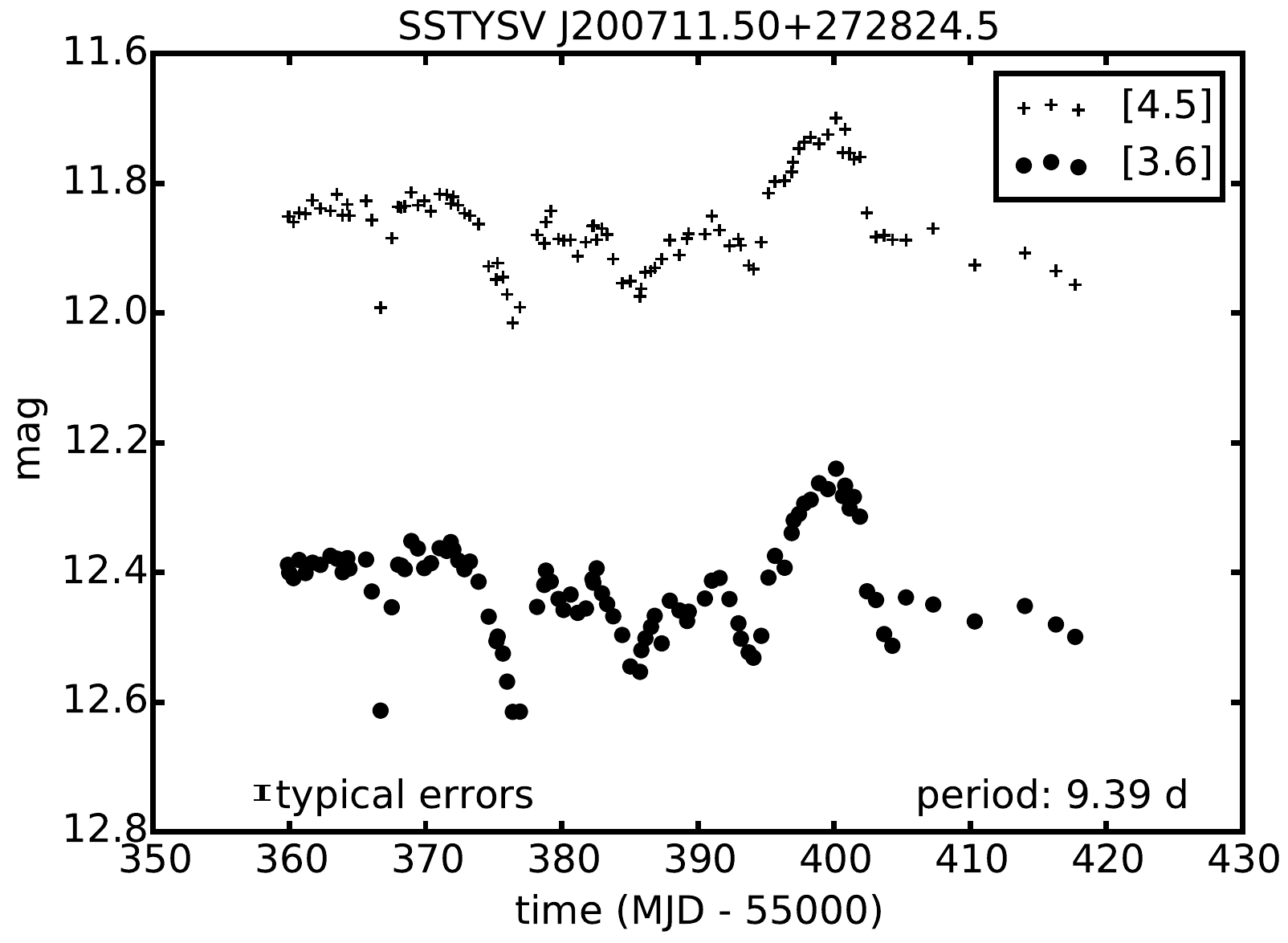}
\includegraphics[width=0.32\textwidth]{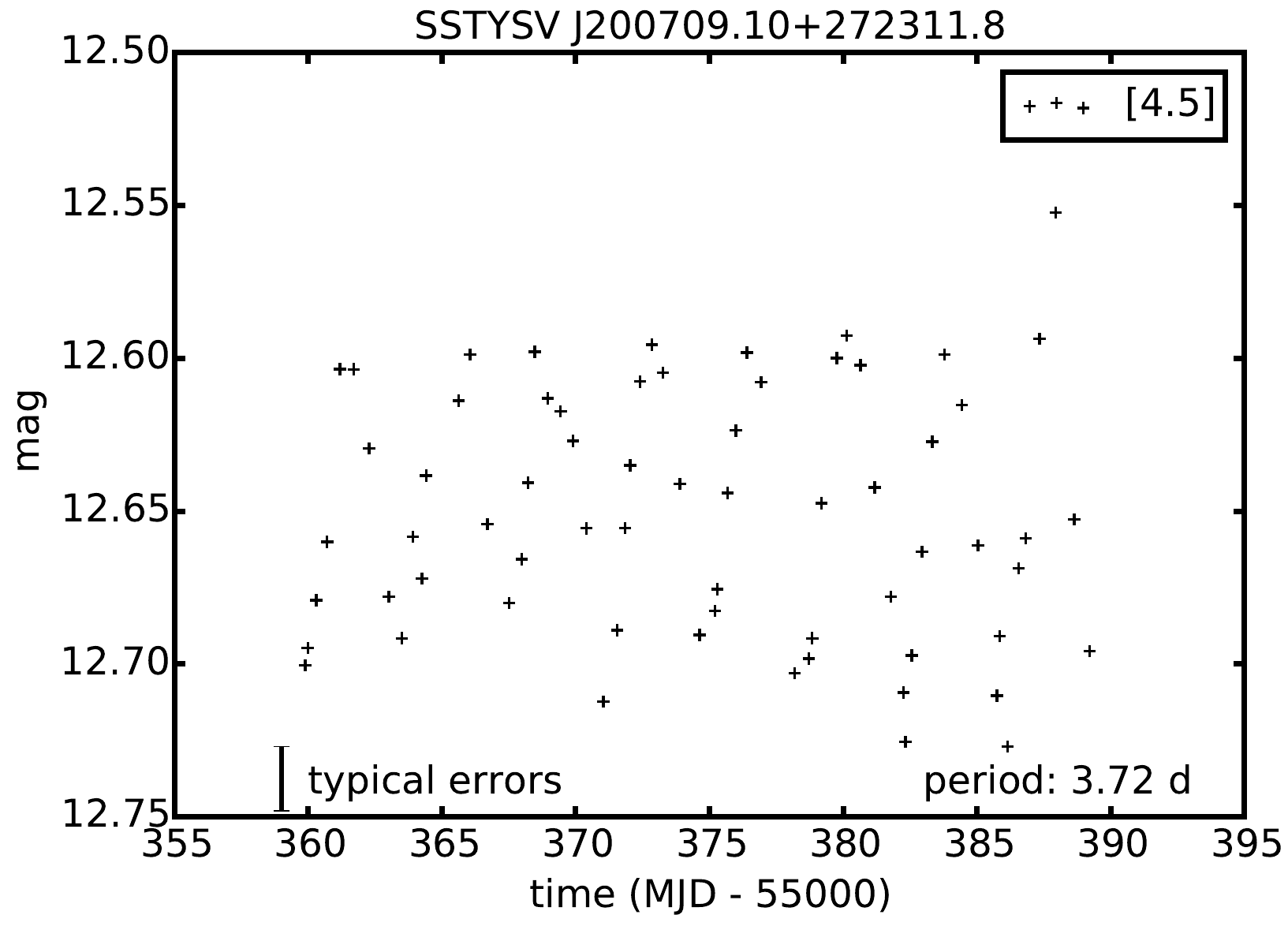}
\caption{Examples of sources displaying a periodic signal in their variability; 
this can be overlaid by other, non-periodic signals like for the source 
in the middle plot. Typical errors on the light curve data points, false alarm 
probaility, and detected period given at the bottom of each plot.}
\label{var_periodic}
\end{figure*}
We test for three types of variability in our sample of mid-IR light curves: periodic variability, ``$\chi^2$-variability'', and correlated variability if simultaneous light curves in more than one band are available. These methods are described in detail in \cite{Rebull2014}; we give a short overview here and present typical cases of those variability behaviors.

\subsubsection{Chi-squared variability test}
\label{sect:chisquaredvariabilitytest}

On any given light curve in our sample, we perform a chi-squared test for variability by calculating the reduced chi-squared statistic with respect to the mean for the time series:

\begin{equation}
\chi_{red}^2 = \frac{1}{N-1} \sum^{N}_{i=1}\frac{(mag_i - \overline{mag})^2}{\sigma_i^2}
\end{equation}

with $N$ being the number of data points in the time series, $mag_i$ the individual measured magnitudes, $\sigma_i$ their individual errors, and $\overline{mag}$ the mean of the measured magnitudes. As a conservative approach, we identify a source as $\chi^2$-variable if its light curve in the 3.6\,$\mu$m or 4.5\,$\mu$m band yields $\chi_{red}^2\geq 5$. As demonstrated by \cite{Rebull2014} in their section 5.2, this corresponds to a significance larger than $3\sigma$ for source variability.

We show light curves of several sources that exceed this $\chi_{red}^2$ variability threshold in Figure~\ref{var_chisq}.

\subsubsection{Correlated variability}
\label{sect:correlatedvariability}

\begin{figure*}[ht!]
\hspace{1cm}
\includegraphics[width=0.4\textwidth]{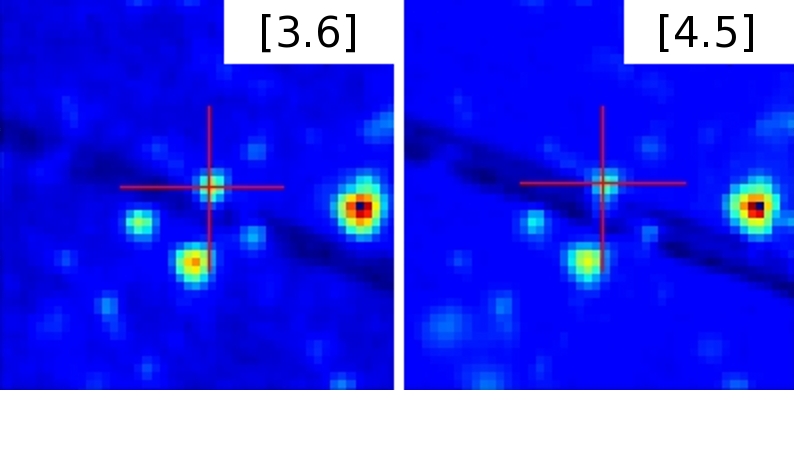}
\hspace{1cm}
\includegraphics[width=0.35\textwidth]{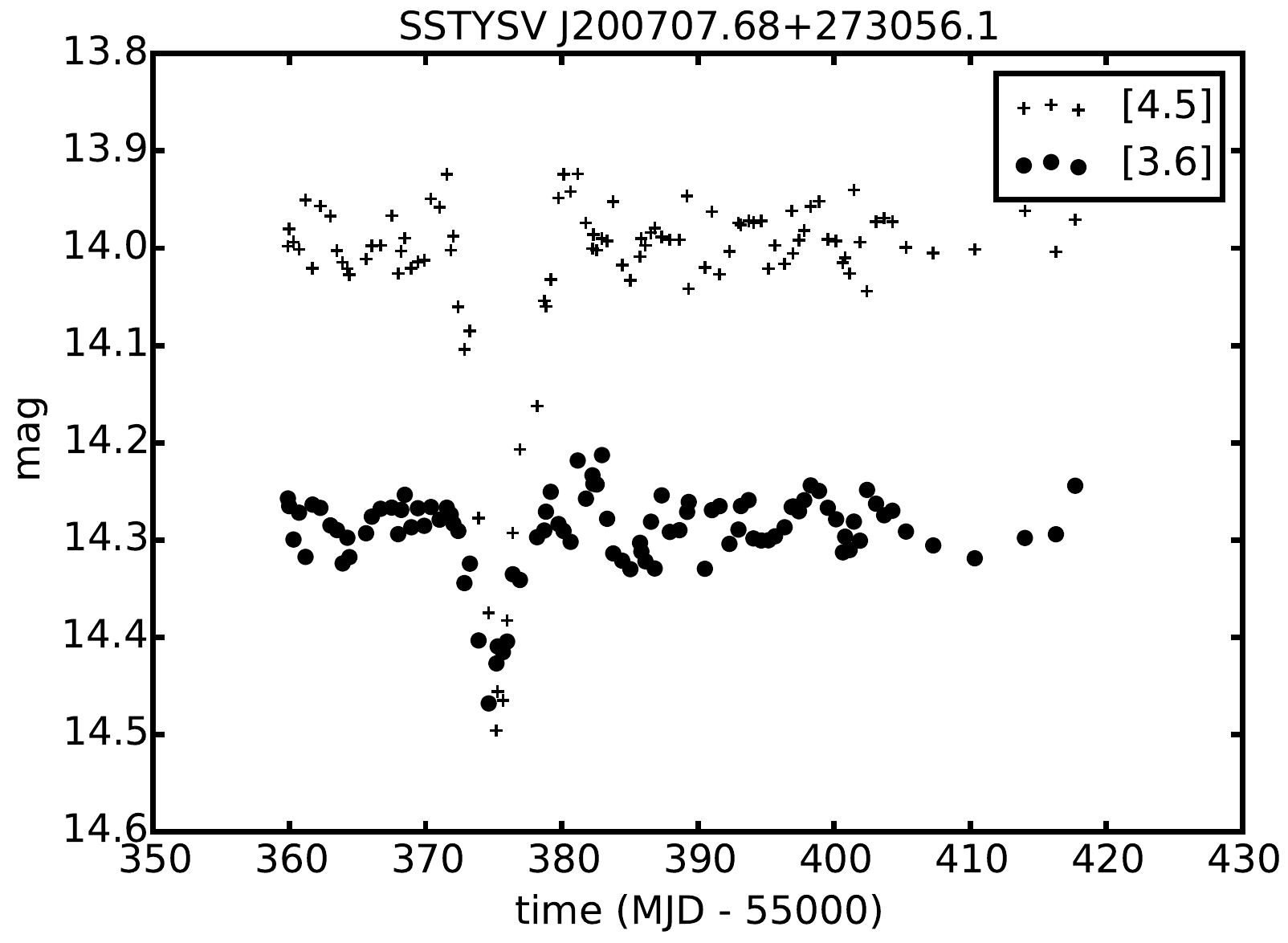}
\caption{Example of a source with spurious variability in the light curve around $MJD = 55\,375$, 
induced by column pulldown moving across the source due to the rotation of the field of view over 
time (see section 2.5 in \cite{Rebull2014} for a detailed discussion of this instrumental effect).}
\label{var_spurious}
\end{figure*}
If simultaneous light curves in two bands are available, the Stetson index can be used as a measure for variability \citep{Stetson1996}. It is calculated as

\begin{equation}
S = \frac{1}{N} \sum^N_{i=1} sgn(P_i) \sqrt{|P_i|}
\end{equation}

with $N$ being the number of paired observations in the two bands and $sgn(P_i)$ is the sign of the quantity $P_i$. $P_i$, the product of the normalized residuals for the i-th two-band pair of observations is given by

\begin{equation}
P_i = \frac{N}{N-1}\frac{mag_{i, [3.6]} - \overline{mag_{[3.6]}}}{\sigma_{i, [3.6]}}\frac{mag_{i, [4.5]} - \overline{mag_{[4.5]}}}{\sigma_{i, [4.5]}}
\end{equation}

A large positive Stetson index occurs for light curves in which correlated variability is present in both bands, while a negative Stetson index indicates anticorrelated behavior in the two bands. As shown by \cite{Rebull2014} in their section 5.1, a value of $S>0.9$ can be used as a robust indicator for variability of a given source. In addition, \cite{Rebull2014} showed in their section 5.4 that the Stetson test is, as expected, generally more sensitive to variability features in YSOs than a one-band $\chi^2$ test.

We show a set of sources with Stetson-variable light curves in Figure~\ref{var_stetson}.

\subsubsection{Periodic variability}
\label{sect:periodicvariability}

\label{periodicity}
Finally, we test sources for periodic variability. Sources with low-amplitude periodic behavior may not be picked up by our (by design) conservative $\chi^2$ threshold for variability, and they might fail the Stetson test for variability because the variability amplitude is too low or only single-band data is available. To search for periodic changes in unevenly spaced data, we decided to use the periodogram analysis provided by the NASA Exoplanet Archive Periodogram Service\footnote{http://exoplanetarchive.ipac.caltech.edu/\newline cgi-bin/Periodogram/nph-simpleupload} \citep{Akeson2013}. Three different period search algorithms are provided by this service: Lomb-Scargle \citep{Scargle1982}, Box-fitting Least Squares \citep{Kovacs2002}, and the Plavchan Algorithm \citep{Plavchan2008periods}. They all have different strengths and weaknesses with repect to sinusoidal vs.\ non-sinusoidal variations as well as periodic behavior on top of other light curve trends. After testing all three, the Lomb-Scargle approach turned out to be most suitable for our purposes; see discussion in \cite{Rebull2014} for details. 

For the period search, we require the light curve of a given source to contain at least 20 epochs. We ran a Lomb-Scargle period search for the $3.6\,\mu\mathrm{m}$ and $4.5\,\mu\mathrm{m}$ light curves, and, if both are available for the same source, also for the [3.6]-[4.5] color light curves. Because our total epoch coverage is 40 days, we restricted our search to periods between 0.1 and 15 days and generally required a false alarm probability (FAP) of less than 0.03; for details see \cite{Rebull2014}. Additionally, the phased light curves were checked by eye for consistency. If a significant period was detected, we gave preference to the period detected in the $3.6\,\mu\mathrm{m}$ band, because this band is generally less affected by long-term trends (such as changes in absorption by disk material in the line of sight); long-term trends make it hard for the Lomb-Scargle algorithm to identify underlying periods. If no significant period in $3.6\,\mu\mathrm{m}$ was detected, we proceded as follows: if a significant period is detected in the $4.5\,\mu\mathrm{m}$ band, we report that period. If that band also does not show a significant periodicity, we test if the [3.6]-[4.5] color light curves show periodicity and report the period if it is significantly detected. In Table 2 we specify from which channel the reported period is derived. In total, we find 33 sources among the standard set members that display a significant periodicity. 21 sources show periodicity in the [3.6] band. Out of those, 11 sources also show periodicity in the [4.5] band. An additional 9 sources show periodicity \textit{only} in the [4.5] band. 8 sources show periodicity in the [3.6]-[4.5] color light curves, among those 3 objects for which the periodicity in exclusively found in the color light curve.

We show light curves of several sources with detected periodic changes in Figure~\ref{var_periodic}.

\begin{figure}[ht!]
\includegraphics[width=0.5\textwidth]{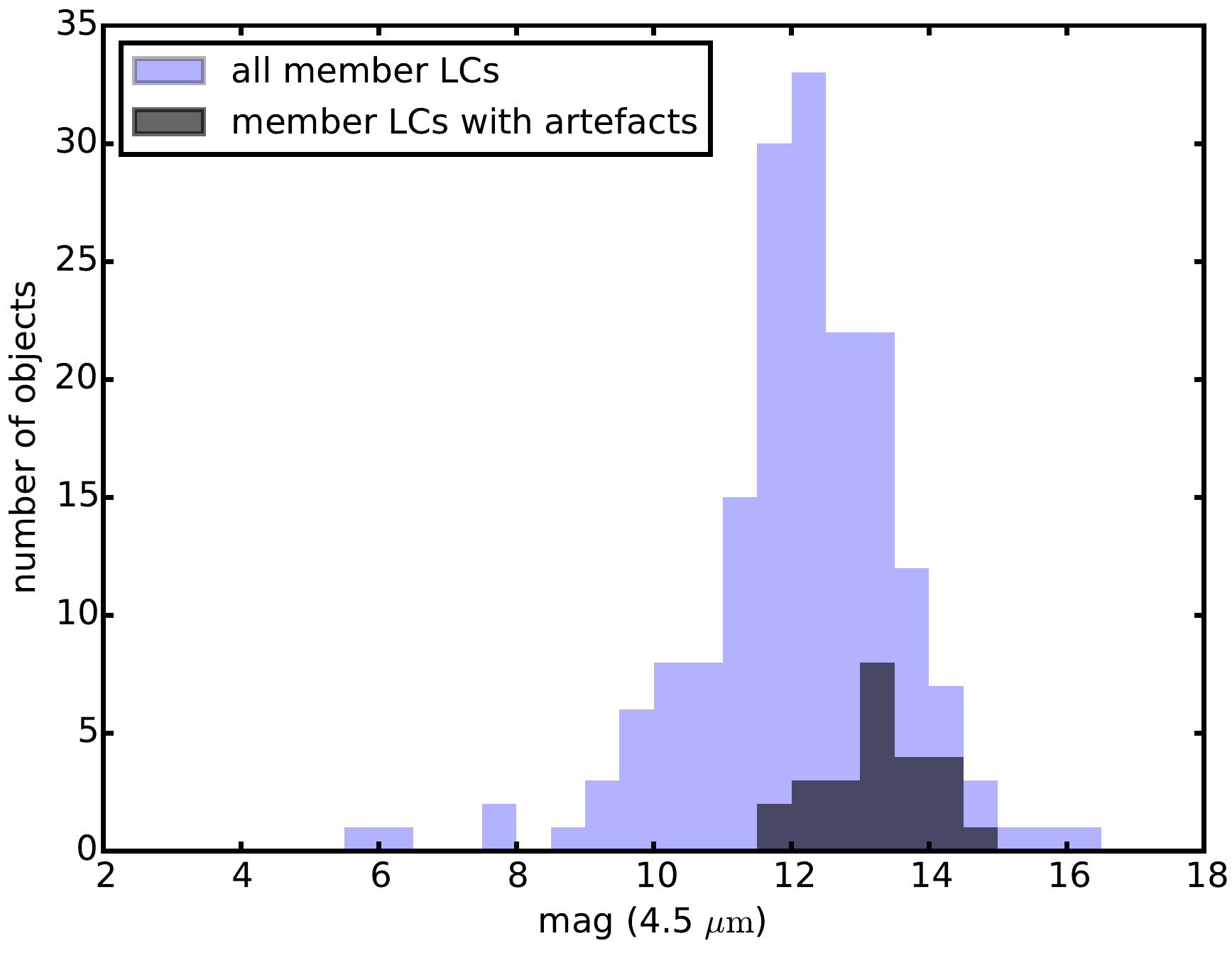}
\caption{Brightness distribution of member sources with significant light curve artifacts (dark grey) 
versus all member sources (light blue). As expected, mostly faint sources around \mbox{$m[4.5] = 13-14\mbox{mag}$} 
are significantly affected by readout artifacts. }
\label{hist_spurious}
\end{figure}

\subsubsection{Spurious variability signatures}
\label{sect:spuriousvariabilitysignatures}

\label{artefacts}
For the fainter stars, we found that some light curves in IRAS~20050+2720 display isolated variability features on the scale of a few days. It turned out that some of these features were produced by column pulldown effects of the IRAC arrays or by PSF artifacts of bright stars (which rotate with observation epoch) in the vicinity of the fainter source. We therefore visually checked all light curves and the images for such features and excluded sources with spurious variability from further light curve analysis. 

% from the Spitzer website: A depression of the intensity of the pixels along a column in the detector array, caused by a high pixel value (usually a star or a cosmic ray), is called "column pulldown". Unlike during the cryogenic Spitzer mission, the pulldown is not constant along the column, but is instead better represented by an exponential function. Pixels below and above the location of the triggering pixel are affected in a different way, and need to be corrected independently.

  As an example, we show in Figure~\ref{var_spurious} one source for which a spurious 
  variability feature was produced by column pulldown induced by a bright source elsewhere 
  on that column of the array. The number of cluster member stars for which we found such 
  artifacts is fairly low among the bright sources, but a larger fraction of the fainter 
  sources was affected (see Figure~\ref{hist_spurious}). In total, out of 181 member 
  stars with light curves, 26 objects were 
  significantly affected and fully excluded from the variability analysis. 
  In the following we will only refer to the 155 cluster members without light curve 
  artifacts when discussing variability properties.

\section{Results and Discussion}
\label{sect:resultsanddiscussion}

\label{results}

\subsection{Types of variability and detection biases}
\label{sect:typesofvariabilityanddetectionbiases}

We tested the standard set of members for variability of the types discussed above, i.e.\ $\chi^2$, periodic, and Stetson two-band variability. We can compare the presence of variability signatures in different SED classes. However, the detectability of variability depends not only on the intrinsic source properties, but also on data coverage of that source, whether two-band data is available, and the apparent brightness of the source. A detailed analysis of our detectability thresholds for variable sources of different apparent brightness is given in \cite{Rebull2014}. In addition, we mention here that Stetson indices could only be calculated for the middle field (see Figure~\ref{skyview_fov}), because this is the area where the fields of view of the [3.6] and [4.5] band observations overlap. The outer two fields were only covered in either [3.6] or [4.5], so that only single-band data is available for those. This skews the detection rates of correlated two-band variability for the SED classes, because the fraction of sources with SED class II and III with available two-band light curves is slightly smaller than for the sources with SED class I and F (see Table~\ref{twoband_vs_class}). This is an effect of class III sources being more spread out over the cluster and therefore being covered more often by the outward-lying one-band observation windows. Such differences in spatial extent have been observed in young stellar clusters before and are interpreted as an evolutionary effect due to movement of the older cluster members over time \citep[e.g.][]{Pillitteri2013}.

\begin{figure}[ht!]
\includegraphics[width=0.5\textwidth]{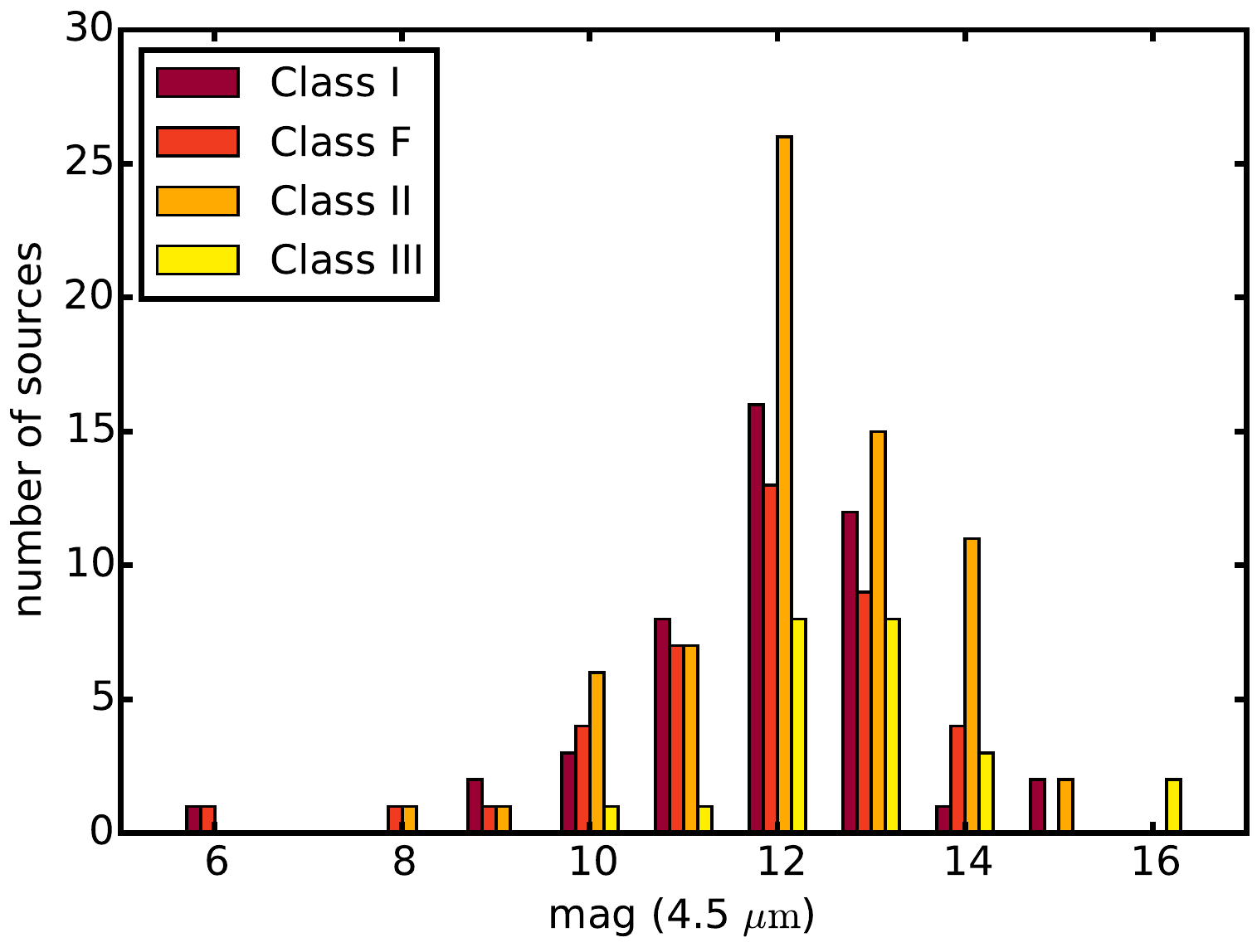}
\caption{Brightness of member sources at 4.5\,$\mu$m, split up by SED classes and binned in 1-mag intervals. 
Class III objects tend to be fainter, as expected for objects without disks compared to disk-bearing 
objects at the same distance.}
\label{mag_by_class}
\end{figure}

In principle, we might also expect to see a skew in detectability for periodicity and $\chi^2$-variability if the brightness of sources in the two observed \textit{Spitzer} bands is very different by SED class. As shown in Figure~\ref{mag_by_class}, the brightness distribution peaks around 12 to 13\,mag for all SED classes, with class III objects peaking at ca.\ 0.6 mag fainter than class II sources. We therefore expect a slightly smaller variability detection rate for class III objects based on IR brightness, but in general differences in detected periodicity and $\chi^2$-variability will mainly be driven by the amplitude and frequency of those variability types, i.e.\ by source-intrinsic properties.

  \begin{table}[t!]
  \caption{Member sources without light curve artifacts with two-band \textit{Spitzer} light curves, split up by SED class.}
  \begin{tabular}{l l l l }
  \hline\hline
  SED class	& \# of 	& \# of             	& fraction of 	\\ 
            & members   & two-band LCs          & two-band LCs \\ \hline \vspace{0.1cm}
  I		& 43		& 35			& $0.81^{+0.05}_{-0.06}$		\\ \vspace{0.1cm}
  F		& 33		& 29			& $0.88^{+0.05}_{-0.06}$		\\ \vspace{0.1cm}
  II		& 59		& 38			& $0.64^{+0.06}_{-0.06}$		\\ \vspace{0.1cm}
  III		& 20		& 11			& $0.55^{+0.10}_{-0.11}$		\\ \hline
  \end{tabular}
  \label{twoband_vs_class}
  \end{table}

\begin{figure*}[ht!]
\includegraphics[width=0.48\textwidth]{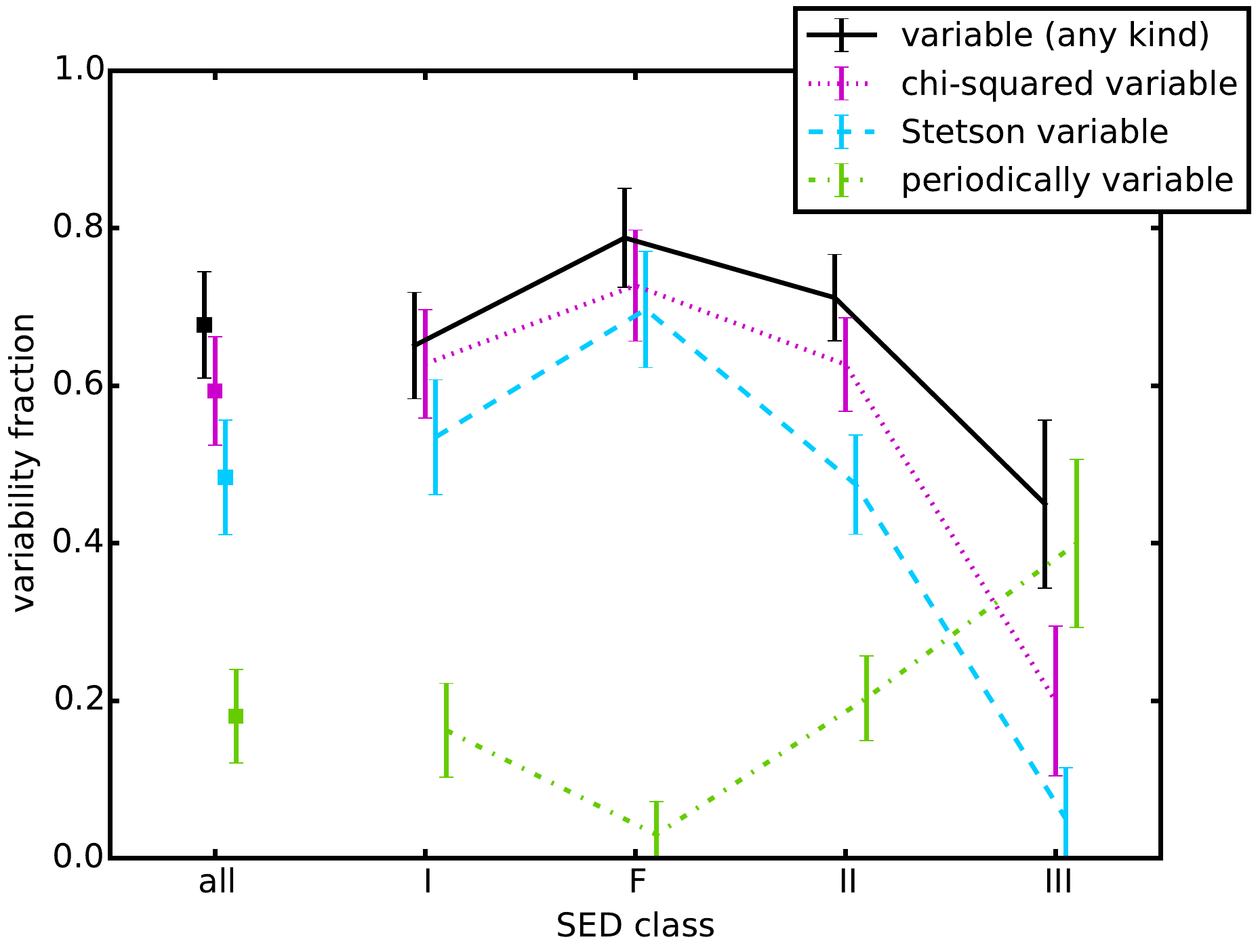}
\hspace*{1.0cm}
\includegraphics[width=0.4\textwidth]{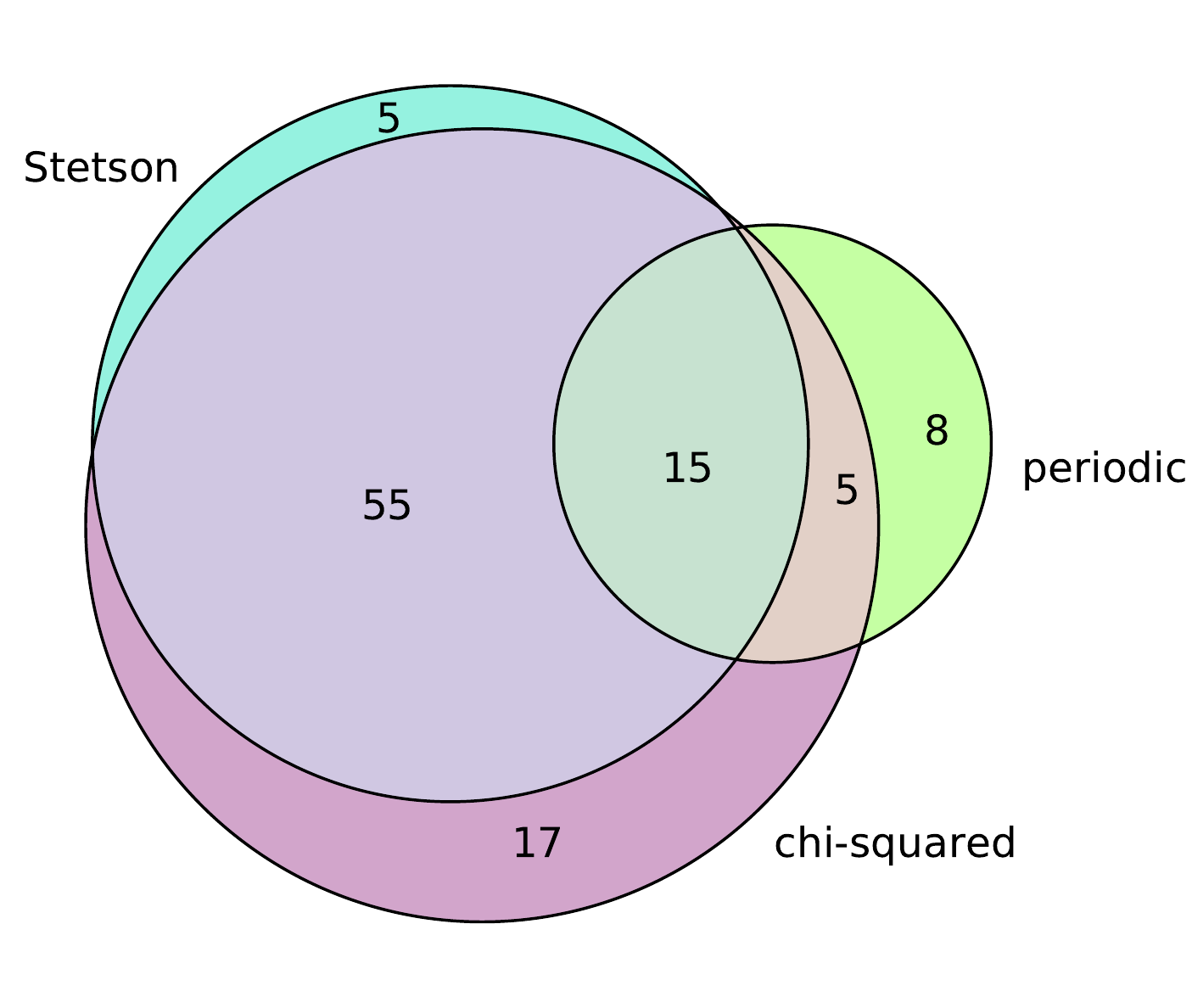}
\caption{\textit{Left:} Fraction of variable sources in the standard set of members with light curves, split up by SED class and variability detection type.  \textit{Right:} Venn diagram (approx.) showing the overlap between different variability types for the standard set of members with light curves.}
\label{var_type_members}
\end{figure*}

  \begin{table*}[ht!]
  \caption{Variability fractions of member sources without light curve artifacts by SED class and variability type.}
  \begin{center}
  \begin{tabular}{l l l l l l}
  \hline\hline
  variability type                    & all Classes     & Class I          & Class F          & Class II         & Class III        \\ \hline
  any kind                            & 0.68 (105/155) &  0.65 (28/43) &  0.79 (26/33) & 0.71 (42/59) & 0.45 (9/20) \\
  periodic                            & 0.18 (28/155) &  0.16 (7/43) &  0.03 (1/33) & 0.20 (12/59) & 0.40 (8/20) \\
  Stetson                             & 0.48 (75/155) &  0.53 (23/43) &  0.70 (23/33) & 0.47 (28/59) & 0.05 (1/20) \\
  $\chi^2$                            & 0.59 (92/155) &  0.63 (27/43) &  0.73 (24/33) & 0.63 (37/59) & 0.20 (4/20) \\ \hline
  periodic \& Stetson \& $\chi^2$     & 0.10 (15/155) &  0.14 (6/43) &  0.03 (1/33) & 0.14 (8/59) & 0.00 (0/20) \\
  periodic \& Stetson, not $\chi^2$   & 0.00 (0/155) &  0.00 (0/43) &  0.00 (0/33) & 0.00 (0/59) & 0.00 (0/20) \\
  periodic \& $\chi^2$, not Stetson   & 0.03 (5/155) &  0.02 (1/43) &  0.00 (0/33) & 0.02 (1/59) & 0.15 (3/20) \\
  periodic, not Stetson, not $\chi^2$ & 0.05 (8/155) &  0.00 (0/43) &  0.00 (0/33) & 0.05 (3/59) & 0.25 (5/20) \\
  Stetson, \& $\chi^2$, not periodic  & 0.35 (55/155) &  0.37 (16/43) &  0.61 (20/33) & 0.31 (18/59) & 0.05 (1/20) \\
  Stetson, not periodic, not $\chi^2$ & 0.03 (5/155) &  0.02 (1/43) & 0.06 (2/33) &0.03 (2/59)&0.00 (0/20) \\
  $\chi^2$, not periodic, not Stetson & 0.11 (17/155) & 0.09 (4/43)&0.09 (3/33)&0.17 (10/59)&0.00 (0/20) \\ \hline
  \end{tabular}
  \label{var_table}
  \end{center}
  \end{table*}

%\pagebreak

\subsection{Variability fractions}
\label{sect:variabilityfractions}

We proceed by listing the number of detected variable member sources and 
specifying the variability type by SED class in Table~\ref{var_table}. A graphical 
representation of the variability fractions is given in Figure~\ref{var_type_members}, 
left. We find that 68\% of all member sources with clean light curves 
(i.e.\ without artifacts) are detected to be variable with at least one type of 
variability. The overwhelming fraction of those are detected to be Stetson- or 
$\chi^2$-variable. Note that this is not an exclusive identification; many sources 
are both Stetson- and $\chi^2$-variable at the same time. In Figure~\ref{var_type_members}, 
right, we show a Venn diagram in which the displayed areas approximate the overlapping 
numbers of sources and their variability type. We also note that we do not find any cluster 
members for which the [3.6] and [4.5] light curves are significantly Stetson-anticorrelated,
as expected \textit{a priori}.

Concerning the variability fractions and types in the individual SED classes 
of member sources, we find that the total variability fraction is around 70\% for 
Class I, F, and II sources, with negligible differences. The overall variability 
fraction of Class III sources is lower with 45\%. The strong drop in detected Stetson 
variability is real for class III sources, and not entirely due to the fact that 
a smaller fraction of class III sources have two-band light curves. If we calculate 
variability fractions only for objects with two-band data, we still find that about 70-80\% of class 
I, F, and II sources with two-band light curves display Stetson variability, but 
only ca.\ 9\% of the class III sources with two-band light curves do. We will 
see in the next section that this is because the variability amplitudes are much 
smaller for class III sources, and therefore go mostly undetected except for 
periodic signals which are easier to pick out of the noise.

We find that the fraction of detected periodic variability is much lower for the disk-bearing sources, i.e.\ class I, F, and II objects, than the fraction of detected Stetson or chi-squared variability for the same group of sources. In contrast, most of the variable class III sources display periodic variability. This is due to the fact that the periodic signals are typically caused by starspots, which induce rather small modulation amplitudes at the observed wavelengths; other processes with larger amplitudes, such as disk changes or accretion signatures which take place in disk-bearing objects often dominate the light curve, so that small periodic signals are typically not picked up by the search algorithm. The disk-free objects, on the other hand, lack such large-amplitude processes, as shown in the next section, so that periodic signals are more readily detected. We will also give a more detailed discussion on the different physical processes causing variability in young stellar objects in section \ref{colorspace}.
Our findings are consistent with the typically high variability fractions found in the mid-IR for other clusters. \cite{Morales-Calderon2011} reported a variability fraction of 70\% among observed members of the ONC. Similarly, \cite{Flaherty2013} report a variability fraction of 60\% for members of IC 348. \cite{Cody2014} find ca.\ 90\% of the members in NGC 2264 to be variable, \cite{Guenther2014} report a variability fraction of ca.\ 80\% for members of L1688, and Wolk et al.\ (submitted) find comparable fractions of ca.\ 70\% in GDD-1215.

\subsection{Amplitudes of variability}
\label{sect:amplitudesofvariability}

\begin{figure}[ht!]
\includegraphics[width=0.48\textwidth]{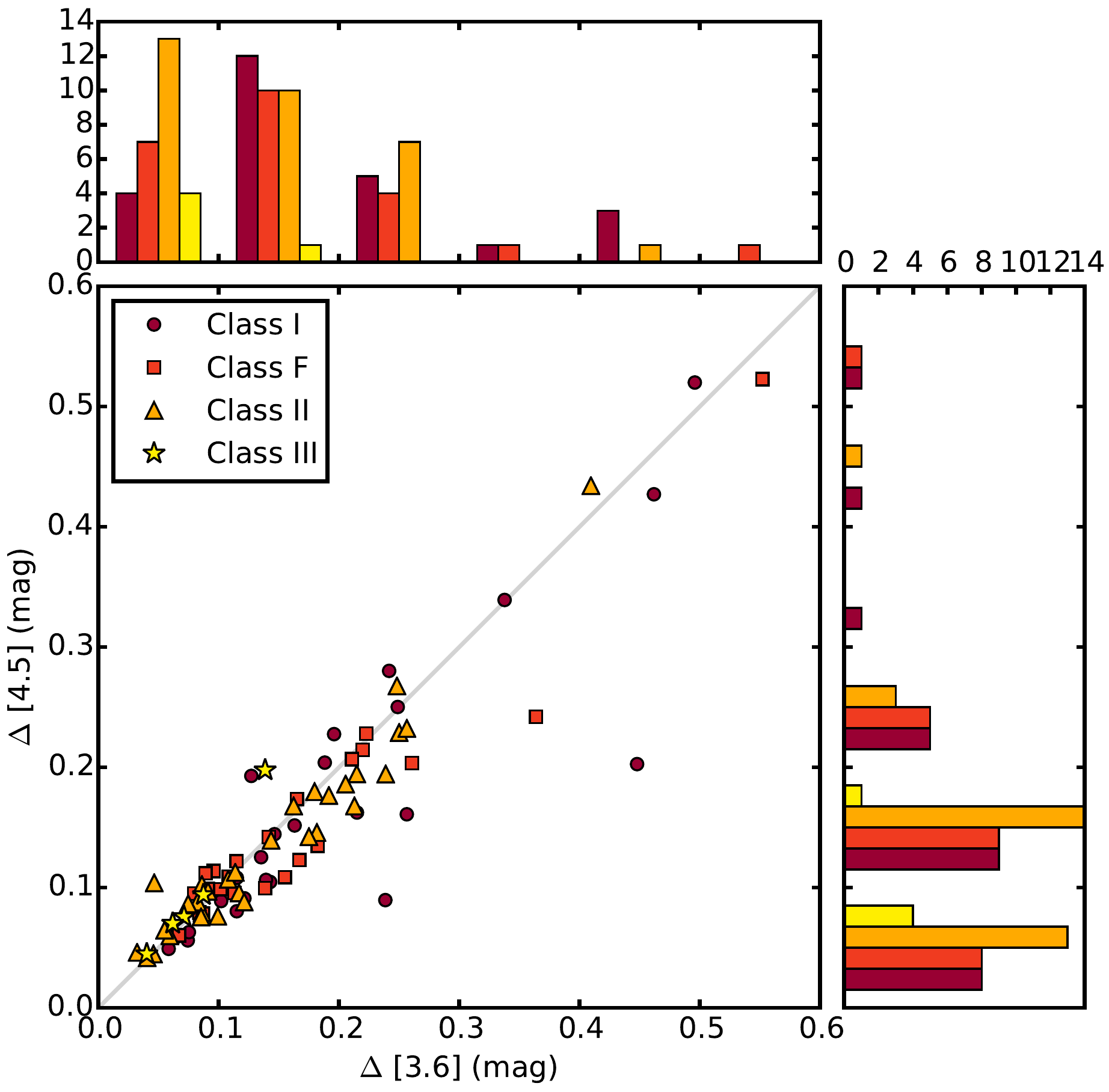}
\caption{The amplitude of variability of light curves of cluster members (90th percentile minus 10th percentile in mag) 
in [3.6] and [4.5] is similar for SED classes I, F, and II, but significantly smaller for SED class III. 
The spread in flux of individual member star light curves tends to be larger in [3.6] than [4.5].}
\label{deltas_scatter}
\end{figure}

To characterize the amplitude of variability in a given light curve, we chose the spread of the individual magnitudes as an appropriate measure. To account for possible single outliers in the light curves, we report the 80\% magnitude spread of a given light curve (i.e.\ the 90th percentile magnitude minus the 10th percentile magnitude). The result, broken up by SED class, is shown in Figure~\ref{deltas_scatter} for the two infrared bands, using all sources of the standard set of members that have been detected to be variable.

We find that sources with disks, i.e.\ SED classes I, F, and II, 
show a mean $3.6\,\mu\mathrm{m}$ variability amplitude of 0.19~mag, 
0.17~mag, and 0.16~mag, respectively. The distribution is broad and 
reaches out to ca.\ 0.5 magnitudes for disk-bearing sources. In 
contrast, diskless stars, i.e.\ objects with SED class III, show a 
markedly different variability amplitude pattern. They are preferentially 
detected with low variability amplitudes with a mean of 0.08 magnitudes 
at $3.6\,\mu\mathrm{m}$ (compared to a typical standard deviation of 0.02 
displayed by an individual light curve),
and the histogram quickly falls off for larger 
amplitudes. We list the means of the variability amplitude distributions and 
the standard deviations of the respective distribution per SED class in Table~\ref{ampl_tab}.

To test if those distributions are drawn from the same parent distribution, we 
use the Anderson-Darling two-sample test 
\citep{AndersonDarling1952, ScholzStephens1987} 
for each of the possible combinations of two SED classes; 
this test is similar to a two-sample Kolmogorov-Smirnov test, 
but has a higher sensitivity to differences of the tails of the tested distributions.
The result of this statistical test 
is given in the form of $p$-values, i.e.\ the chance to 
obtain the observed difference in variability amplitudes, or a more extreme difference,
if the parent distributions were in fact identical.

We find that the distribution 
of variability amplitudes is statistically indistinguishable for 
class I, F, and II sources, i.e.\ the sources with disks, whereas 
the disk-free class III sources show a distribution which is unlikely 
to be drawn from the same parent distribution as the class I and F sources 
($p<0.05$ for both the [3.6] and [4.5] band amplitudes).

\begin{table}[ht!]
\begin{center}
\caption{Means and standard deviations of the variability amplitude distributions per SED class.}
\begin{tabular}{l | c c}
\hline\hline
  & mean [3.6] ampl.\ ($\sigma$[3.6]) & mean [4.5] ampl.\ ($\sigma$[4.5]) \\ \hline
Class I  & 0.19\,mag (0.12\,mag) & 0.17\,mag (0.11\,mag)  \\
Class F  & 0.17\,mag (0.11\,mag) & 0.15\,mag (0.09\,mag)  \\
Class II & 0.16\,mag (0.11\,mag) & 0.13\,mag (0.08\,mag) \\
Class III & 0.08\,mag (0.03\,mag) & 0.09\,mag (0.05\,mag)  \\ \hline
\end{tabular}
\label{ampl_tab}
\end{center}
\end{table}

The variability amplitudes we find for the IRAS~20050+2720 members are consistent with findings for other young stellar clusters. \cite{Guenther2014} reported median [4.5] variability amplitudes of 0.26, 0.15, 0.14, and 0.05 mag for class I to III members of the cluster L1688; Wolk et al.\ (submitted) find similar mean variability amplitudes of 0.17, 0.15, and 0.06 mag for class I, II, and III members of the cluster GGD12-15. 

For several individual sources among the standard set of members, we find that the light curve spread in [3.6] tends to be larger than in [4.5], as seen in Figure~\ref{deltas_scatter}, where many sources fall towards the [3.6] side of the diagonal line representing a 1:1 ratio. Generally, a larger spread in the [3.6] band is consistent with variability caused mainly by extinction. However, the difference in the light curve spread distributions turns out to be of low statistical significance for the sample of members in this cluster.
%\pagebreak

\subsection{Time scales of variability}
\label{sect:timescalesofvariability}

\begin{figure*}[ht!]
\includegraphics[width=0.48\textwidth]{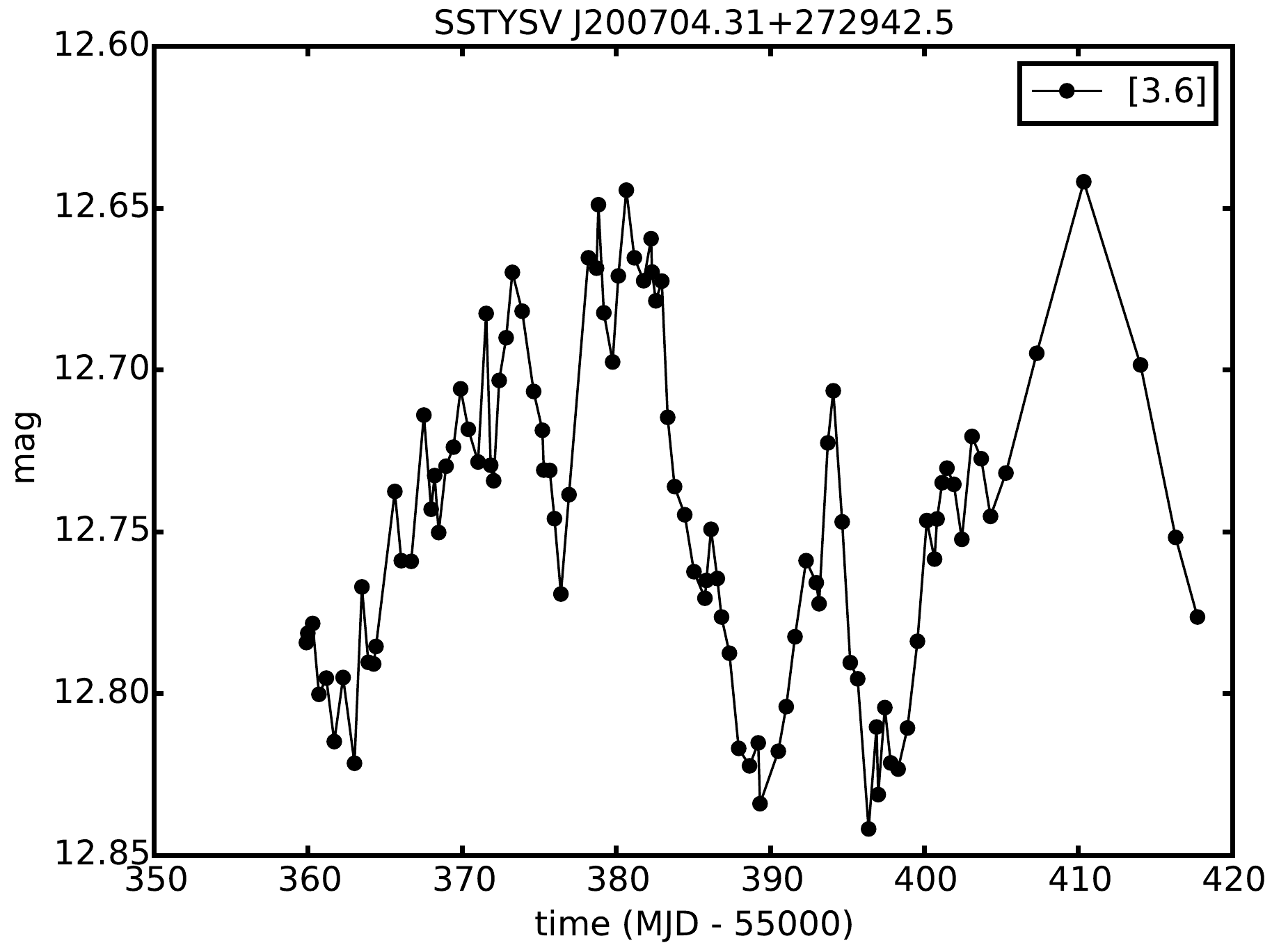}
\includegraphics[width=0.47\textwidth]{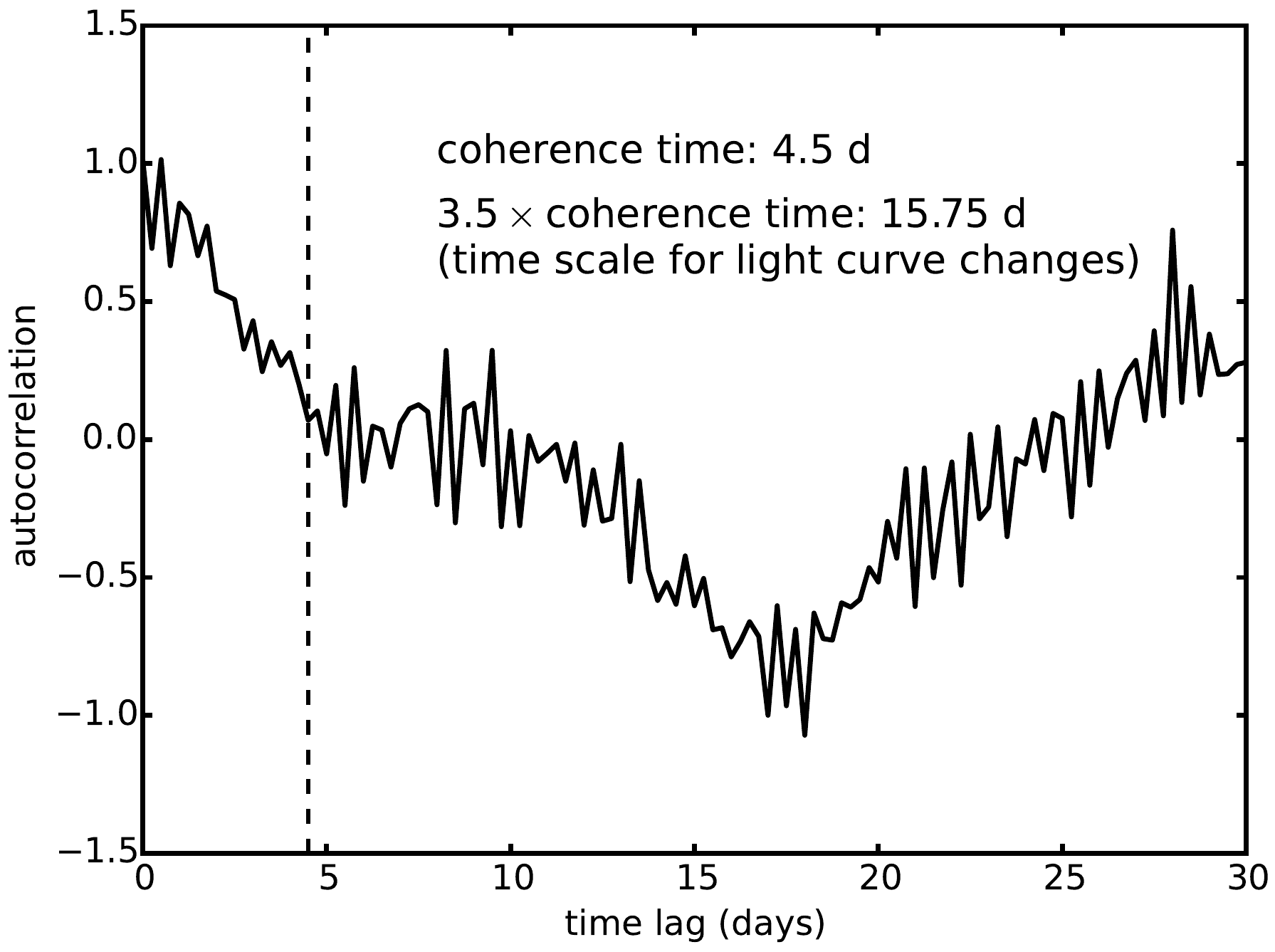}
\caption{Light curve at $3.6\,\mu\mathrm{m}$ of a non-periodic source with SED class II (left) and the autocorrelation function of the light curve (right). The dashed vertical line represents where the autocorrelation falls below a value of 0.25 for the first time; we refer to this time scale as the coherence time.}
\label{autocorr_example}
\end{figure*}

\begin{figure*}[ht!]
\includegraphics[width=0.41\textwidth]{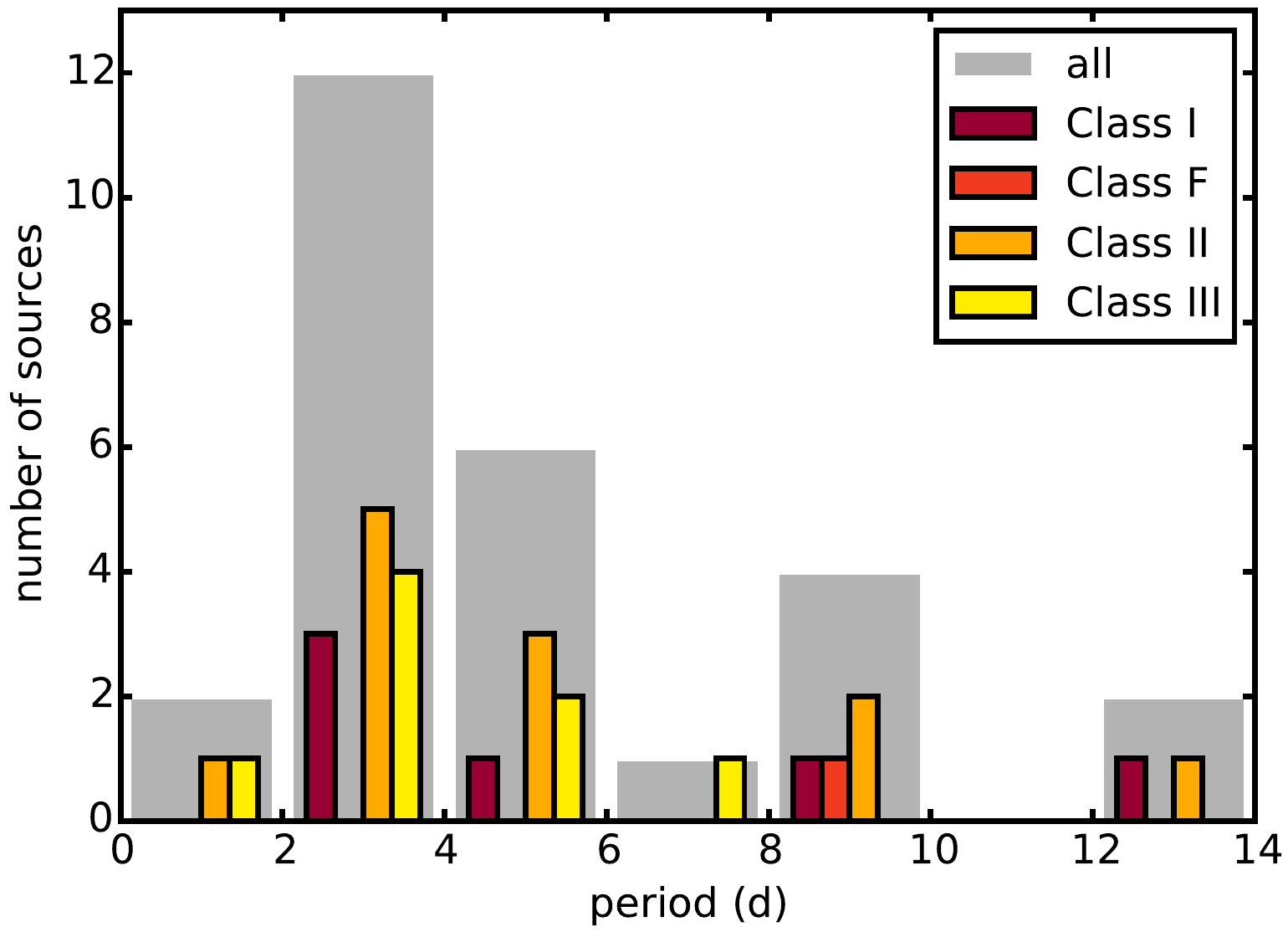}
\includegraphics[width=0.59\textwidth]{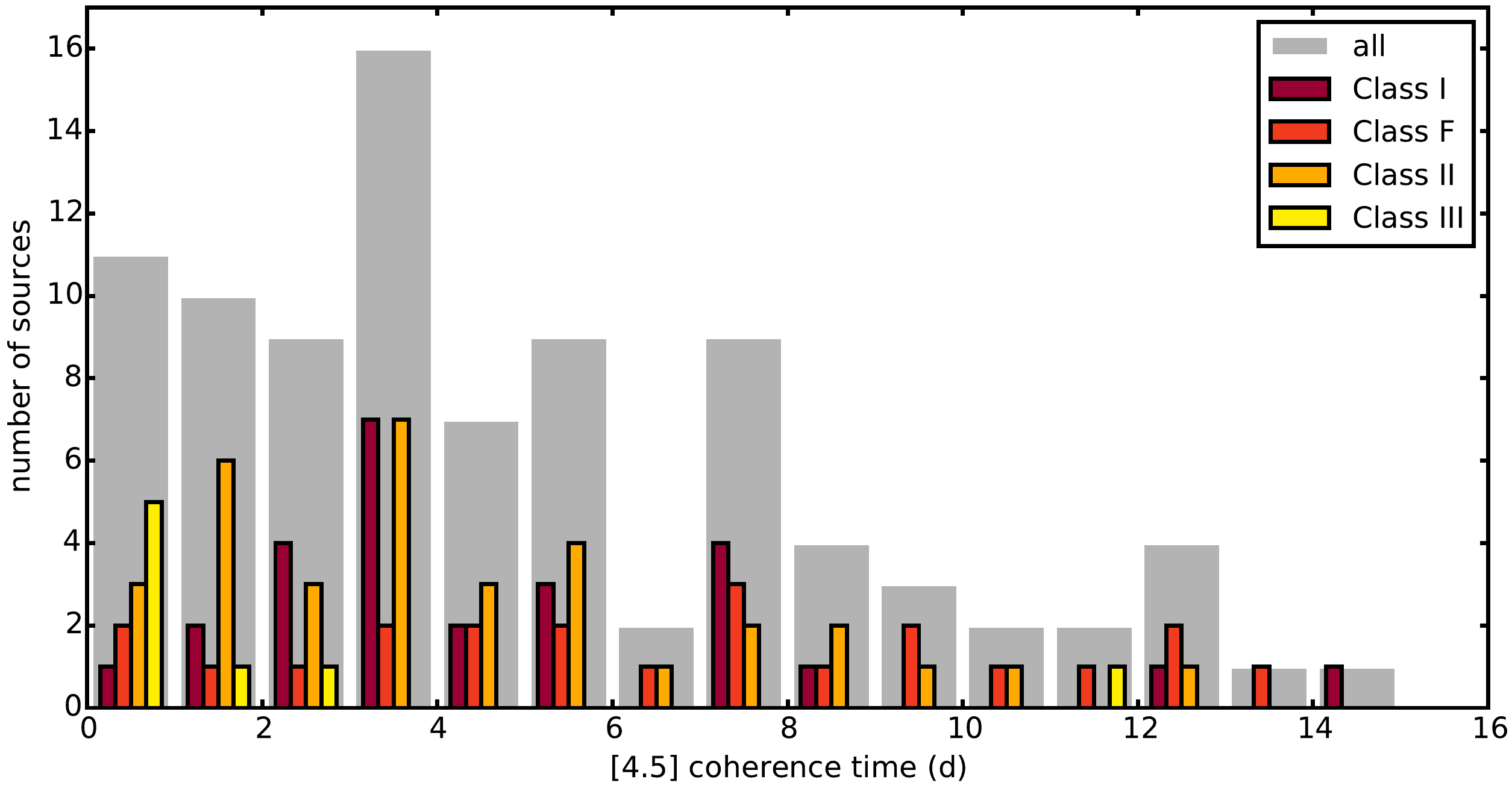}
\caption{Distribution of detected periods (left) and detected coherence time scales in the $4.5\,\mu\mathrm{m}$ band (right) for member sources without light curve artifacts.}
\label{var_timescales}
\end{figure*}

Another quantity we want to test for SED-class specific behavior is the time scale of detected variability. In the case of periodic variation, this time scale is easily identified, namely as the detected period. However, the fraction of periodic sources in our sample of members is small ($<20\%$), as shown in Figure~\ref{var_type_members}. We therefore chose to include sources with non-periodic variability in this time scale analysis.

For these sources, we use the coherence time of the autocorrelation function of the individual light curves as a time scale measure. The autocorrelation function, i.e.\ the cross-correlation of a time series with itself, can be used to test if some sort of recurring pattern is present in the time series. For periodic (or near-periodic) light curves, the autocorrelation displays one or more peaks at time lags corresponding to multiples of the period, and this has been used to infer stellar rotation periods, for example \cite{McQuillan2014}. However, since we are also interested in phenomena that are not necessarily periodic (as we would have picked those up in our periodicity search already), we use the coherence time of the autocorrelation function as a measure for the characteristic time scale of changes in a light curve. As our time sampling is uneven and the autocorrelation function requires an evenly sampled input by definition, we linearly interpolated the light curves on time steps of 0.1\,d. In principle, such an interpolation can change the variability properties of a light curve; however, since we are looking for variability signatures on much larger time scales than 0.1\,d, this is not an issue here. For a discussion of other timescale metrics that do not use linear interpolation, see \cite{Findeisen2015}.

The coherence time refers to the time lag at which the autocorrelation function falls below a set value between $1$ and $-1$ for the first time. This value is often chosen to be 0.5, but this is not optimal for the rather sparse time sampling of our light curves. This is due to the fact that the coherence time scales for our sources are rather short (of the order of a 0.5-1 days if using a threshold of 0.5), and that means that there are only very few steps necessary for the autocorrelation function to fall below the threshold, making it vulnerable to the effects of noise. We found that using a threshold of 0.25 for the autocorrelation function yields more robust results for our data.  

In any case, the coherence time is not the characteristic time scale for changes in the light curve itself; it is a relative measure, i.e.\ a light curve with a shorter coherence time shows changes on a shorter time scale than a different light curve with a longer coherence time. As an example, we show a single light curve of a member source and its autocorrelation function, coherence time and actual time scale on which changes occur in Figure~\ref{autocorr_example}. As discussed in more detail by Wolk et al. (submitted), one can show for sources where a periodical variation is detected that the coherence time and the detected period of the sources generally follow a linear relationship for our targets, with the period of a given source being ca.\ 3.5 times longer than the coherence time.
% There are couple of cases in which the coherence time picks up longer-term signals than the periodicity search, but in general the coherence time times 3.5 yields similar values to the detected periods (see Figure~\ref{periods_coherencetime}). 

\begin{table}[ht!]
\begin{center}
\caption{Means and standard deviations of the coherence time $t_{coh}$ distributions per SED class.}
\begin{tabular}{l | c c}
\hline\hline
  & mean [3.6] $t_{coh}$ \ ($\sigma$[3.6]) & mean [4.5] $t_{coh}$  ($\sigma$[4.5]) \\ \hline
Class I  & 5.46\,d (3.75\,d) & 4.97\,d (3.30\,d)  \\
Class F  & 7.22\,d (4.03\,d) & 7.50\,d (4.49\,d)  \\
Class II & 3.51\,d (2.67\,d) & 4.36\,d (2.90\,d) \\
Class III & 1.98\,d (2.04\,d) & 2.41\,d (3.41\,d)  \\ \hline
\end{tabular}
\label{coh_tab}
\end{center}
\end{table}

\begin{figure*}[ht!]
\includegraphics[width=0.32\textwidth]{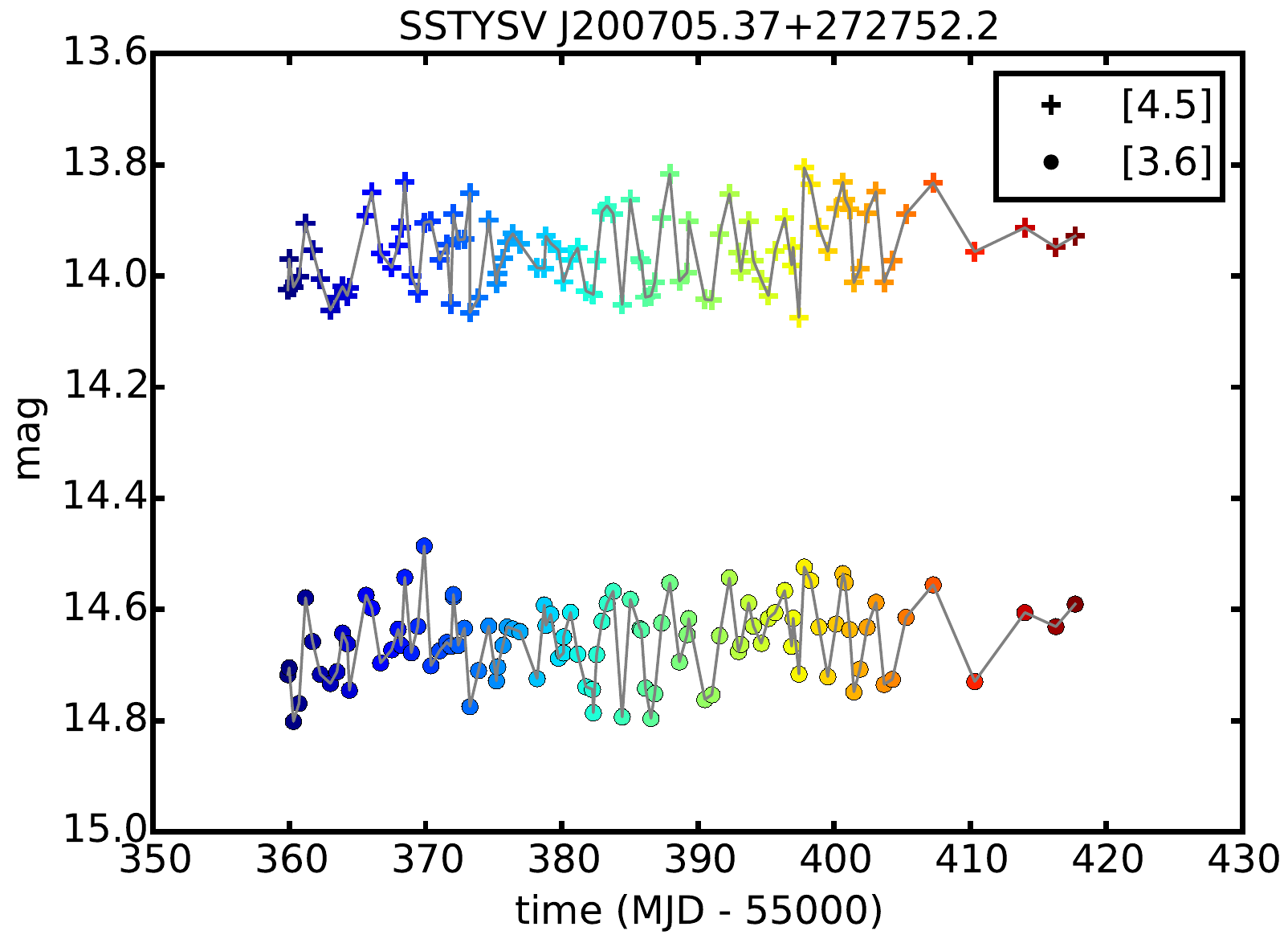}
\includegraphics[width=0.32\textwidth]{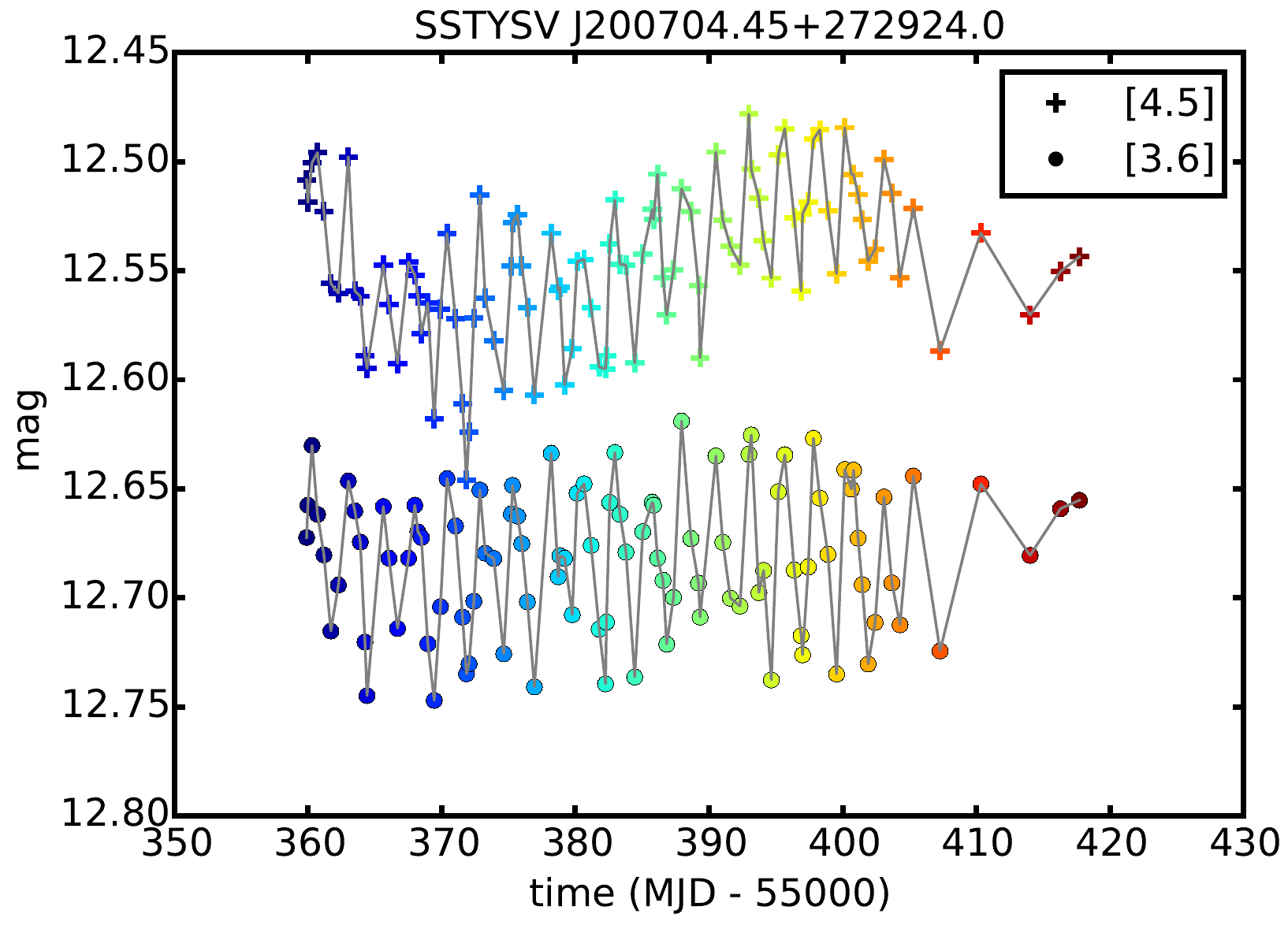}
\includegraphics[width=0.32\textwidth]{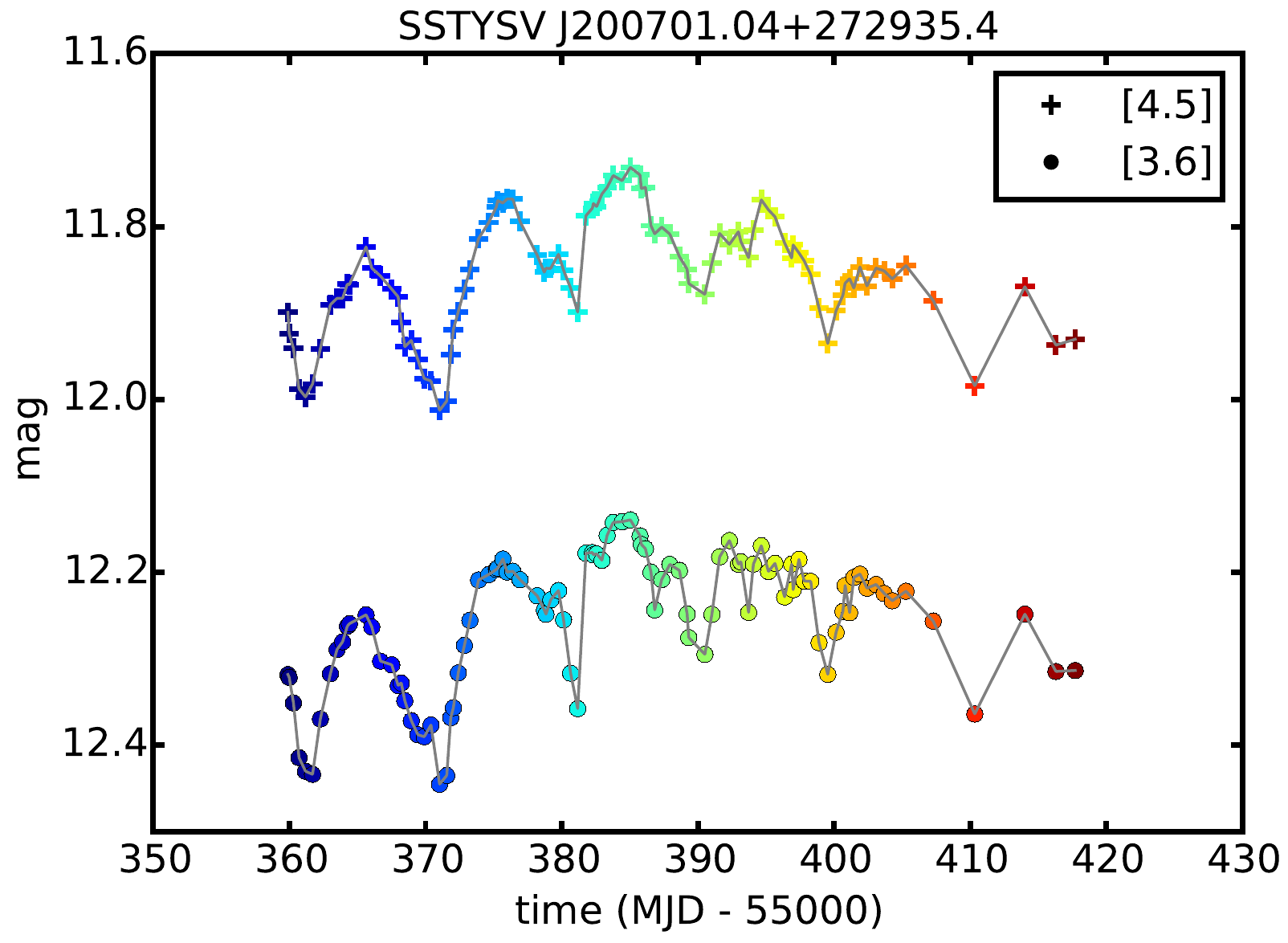}

\includegraphics[width=0.32\textwidth]{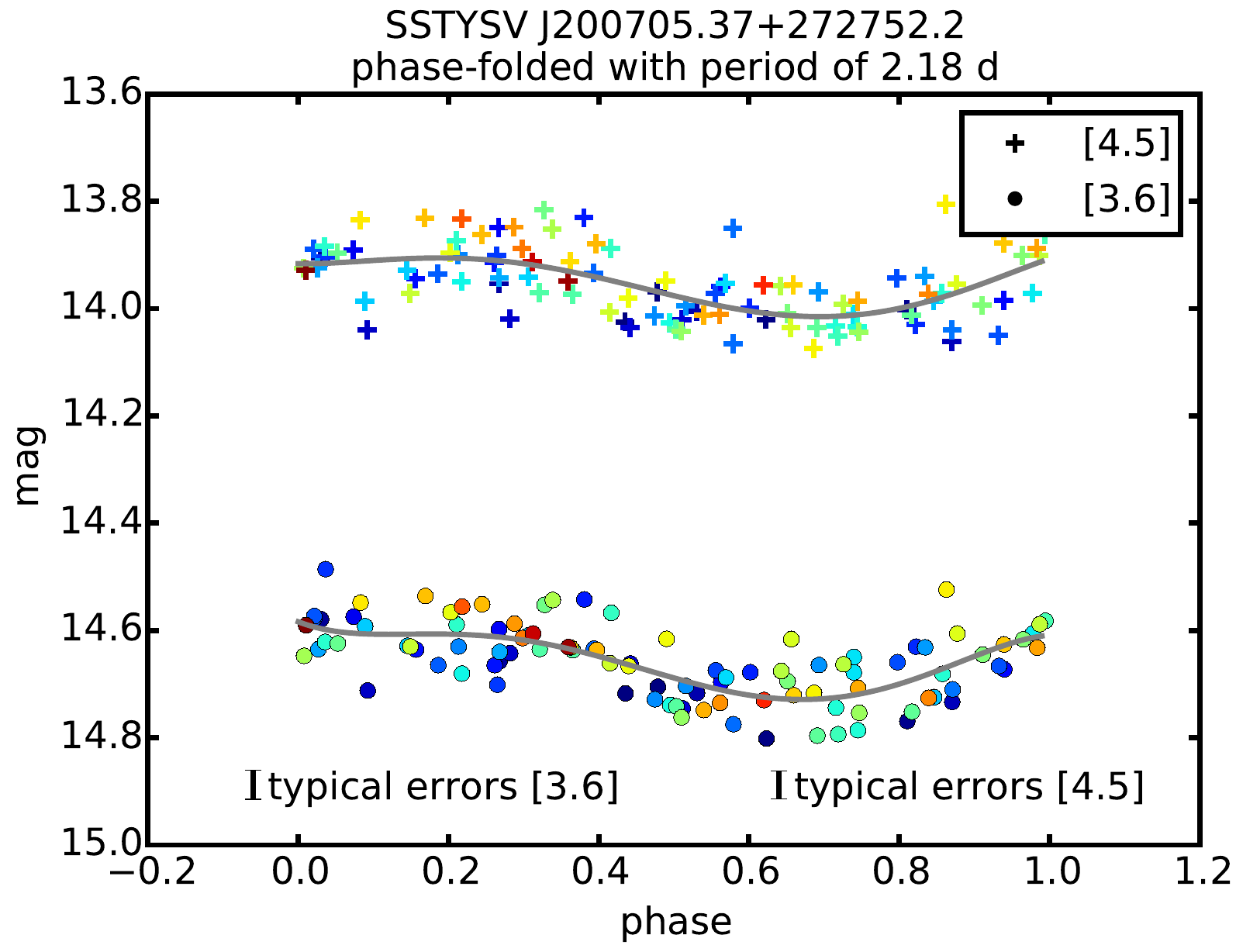}
\includegraphics[width=0.32\textwidth]{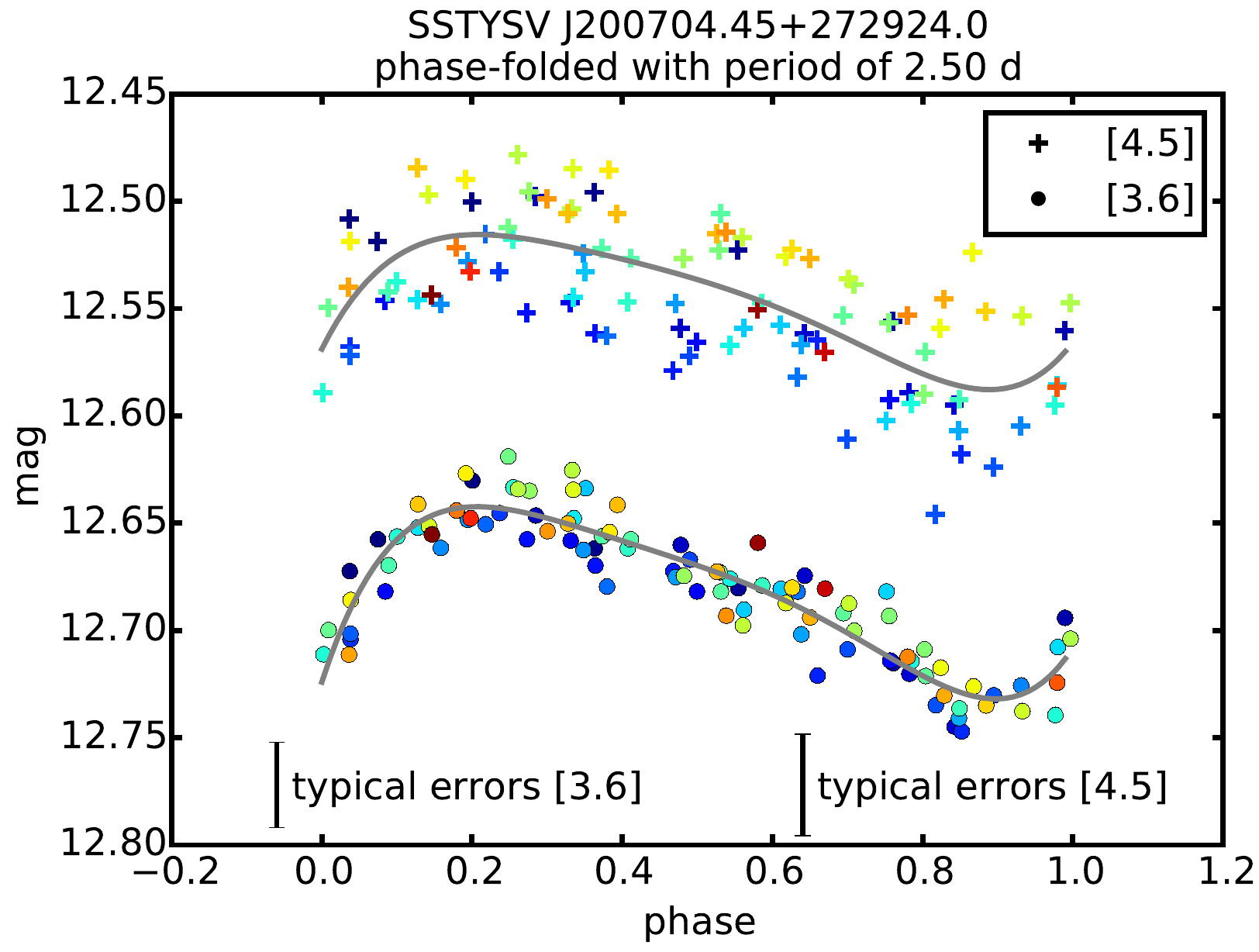}
\includegraphics[width=0.32\textwidth]{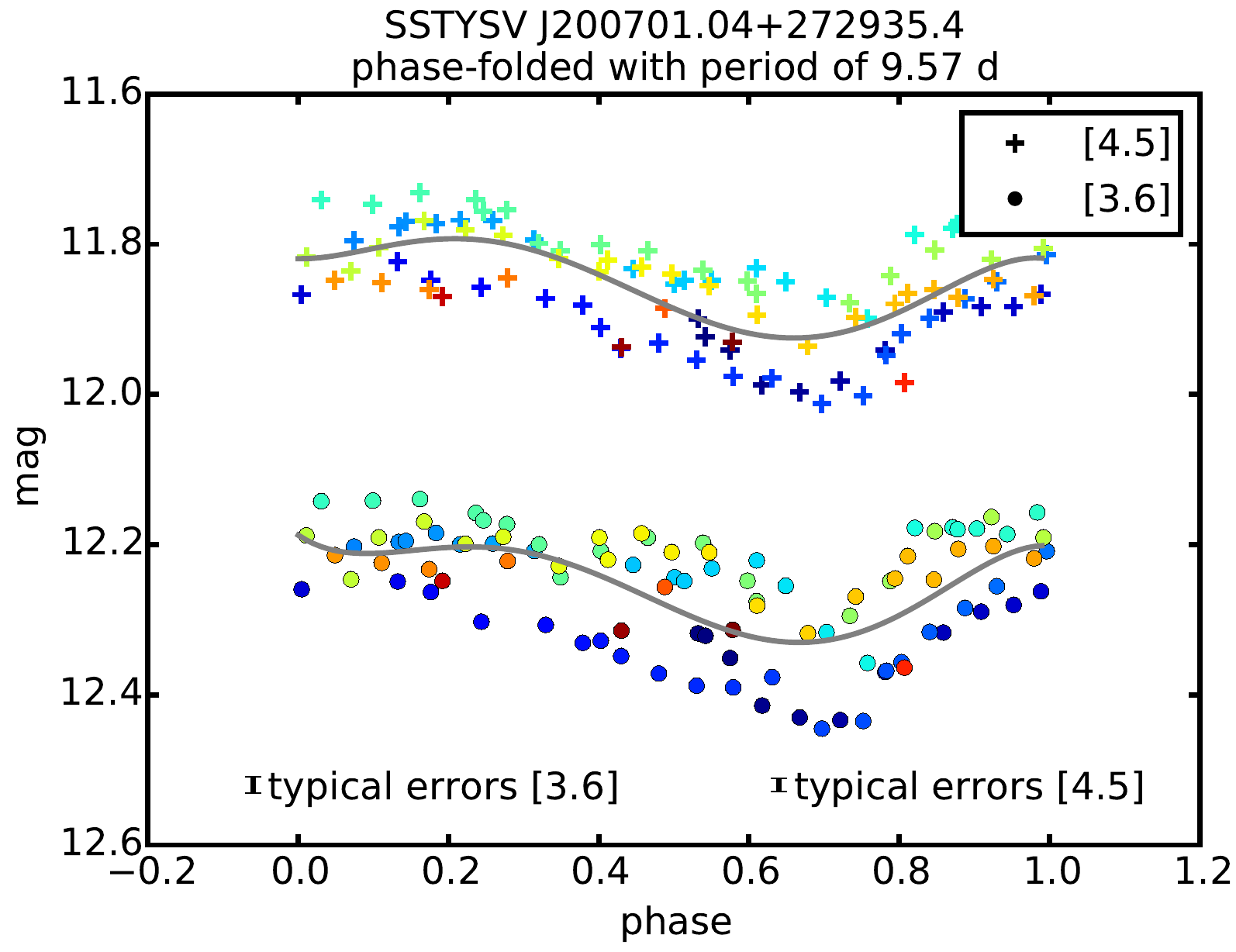}

\caption{Three example sources with periodic variability; light curves in top panel, phase-folded light curves of the same sources in bottom panel. Color refers to time ordering (not phase ordering) of the data points in both panels. In the chosen examples, the amplitude of additional variability processes on top of the periodicity increases from left to right; the source in the middle shows long-term variability on top of the periodicity predominantly in the 3.6\,$\mu\mathrm{m}$ band. The left and the right objects have SED class II, the middle object has SED class III.}
\label{lcs_phased}
\end{figure*}
We next compare the detected periods and coherence times across SED classes for our cluster members, again using only sources without light curve artifacts. We give a graphical representation of the result in Figure~\ref{var_timescales}, where the left plot shows the distribution of detected periods per SED class and the right plot shows the coherence times. As mentioned in section~\ref{periodicity}, the [4.5] data often show longer-term trends overlaying the periodic short-term changes; since we are interested in the time scales dominating the behavior of a light curve, periodic or not, we focus on the coherence times in the [4.5] band here. However, as will be shown below, the results for the [3.6] band are qualitatively similar.
For the periodic sources, we find preferentially periods between 2 and 6 days across all SED classes, with some longer periods $>8$ days detected for disk-bearing objects. The number of detected periodic sources is too small to test for differences in the distributions per class on a statistically significant level. 

For the coherence times, we find that class I and II objects peak around 4 days with a significant tail to longer times; i.e.\ the actual time scales for light curve variability peak at $3.5\times 4 = 14$\,d. We have a smaller number of class F objects and their distribution is therefore more prone to sampling uncertainties; that aside, we find a peak at short coherence times ($<1$\,d) and a flatter peak around coherence times of $4-8$\,d. For disk-free class III sources, we find a noticeably different behavior with a strong peak at short coherence times $<1$\,d and a steep decline towards longer coherence times. We list the means and standard deviations of the coherence time distributions per SED class in Table~\ref{coh_tab}. We test if the distributions found for the different SED classes are compatible with being drawn from the same parent distribution, using the two-sided Anderson-Darling test. We find that nearly all pairings of coherence time distributions in two different SED classes are unlikely to stem from the same distribution ($p<0.05$), except for the comparison between class I and class F in the [3.6] band and between class I and class II in the [4.5] band which do not pass this threshold. 

% To quantify how significant the differences in these behaviors are, we have compared the coherence time distributions with the two-sided Anderson-Darling test. The result is given in Table~\ref{coherence_tab}. The differences between class I and II are not significant, but class F and III are significantly different from each other as well as from class I and II.

\subsection{Periodic variability and additional scatter}
\label{sect:periodicvariabilityandadditionalscatter}

\label{additional_scatter}

For the 28 members of the standard set for which we have detected periodic variability, 
the noise in the phase-folded light curves differs strongly from source to source. 
In some of the cases, this is due to an underlying long-term trend; in other cases 
there seem to be additional short-timescale processes happening on top of the periodic 
variability. We show three examples of the original light curves and their phase-folded 
versions in Figure~\ref{lcs_phased}.

To quantify the presence and magnitude of processes on top of strictly periodic variability, we looked at the scatter in the original light curves with respect to the median compared to the scatter in the phase-folded light curves with respect to a fit to that light curve. Many of our sources do not display sinusoidal variations, but rather some complicated and often non-symmetric profile over the detected period. We therefore chose to fit the phase-folded light curve with a fifth-order polynomial, plotted with a grey line in the examples shown in Figure~\ref{lcs_phased}. The quantity we examine in the following paragraphs is the ratio of the reduced $\chi^2$ values for the phase-folded light curves with respect to the polynomial fit, compared to the reduced $\chi^2$ values of the original light curve with respect to the median.

\begin{figure}[t!]
\includegraphics[width=0.48\textwidth]{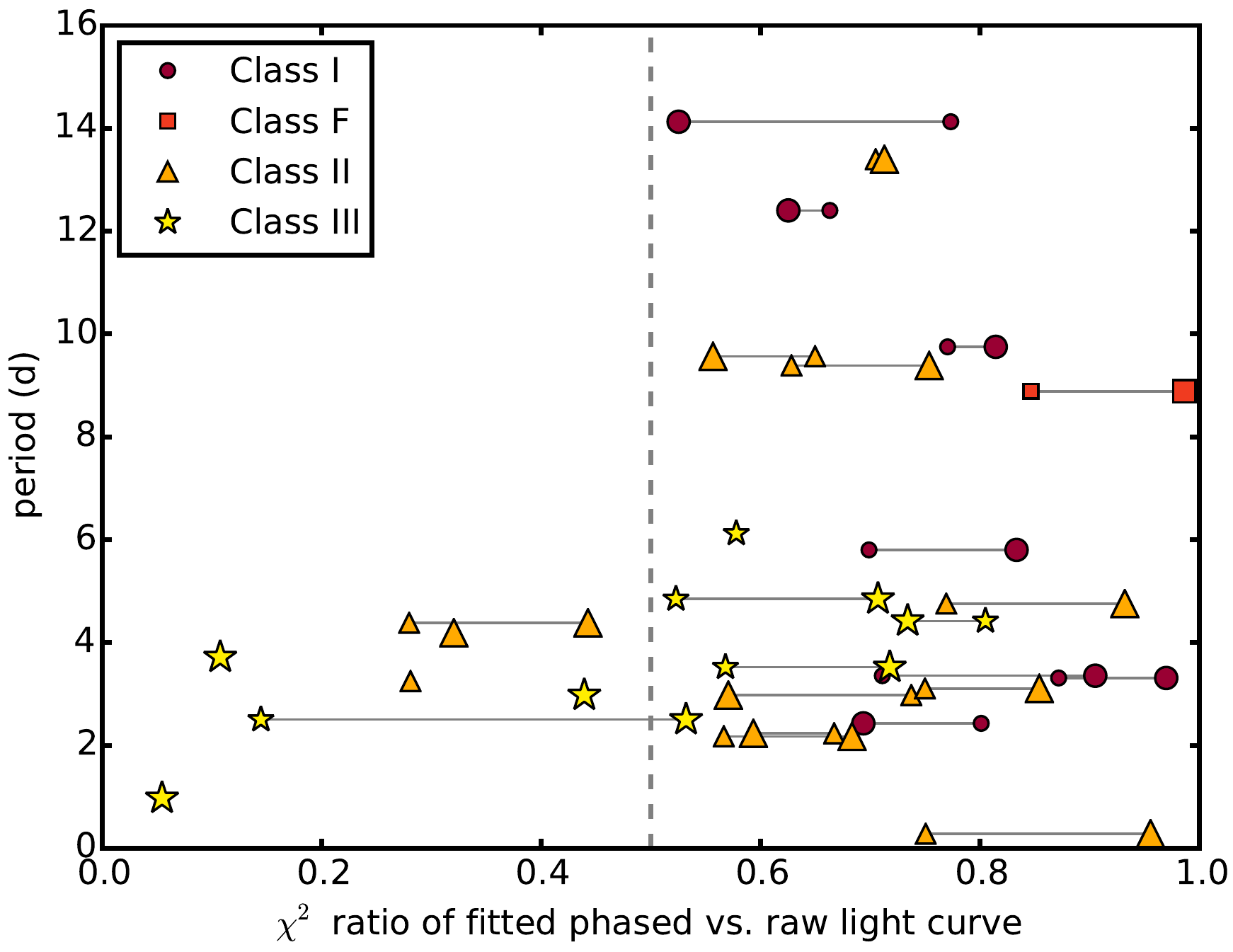}
\includegraphics[width=0.48\textwidth]{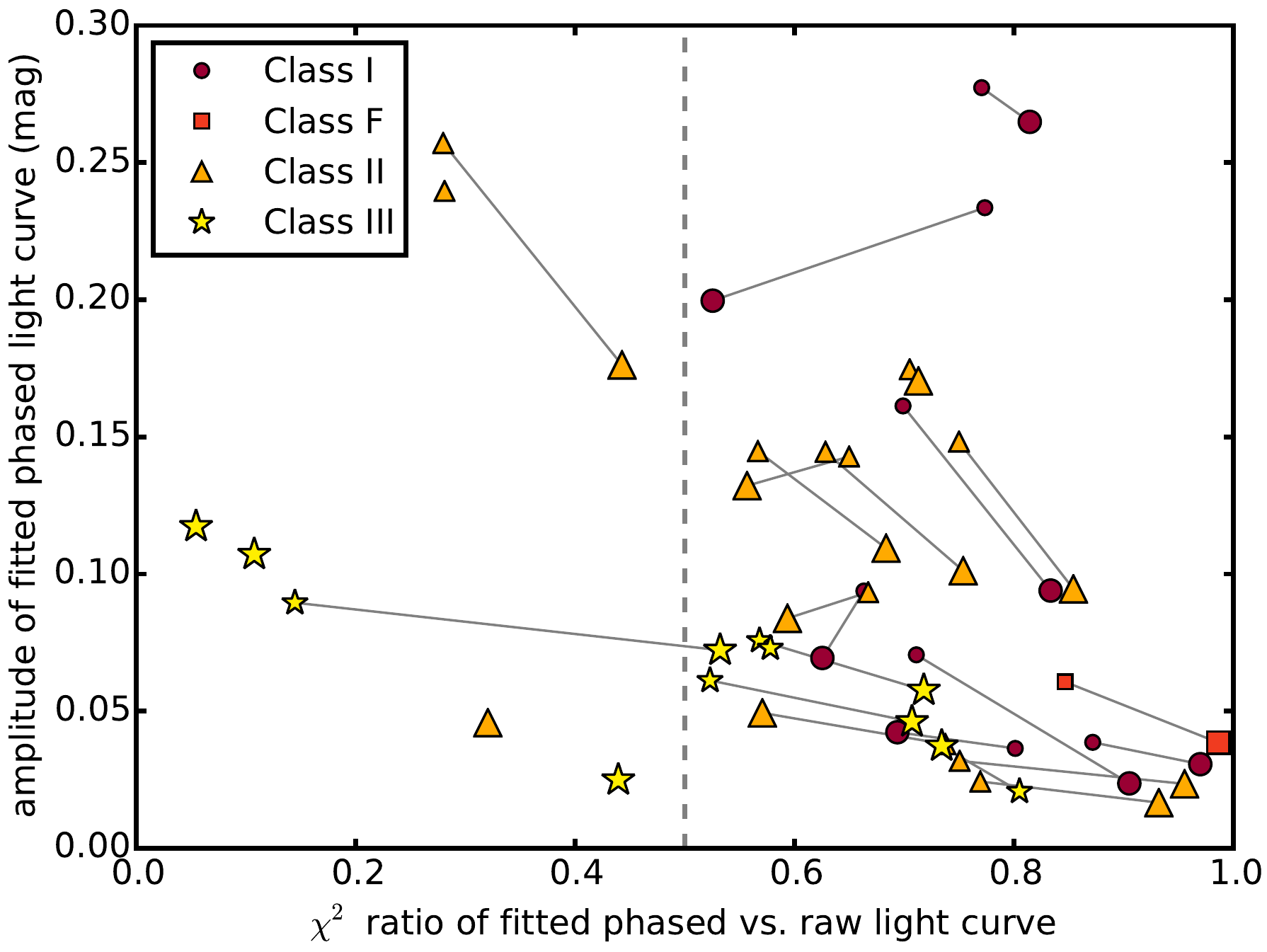}
\caption{Top: The detected period versus the scatter ratio (phase-folded and fitted light curve vs.\ raw light curve). SED classes of the objects are indicated by different symbols. The scatter ratios of a single source are generally different for the [3.6] and [4.5] band, and we show both in the plot, connected by grey lines. The value for the [4.5] band is indicated by the larger symbol. Sources in the left hand side of the plot show mainly periodic behavior with little extra scatter. Sources on the right hand side show large scatter remaining even after the periodic behavior is subtracted.\newline
Bottom: The peak-to-peak amplitude of the fit to the phase-folded light curve versus the scatter ratio (phase-folded and fitted light curve vs.\ raw light curve). Again, data points for the same source in the [3.6] and [4.5] band are connected with grey lines. Apart from the two Class II sources in the upper left, we find the same population pattern as in the top panel, indicating that it is truly the amplitude of the additional scatter and not the amplitude of the periodic signal which drives the pattern in the top panel (see text for details).}
\label{scatter_vs_period}
\end{figure}

\cite{Cody2014} used a slightly different approach to identify deviations from periodicity in CoRoT and Spitzer light curves in their section 5.2. They remove long-term trends first, then subtract a smoothed version of the phased light curve, and finally evaluate the root mean square (r.m.s.) scatter of the residuals against the r.m.s.\ scatter of the original light curve. They label this ratio the ``Q value''. 
% and give comparisons to the approach of \cite{Cody2014} where light curve quality is sufficient for that. 
% Figure~\ref{scatter_and_qs} shows that these two approaches yield comparable results for the light curves where both have been applied; 
We have verified that the relationship of the two measurements of additional scatter (Q values and residual $\chi^2$) is roughly linear for the sources where both quantities could be reasonably well determined. Given that we are dealing with significantly fewer data points in our light curves compared to optical CoRoT data, we adopt the approach described above.
% To quantify this, we performed two tests for correlation, the nonparametric Spearman's R test and the linear Pearson's R test. For the [3.6] band, we find a p-value from Spearman's R of $7\times10^{4}$, i.e.\ the samples are strongly correlated; the p-value from Pearson's R is even smaller with $4\times10^{-5}$. For the [4.5] band we find significant, but less pronounced correlation with p-values of $0.014$ and $0.034$. 

We next compare the ratio of $\chi^2$ values in the fitted phase-folded light curves and the original light curves (the scatter ratio) to two other quantities, namely the detected period and the peak-to-peak amplitude of the detected periodicity. 

The scatter ratio vs.\ period plot (Figure~\ref{scatter_vs_period}, top panel) shows a strong dependence on SED class. The amount of additional scatter on top of periodicity increases in the plot from the left to the right. We find that sources that are mostly periodic with low additional scatter (i.e.\ sources in the left half) are objects with SED class III or II, and display short periods of less than five days. Class III sources also extend into the right side of the plot, but stay at short periods. In contrast, sources with disks, i.e.\ class II, F, and I, populate the full height of the right half of the plot, meaning that these sources can display large amounts of additional scatter at both short and long periods.

It is noteworthy that objects with SED classes I and F never display clean periodicity, but are always located on the right hand side of this plot.

It is in principle possible that this trend is caused by a peculiar distribution of amplitudes of periodic behavior across the SED classes. For example, if a source displays periodicity with a large amplitude and a small-amplitude extra scatter, the source will show up on the left side of the plot. However, if the periodic amplitude is smaller than the amplitude of the extra scatter, it will show up on the right hand side. We therefore also looked at the extra scatter versus the peak-to-peak amplitude of the periodicity, inferred from the polynomial fit as shown in Figure~\ref{lcs_phased}. The result is shown in the bottom panel of Figure~\ref{scatter_vs_period}. If the stratification in the top panel of Figure~\ref{scatter_vs_period} was caused by different periodic amplitudes and not by differences in the additional scatter, we would expect to see sources with low scatter ratio to display high periodic amplitudes, i.e.\ to be located in the upper left, and sources with high scatter ratio to display low periodic amplitudes, i.e.\ be in the lower right. However, with the exception of two Class II sources which have moved significantly upward in this plot, the general distribution is the same as in the previous plot. Class III sources display low periodic amplitudes and little extra scatter, and Class I and F sources display high extra scatter, no matter if their periodic amplitudes are large or small. Class II sources populate both parts of the diagram.

This means that there is a true dependence of the ``cleanliness'' of periodic processes on the SED class of objects. This is not extremely surprising for Class III sources, since they have mostly lost their disks and should be dominated by periodic starspot modulations. We see that some Class II objects can mimic this behavior of periodicity without other apparent processes over the 40-day time scale of our observations. These Class II objects display short periods similar to the Class III sources; however, other Class II objects display high levels of additional scatter on top of the periodicity. It is noteworthy that \textit{all} detected periodic Class I and F objects also have large additional scatter, and these objects display a wide range of periods. The additional scatter seems to be intrinsically linked to the presence of a disk; density fluctuations and fluctuations of the disk scale height may be explanations for this scatter.

\subsection{Physical processes causing variability: clues from color space}
\label{sect:physicalprocessescausingvariabilitycluesfromcolorspace}

\label{colorspace}
\begin{figure}[t!]
\includegraphics[width=0.48\textwidth]{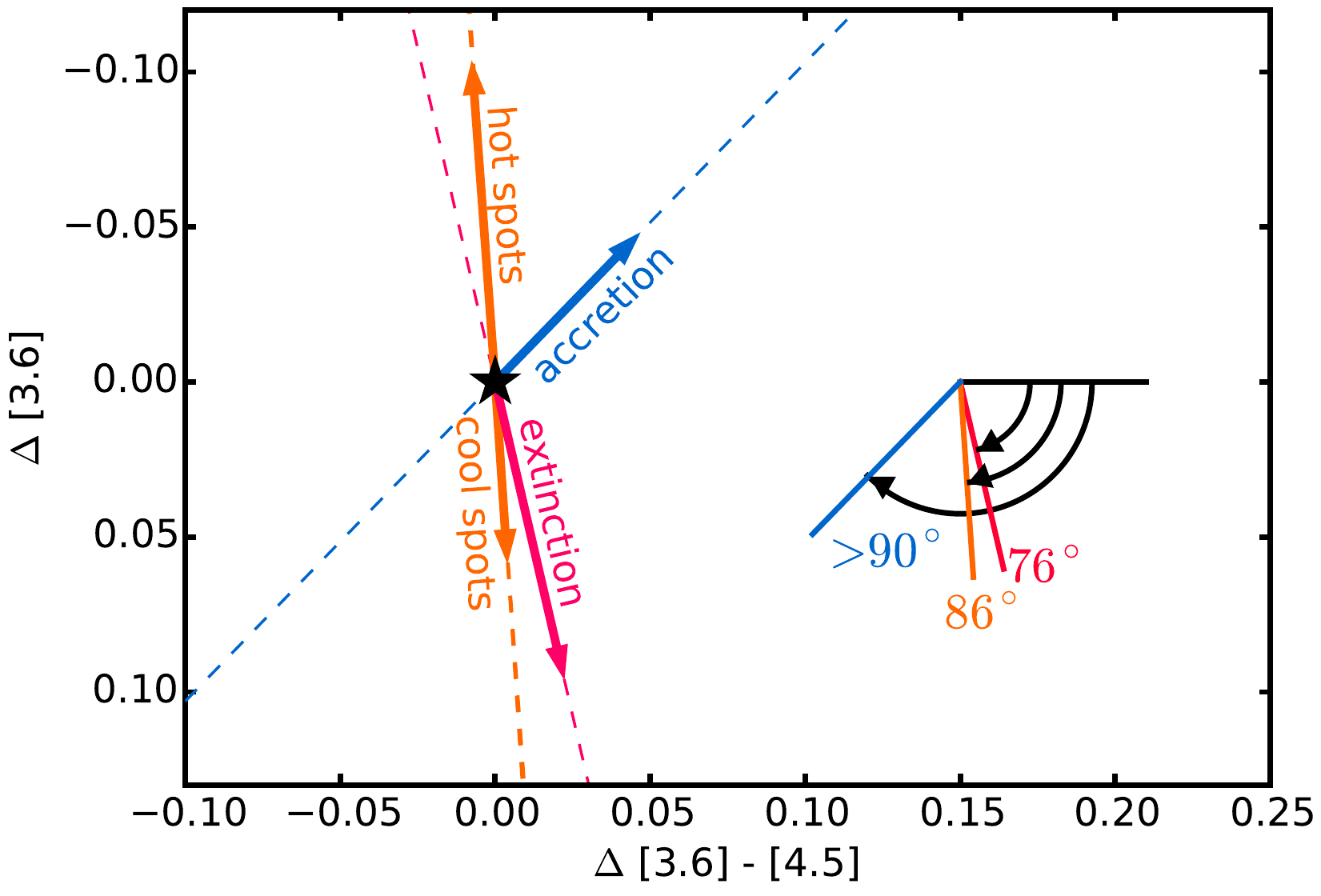}
\caption{Color-magnitude diagram (CMD) for IRAC's 3.6\,$\mu$m and 4.5\,$\mu$m bands. A given source (black star) will produce
different walks through the CMD depending on the physical processes occurring. Increasing extinction by gas and dust, as well as surface
coverage by hot or cool spots will cause walks that will make the source become redder when fainter and bluer when brighter,
while increasing accretion in the plotted model will cause the source to become redder when brighter (and bluer when fainter for decreasing accretion). The specific
models used for the tracks in this CMD are explained in the text. The direction in which we measure CMD angles is indicated by
the schematic on the right hand side of the plot.} 
\label{cmd_walks}
\end{figure}

Light curve morphologies can be a very useful tool to classify different types of behavior of young stars, as demonstrated by \cite{Cody2014}. An additional dimension of color information comes into play if time-resolved photometry from more than one wavelength is available. In our case, we have the [3.6]-[4.5] color as a function of time as an additional parameter which we can investigate. Other colors derived from the $JHK$ bands have been used for this purpose as well \citep{Carpenter2001, AlvesdeOliveira2008, Wolk2013}. Single-epoch color-magnitude diagrams in the near-infrared are also used to compare YSOs to stellar evolution isochrones and get a hande on the mass range of the objects in a cluster. This has been done for the members of IRAS~20050+2720 that have been detected in the $r'$ and $i'$ photometric bands (see \citealt{Guenther2012} and their Figure~7); these detected YSOs have masses below $3\,M_\odot$, the majority of them having masses between $0.2\,M_\odot$ and $1\,M_\odot$.

Different physical processes will cause different color changes over time. In Figure~\ref{cmd_walks} we show the influence of cool spots, hot spots, gas and dust absorption, and an accretion model following \cite{Espaillat2010} on the [3.6] vs.\ [3.6]-[4.5] color magnitude diagram (CMD). 

The slopes were derived as follows: For the spots, we assumed a stellar photospheric temperature of 5000 K, and cool and hot spots with temperatures of 4000 K and 7000 K, respectively. For simplicity, we used a blackbody spectrum for the photosphere and spots. For illustrative purposes, we used large filling factors from 5\% to 20\% for both cool and hot spots. As can be seen from Figure~\ref{cmd_walks}, the changes due to spot coverage are almost vertical, i.e.\ colorless\footnote{The resulting slopes for cool and hot spots modelled here are -14.6 and -14.2 in units of (-[3.6] mag)/(([4.5] - [3.6]) mag), i.e.\ steeply downwards in a CMD where the brighter [3.6] magnitudes are upwards on the y axis.}. This is because we are far out in the Rayleigh-Jeans tail of the blackbody spectrum with the 3.6 and 4.5\,$\mu$m bands. The predicted amplitudes of variability due to spots are relatively small for realistic spot coverages (ca.\ 0.15 mag or 15\% in [3.6] and [4.5] for cool spots). The expected spot coverage fraction derived from actual observed modulations is much smaller than our illustrative values; in the near-infrared, where the induced modulation is stronger than in the mid-infrared, typical spot modulation amplitudes suggest spot coverage fractions of a few percent \citep{Carpenter2001, Scholz2005, Wolk2013}.

For extinction we show the typical effect of an ISM-like gas and dust mix on the CMD. Both the environmental density of gas and dust in the cluster and the material in the protoplanetary disk contribute to the extinction of our objects. However, since we observe short-term extinction events in our YSOVAR light curves those most likely caused by changes in the line of sight column density of the protoplanetary disk material. Because gas and dust absorb more strongly at shorter wavelengths, reddening of the spectrum occurs when the source gets fainter. The exact slope of the reddening depends on the gas-to-dust ratio of the extintion material, and on the grain size distribution of the dust. However, to give a general idea, we show the reddening vector measured for the ISM by \cite{Indebetouw2005}\footnote{The resulting extinction slope is -4.3 in units of (-[3.6] mag)/(([4.5] - [3.6]) mag).}.

For the accretion effects a variety of models exist. Generally, accretion causes a redder color when the source gets brighter and a bluer color when the source gets fainter. For illustration, we show a \cite{DAlessio2006} disk model following  \cite{Espaillat2010}. In this model the effect of accretion is the formation of a hot spot on the stellar surface, which increases the irradiation of the disk. This additional flux heats the inner rim of the disk beyond the dust sublimation temperature, so that the inner hole of the disk widens and the surface temperature profile of the disk changes. With increasing mass accretion rates, a source seen at low inclination angles moves to the upper right of the CMD (sources seen edge-on will show deviations due to inclination effects)\footnote{The example of a possible accretion slope shown in Fig.~\ref{cmd_walks} is 1.2 in units of (-[3.6] mag)/(([4.5] - [3.6]) mag).}. However, there are other scenarios such as scattered light which may also cause a source to become bluer as it gets fainter (see \citealt{Bibo1990, Waters1998} for examples of such behavior in UX Ori stars, \citealt{DeMarchi2013} for a detailed geometric explanation for protoplanetary disks in general).

\subsubsection{
Color-magnitude slopes as a function of SED class and variability}
\label{sect:colormagnitudeslopesasafunctionofsedclassandvariability}

\begin{figure}[t!]
\includegraphics[width=0.48\textwidth]{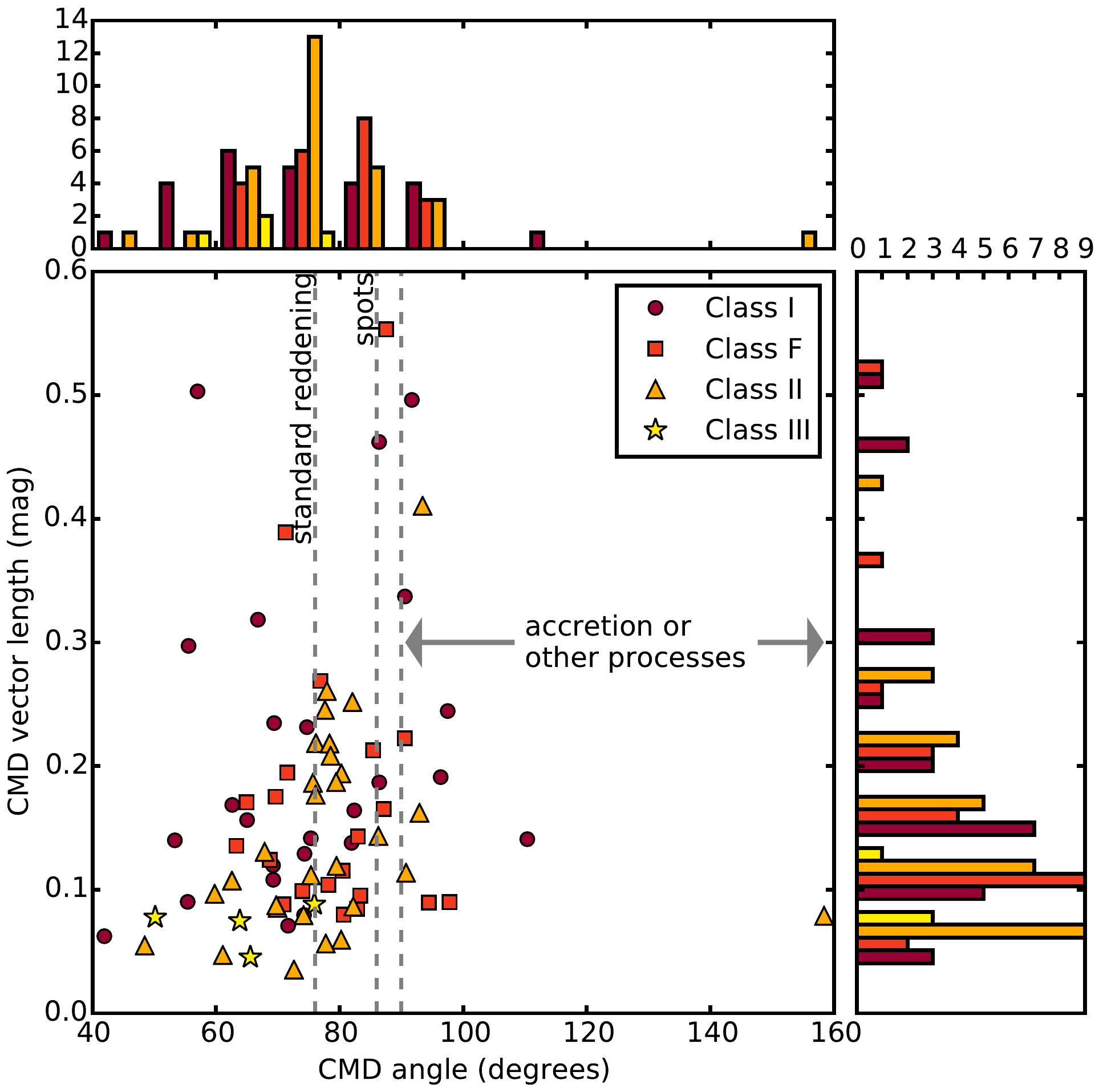}
\caption{Distribution of CMD slope angles and length of the spanned vector in the CMD for the standard set of members. 
The majority of the sources show CMD angles compatible with extinction and spot modulation. Sources with large CMD 
angles caused by other processes such as accretion or scattered light tend to display longer CMD vectors.}
\label{CMD_angle_length}
\end{figure}

To investigate the dependence of CMD slopes on SED class, we performed linear fits to the CMDs of all sources in our standard set. We furthermore require that the sources do not possess light curve artifacts, and that they are flagged by our algorithm as variable sources (because if there is only statistical fluctuation in the light curves, the color slope is meaningless). The data points in the CMDs have non-negligible errors in both the x and y direction. To take these errors into account simultaneously, we performed a least-squares orthogonal fit to a straight line, using the python package \texttt{scipy.odr} which performs an Orthogonal Distance Regression with a Levenberg-Marquardt-type algorithm\footnote{more details on the ODR package can be found here: \mbox{\url{http://docs.scipy.org/doc/scipy/reference/odr.html}} }.
We define our slope angles as the clockwise angle from a horizontal line as shown in Figure~\ref{cmd_walks}. We show all fitted CMD slopes as a function of SED class in Figure~\ref{CMD_angle_length}, together with the length of the vectors spanned in the CMD (we use the 80\% length spread to alleviate the effect of outliers). We have omitted sources with fitted slopes that have very large errors ($>10^\circ$); typical errors are below $5^\circ$. The expected angular slopes for standard (ISM) reddening, cool and hot spots, and the general direction of accretion angles are indicated on the plot as well ($76^\circ$ for ISM reddening, $86.2^\circ$ for hot spots, $85.8^\circ$ for cool spots, and $>90^\circ$ for other processes like accretion or scattered light). We find that only class I, F, and II sources display angles $>90^\circ$ at all, as expected. The majority of those disk-bearing sources, however, show CMD slopes compatible with ISM reddening or spot modulation (or a combination of both). There are only four Class III sources with detected variability and a reasonably good fit to the CMD; this is because the variability amplitudes of the Class III objects are usually too small to yield a well-constrained CMD fit. The four variable Class III sources that do have non-negligible color changes display slopes that are significantly flatter than expected spot modulation. They are all detected to be periodic sources; they are also the sources that display significant additional scatter on top of their periodicity, i.e., sources in the lower right part of Figure~\ref{scatter_vs_period}.

In terms of the amplitude of variability in color space, i.e., the length of the spanned vector in the CMD, we find a dependence on SED class. Similar to the results for the overall variability amplitudes, we again find that class III sources span short vectors in the CMD, while the distribution of vector lengths for disk-bearing sources show a tail towards large CMD spreads; see Figure~\ref{CMD_angle_length}. This trend has also been reported by Wolk et al.\ (submitted) for members of the cluster GDD12-15.

\begin{figure}[ht!]
\includegraphics[width=0.48\textwidth]{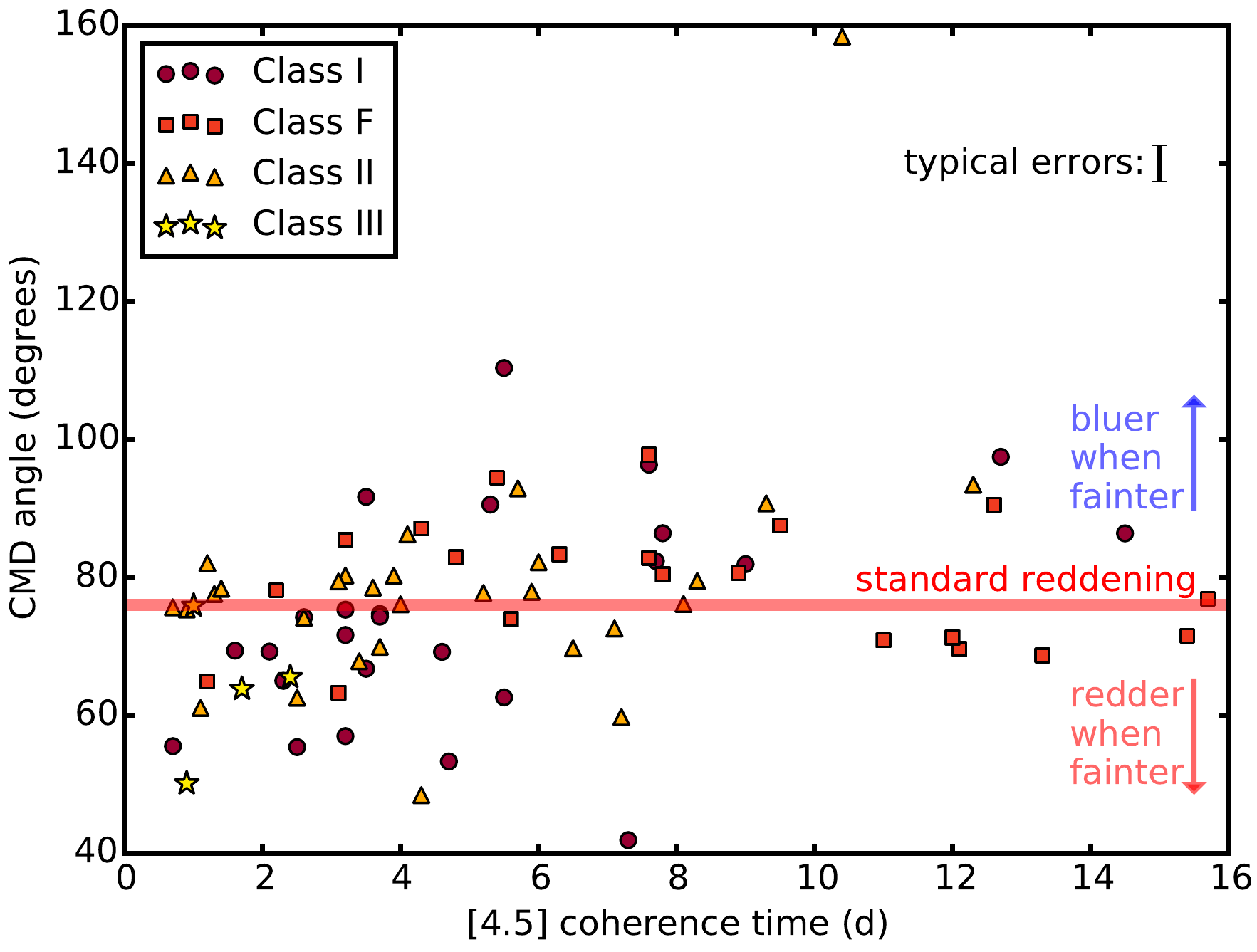}
\caption{CMD slope angle versus coherence time in the $4.5\,\mu\mathrm{m}$ band. Cluster members with long coherence times (i.e., long-term light curve changes) tend to display CMD slopes bluer than ISM reddening.}
\label{timescales_vs_angle}
\end{figure}

\begin{figure*}[ht!]
\includegraphics[width=0.33\textwidth]{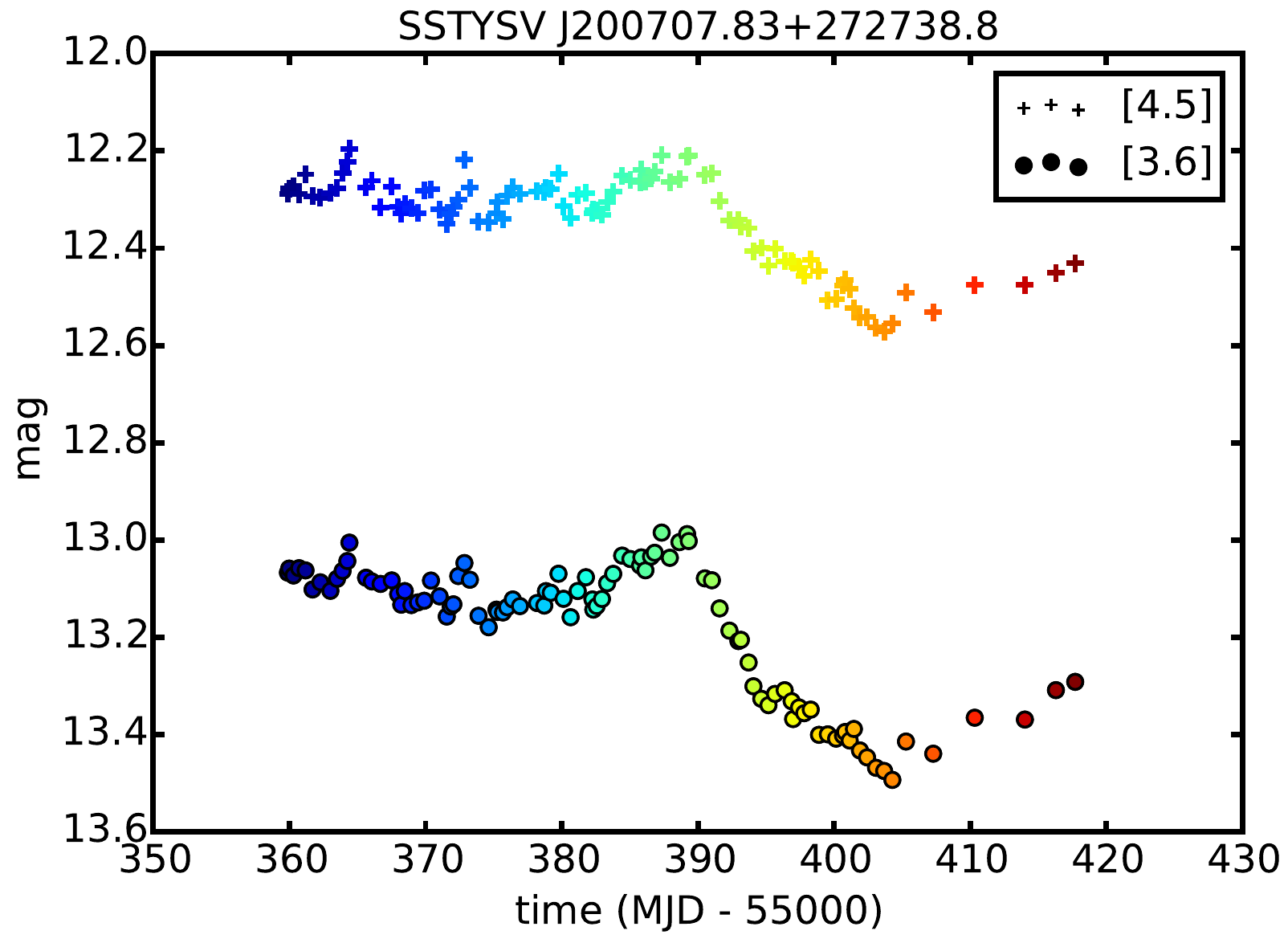}
\includegraphics[width=0.33\textwidth]{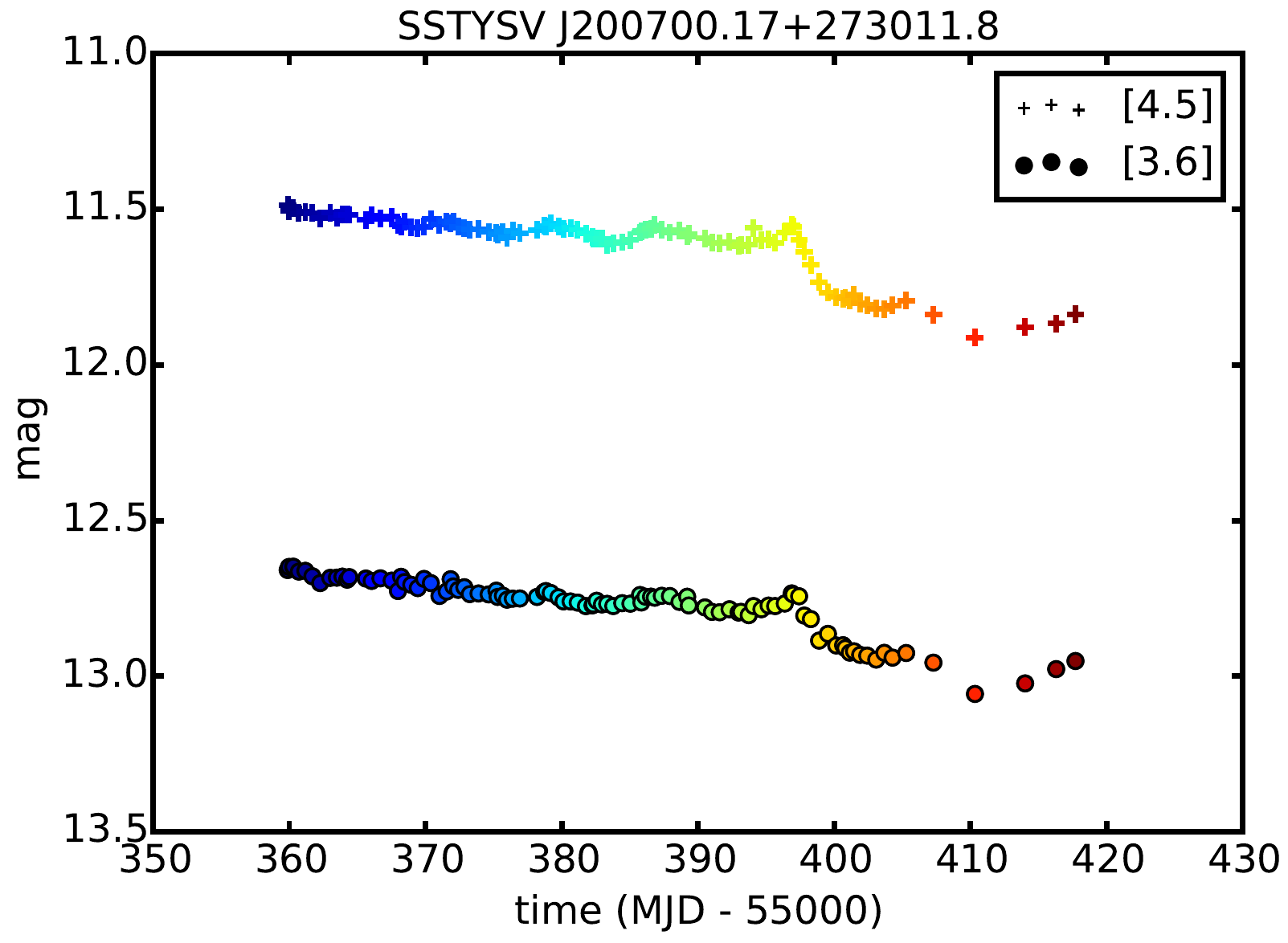}
\includegraphics[width=0.33\textwidth]{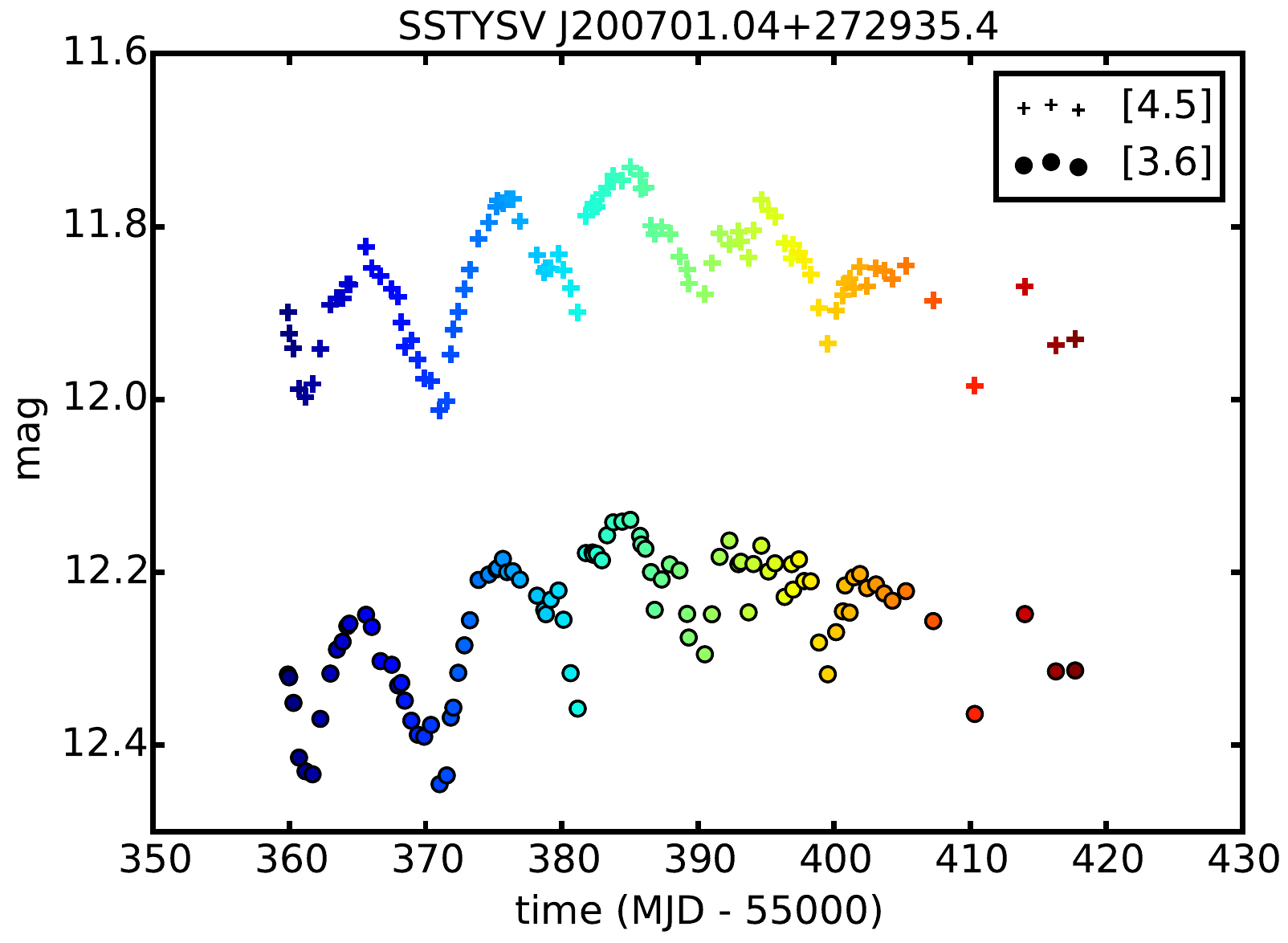}

\includegraphics[width=0.33\textwidth]{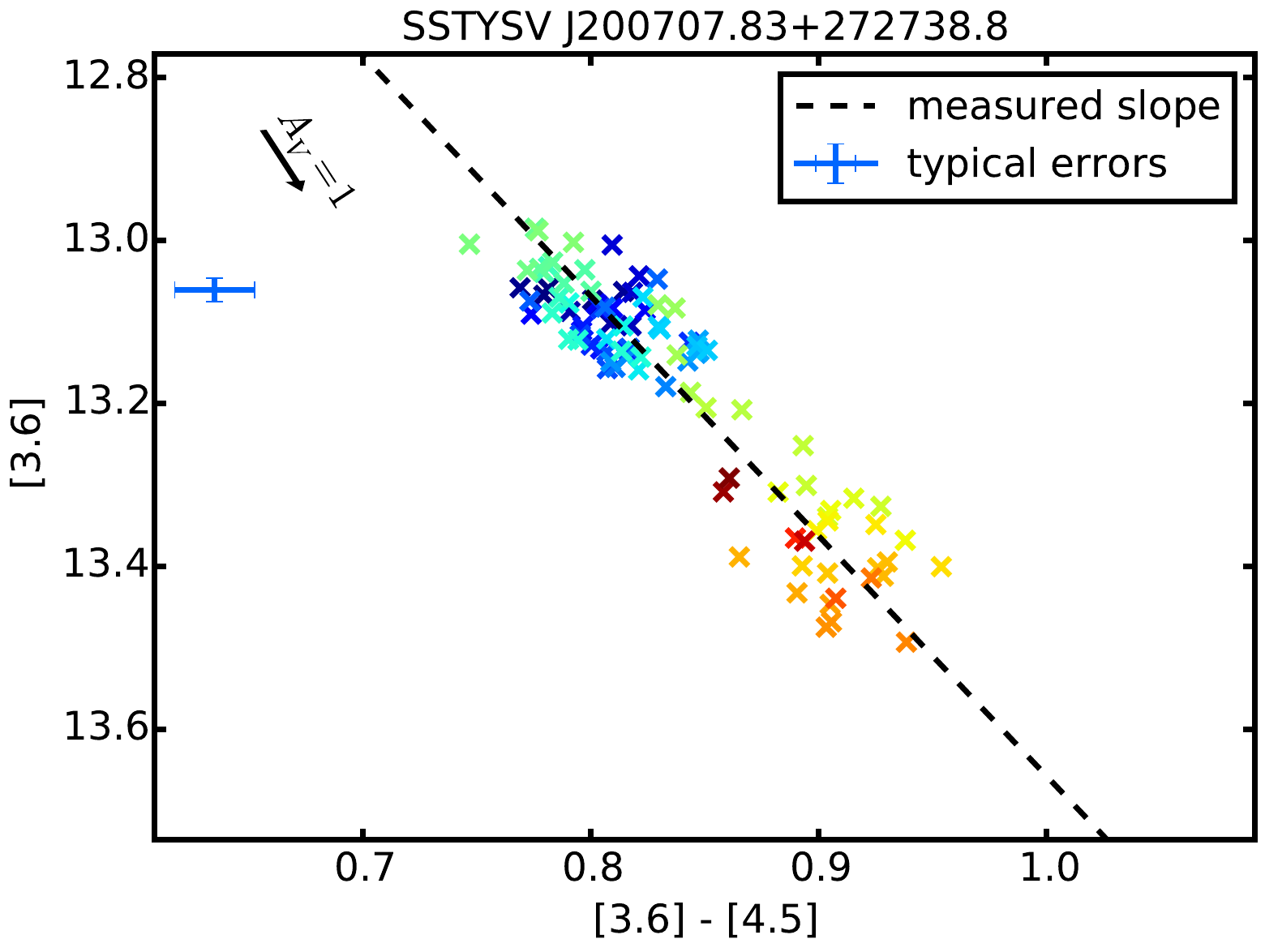}
\includegraphics[width=0.33\textwidth]{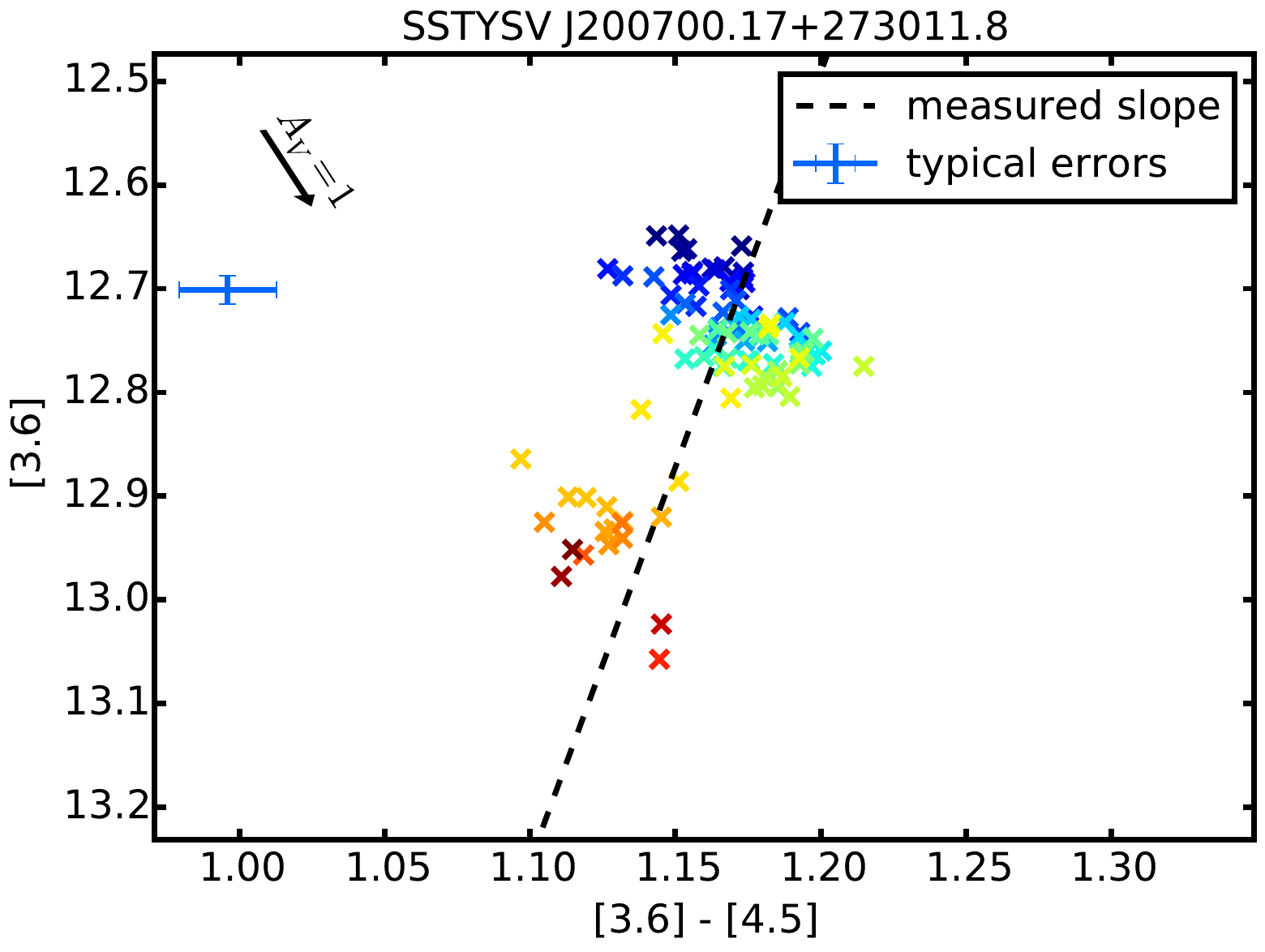}
\includegraphics[width=0.33\textwidth]{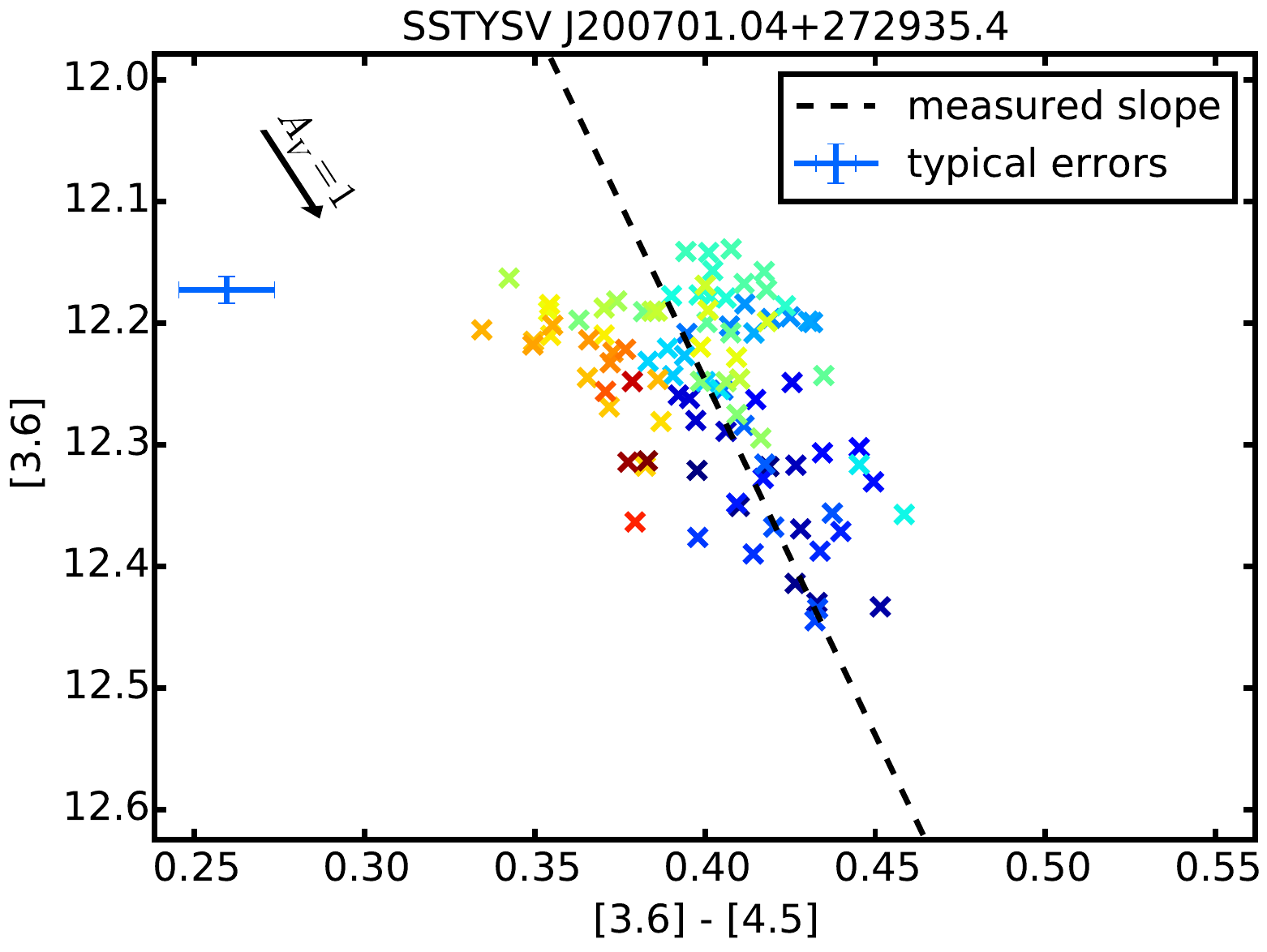}

\caption{Examples of cluster members with distinct properties in the color-magnitude diagram (CMD). Light curves in top row, CMDs of the same source in the bottom row. Typical error bars are shown in the CMD plots. The X/Y ratio in these CMDs is different from the one in Fig.~\ref{cmd_walks} in order to make the slopes less steep and more easily visible; however, the X/Y ratios are the same in all CMDs in this Figure and in Fig.~\ref{cmd_2} to allow intra-source comparisons by eye. See text for discussion of individual sources.}
\label{cmd_1}
\end{figure*}

We find an interesting trend when we compare the CMD slopes with the time scales for variability of the sources, i.e., the periods and, for aperiodic sources, the coherence times. We find a trend of sources with longer variability time scales to display larger CMD slope angles as shown in Figure~\ref{timescales_vs_angle}. The sample of periodic sources that have two-band light curves is too small to allow for a statistically significant result, but the sample of sources where we can calculate the coherence time scales is large enough to test for a correlation with CMD slopes. We use Spearman's rank correlation coefficient $\rho$ for this; it is a non-parametric test (i.e., it does not assume a linear or otherwise specified shape of correlation). It spans values from -1 (perfect anticorrelation) to +1 (perfect correlation), with 0 meaning no correlation between the two variables. The significance of the correlation is again given by the corresponding p-value, i.e., the probability that two random, uncorrelated variables could show the same or a more extreme $\rho$ value than the tested variables. Comparing the coherence time scales in each band and the CMD slopes of our member sources, we find a strong positive correlation of the slope with the [4.5] coherence time ($p=0.00013$, $\rho=0.42$) and with the [3.6] coherence time ($p=0.015$, $\rho=0.27$).

\begin{figure*}[ht!]
\includegraphics[width=0.33\textwidth]{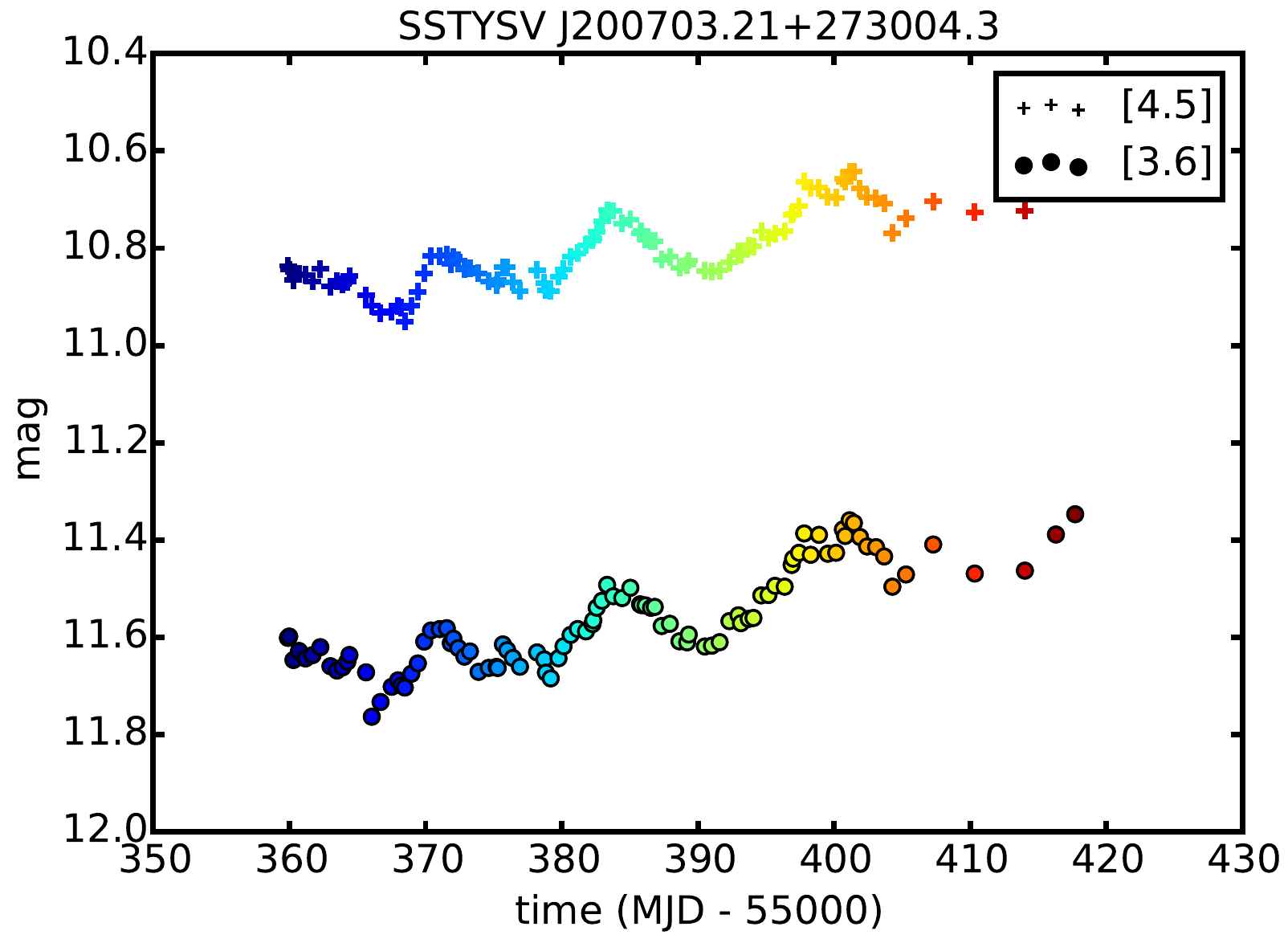}
\includegraphics[width=0.33\textwidth]{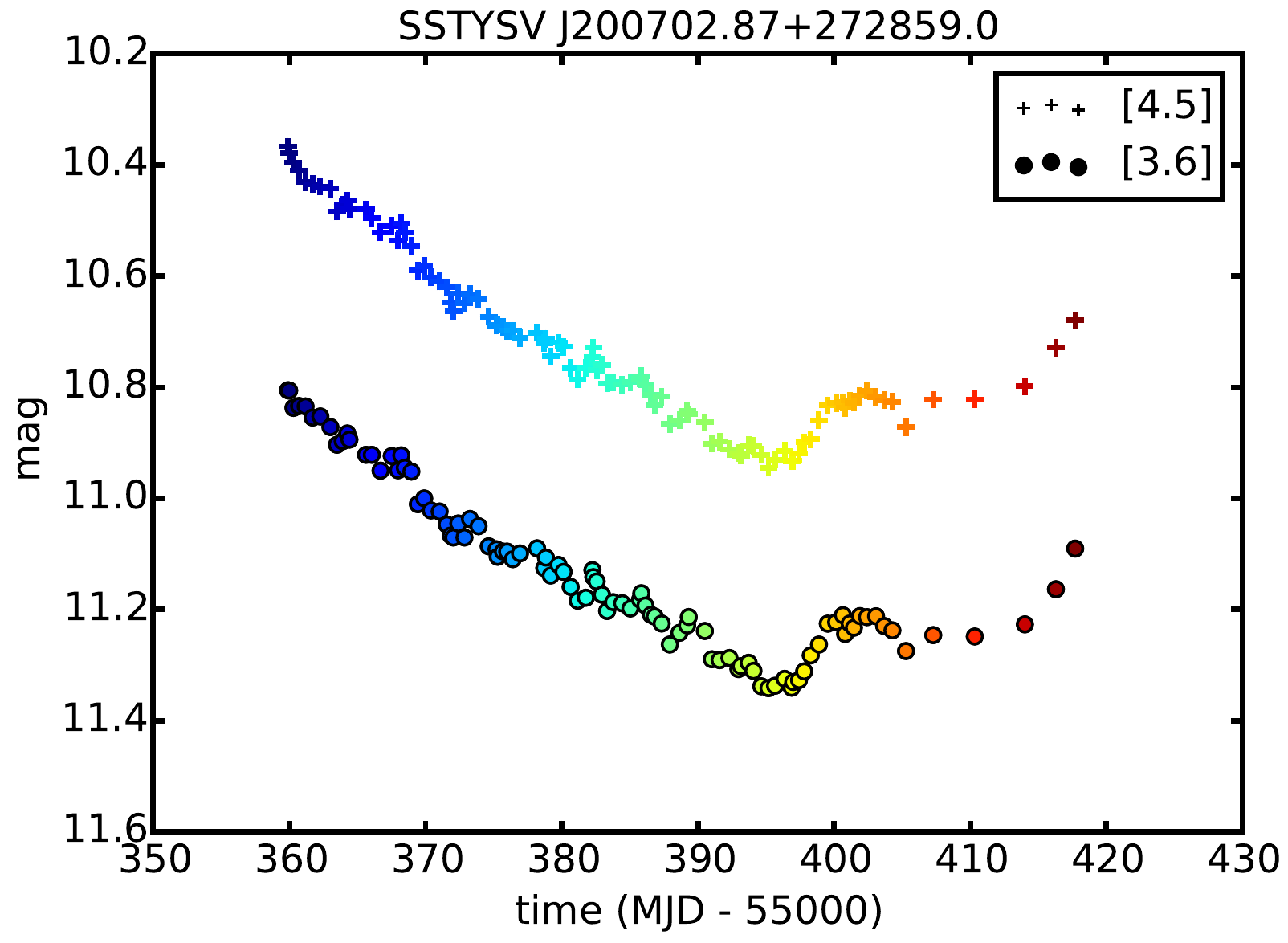}
\includegraphics[width=0.33\textwidth]{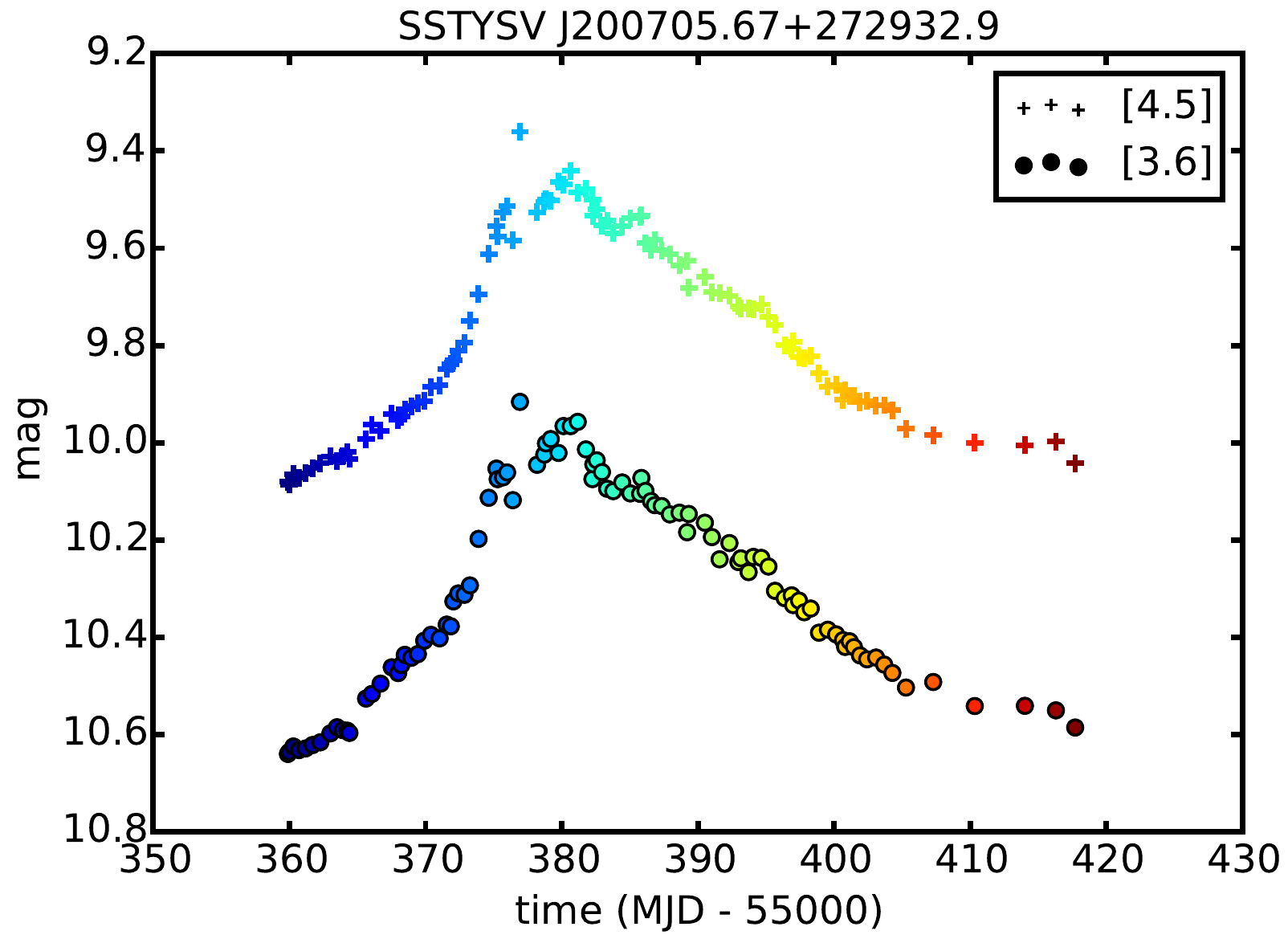}

\includegraphics[width=0.33\textwidth]{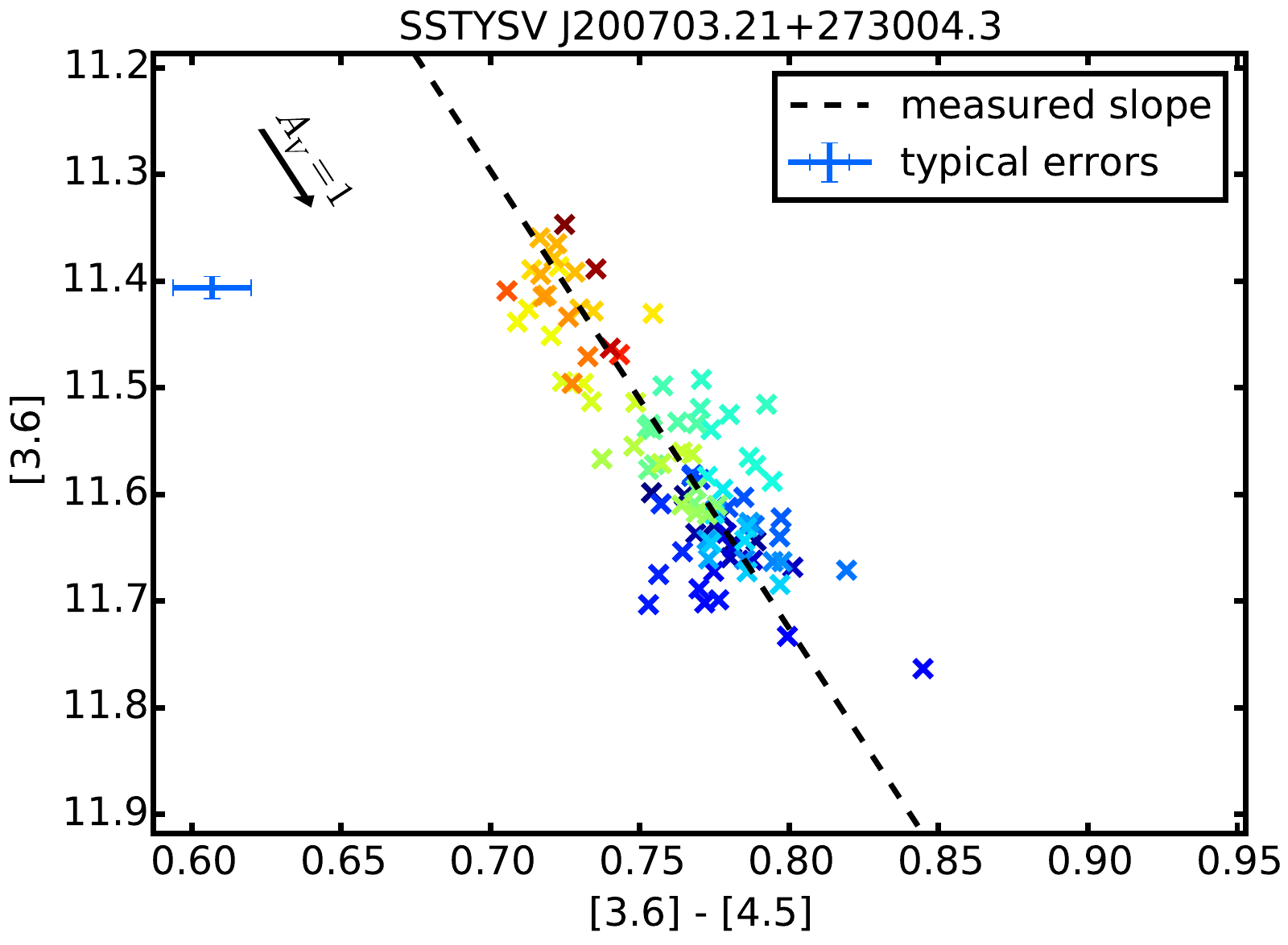}
\includegraphics[width=0.33\textwidth]{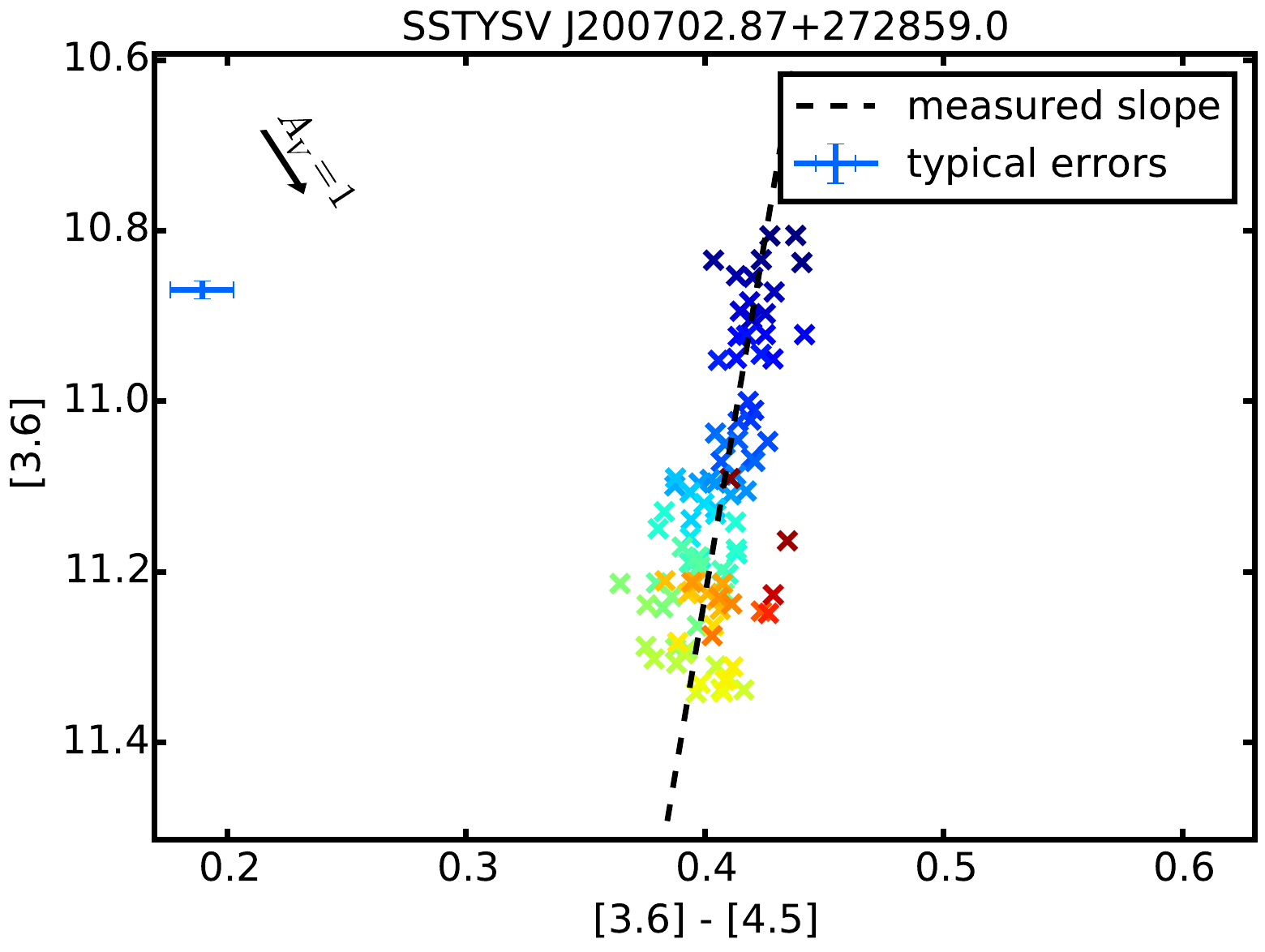}
\includegraphics[width=0.33\textwidth]{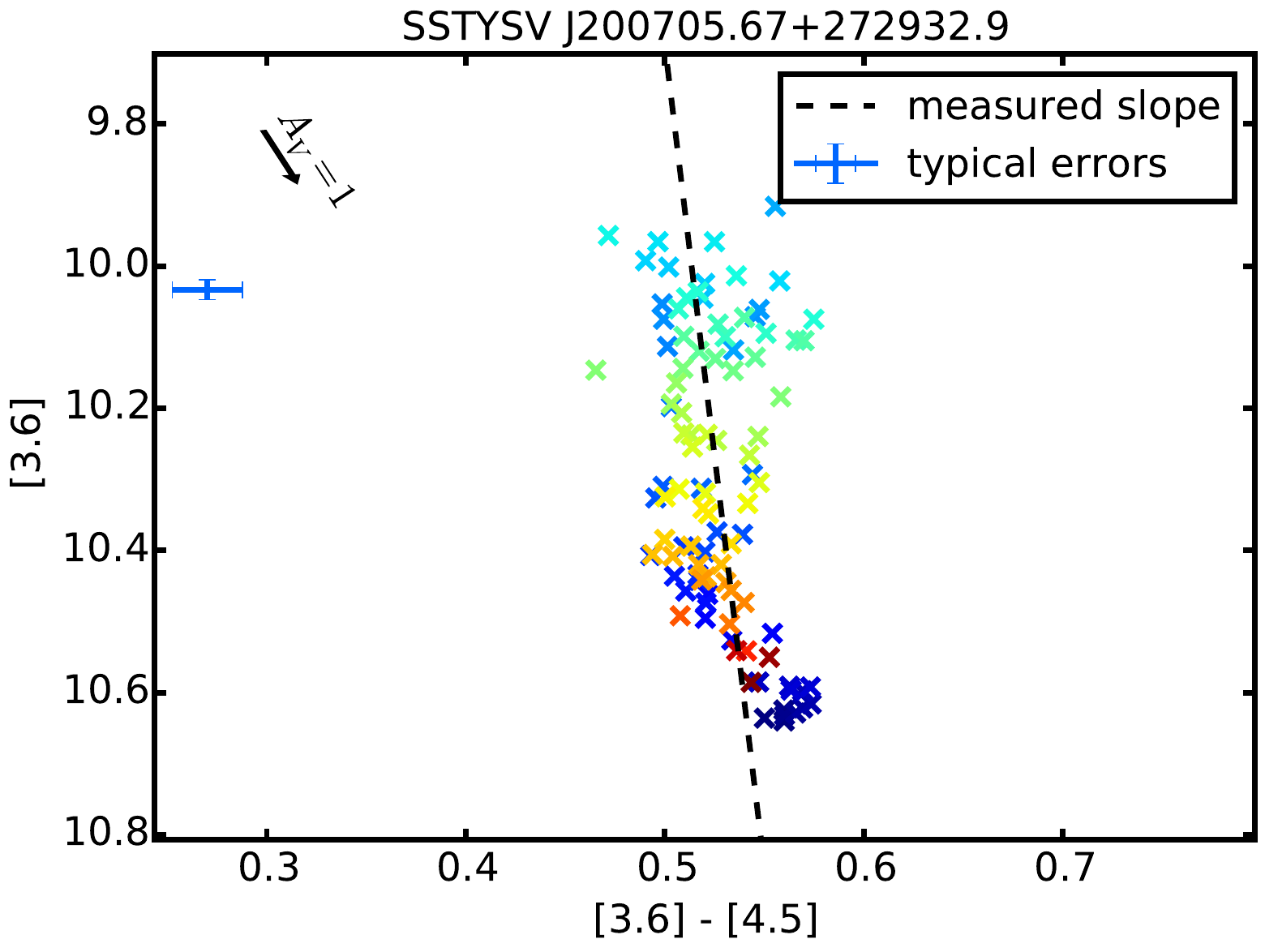}

\caption{Like Fig.~\ref{cmd_1}, for three other sources.}
\label{cmd_2}
\end{figure*}

This means that changes in extinction and spot modulation tend 
to occur on longer time scales than changes in other processes 
like accretion that cause a blueing effect. Specifically, when 
blueing processes dominate the CMD, we find a mean [4.5] 
coherence time of 8.2\,d (i.e.\ light curve changes typically 
occur on time scales of ca.\ $8.2\times 3.5\, \mathrm{d} = 29\,\mathrm{d}$), whereas 
the mean [4.5] coherence time for sources with reddening 
is 5.1\,d.

\subsubsection{Examples for processes in color-magnitude space}
\label{sect:examplesforprocessesincolormagnitudespace}

A prime example where color information is crucial for the physical interpretation of the light curves is given in Figure~\ref{cmd_1}, left and middle. It show two different sources, namely SSTYSV~J200707.83+272738.8 (a Class F source) and SSTYSV J200700.17+273011.8 (a Class I source). These two sources have very similar light curve morphologies: The sources display low-amplitude variability for about 30 days, then a small increase in brightness, followed by a significant dimming of the order of 0.4 magnitudes, lasting for ca.\ 20 days and then increasing slowly again. Despite these striking similarities, the CMDs display very different behavior. For the first source, the CMD is well-described by a straight line with a slope very close to the interstellar reddening law. A likely interpretation would be here that a blob of denser material in the disk has moved into the line of sight. Such dimmings were observed first for the eponymous young stellar object AA Tau, which displays extinction events by its disk on a semi-regular basis \citep{Bouvier1999, Bouvier2013}. The second source's CMD, however, shows that its data points are divided into two subclusters, and the slope between them is in the direction we expect for accretion events. Within the two subclusters, the individual slopes are roughly compatible with the ISM reddending vector again. Here the most likely interpretation is that a sudden drop in the accretion rate occurred which causes the dimming.

The source SSTYSV~J200701.04+272935.4 is what \cite{Cody2014} call a ``dipper'', i.e., it displays regular downward spikes in the light curve. This specific source is of SED class II, and its CMD shows a slope that is consistent with the interstellar reddening law. In addition, we can see a clustering by time in the CMD, indicated by the color coding in the plot in Figure~\ref{cmd_1}, right. The data points in the CMD corresponding to times 400--420, indicated by orange/red crosses, are offset from the rest of the point cloud in a direction roughly orthogonal to the reddening vector. The data for this object generally supports the interpretation of \cite{Cody2014} that dippers are objects with \mbox{(semi-)regular} changes in extinction due to structures in their disk. In our example, this seems to be weakly overlaid with a gradual change in accretion as well, given the clustering of orange/red vs. other data points in the CMD of this source.

Another example for a source with regular changes is SSTYSV~J200703.21+273004.3, shown in Figure~\ref{cmd_2}, left. While the source shows several brightening events, the change in time scales from one peak to the next and the additional upward slope prevent this source from being detected as periodic in our periodicity search. This ``quasiperiodic'' class F object shows a CMD slope perfectly in line with the interstellar reddening law by \cite{Indebetouw2005}. As is evident in visual inspection of the CMD displayed in Fig.~\ref{cmd_2}, the scatter perpendicular to the reddening slope is not much wider than the typical photometric error on the color of this source. This means that both during the individual brightenings and the longer upward trend the brightness changes are dominated by extinction processes. This is consistent with small-scale structures in the disk causing time-variable extinction, as well as a slow decrease in density along the line of sight caused by larger-scale structural differences.

An interesting source with a high-amplitude brightening event is the class F source SSTYSV J200705.67+272932.9, shown in Figure~\ref{cmd_2}, right. It displays an increase in brightness of 0.7 magnitudes in both [3.6] and [4.5], with very small color changes. The overall CMD slope is formally consistent with what one would expect for hot or cool spots. However, the magnitude of the brightness changes is incompatible with that, as realistic spot coverage fractions yield a maximum of ca.\ 0.15 mag in [3.6] or [4.5] brightness as discussed above. The nature of this source's color changes can be interpreted as a mix of processes instead. The rise of the light curve (blue data points in the light curve and the CMD) is still consistent with a decrease in extinction along the line of sight. However, shortly before the peak is reached, the light curve displays a ``knee'' in the upward slope. This is where the CMD slope changes to much steeper, i.e.\ almost colorless track. A possible interpretation is that accretion set in at the knee of the light curve on top of the decreasing extinction; at the peak, the accretion begins to fade, and with a still ongoing decrease in extinction this could lead to the observed high-amplitude colorless change.

A total of 12 sources are present in our sample that have a blueing CMD slope, 
which we associate with accretion being the dominant variability process. Out 
of the 105 variable members that have two-color light curves and 
do not display light curve artifacts, this amounts to a fraction of 11\%. These 
12 sources are listed in Table~\ref{tab:bluers}; all of them have SED classes 
compatible with the presence of disks, as expected. Out of the blueing sources, 
all are detected to be Stetson-variable. Only a single source is flagged as 
periodically variable by our algorithm, the source SSTYSV J200702.00+273058.5. 
The fraction of blueing sources is consistent with the results of 
\cite{Guenther2014} and Wolk et al.\ (submitted), who both find only a few 
blueing sources in their standard set of members for the clusters Lynds 1688
and GGD12-15, corresponding to a fraction of ca.\ 5-15\%.

\begin{table}
\begin{center}
\caption{\label{tab:bluers} Sources in IRAS 20050+2720 that show a blueing CMD slope.}
\begin{tabular}{l c}
\hline\hline
IAU name of source & SED class \\
\hline
SSTYSV J200658.99+273006.9 & F \\
SSTYSV J200706.59+272819.2 & F \\
SSTYSV J200709.91+272755.6 & F \\
SSTYSV J200700.17+273011.8 & I \\
SSTYSV J200702.00+273058.5 & I \\
SSTYSV J200706.60+273200.4 & I \\
SSTYSV J200706.64+272712.7 & I \\
SSTYSV J200707.85+272717.6 & I \\
SSTYSV J200700.57+273120.4 & II \\
SSTYSV J200702.87+272859.0 & II \\
SSTYSV J200705.37+272904.4 & II \\
SSTYSV J200706.83+272715.0 & II \\
\hline
\end{tabular}
\end{center}
\end{table}

Apart from the blueing source with the sudden drop in brightness that we already discussed above, another blueing source is particularly interesting: SSTYSV J200702.87+272859.0, shown in Figure~\ref{cmd_2}, middle. This source is a class II source, and it displays a substantial long-term dimming of 0.5 mag in both [3.6] and [4.5] over ca.\ 35 days. The slope of the CMD shows that the source is blueing when dimming, corresponding to a long-term change in accretion. Also the shorter episode at the end of the light curve where the source brightens in two steps seems to be governed by changing accretion, since the CMD does not show a sudden change of direction.

  \begin{table*}[ht!]
  \caption{Variability fractions of IR-bright member sources, comparison of X-ray sources to full sample.}
  \begin{center}
  \begin{tabular}{l l l l l l}
  \hline\hline
                                              & all Classes     & Class I          & Class F          & Class II         & Class III        \\ \hline
  variable among IR-bright                    & 0.75 (52/69) &  0.71 (10/14) &  0.82 (14/17) & 0.89 (25/28) & 0.30 (3/10) \\
  X-ray detected among IR-bright              & 0.57 (39/69) &  0.14 (2/14) &  0.59 (10/17) & 0.64 (18/28) & 0.90 (9/10) \\
  variable among X-ray detected and IR-bright & 0.77 (30/39) &  1.00 (2/2) &  0.90 (9/10) & 0.89 (16/18) & 0.33 (3/9) \\
  \hline
  \end{tabular}
  \label{tab:xrayvar}
  \end{center}
  \end{table*}

\begin{figure}
\includegraphics[width=0.48\textwidth]{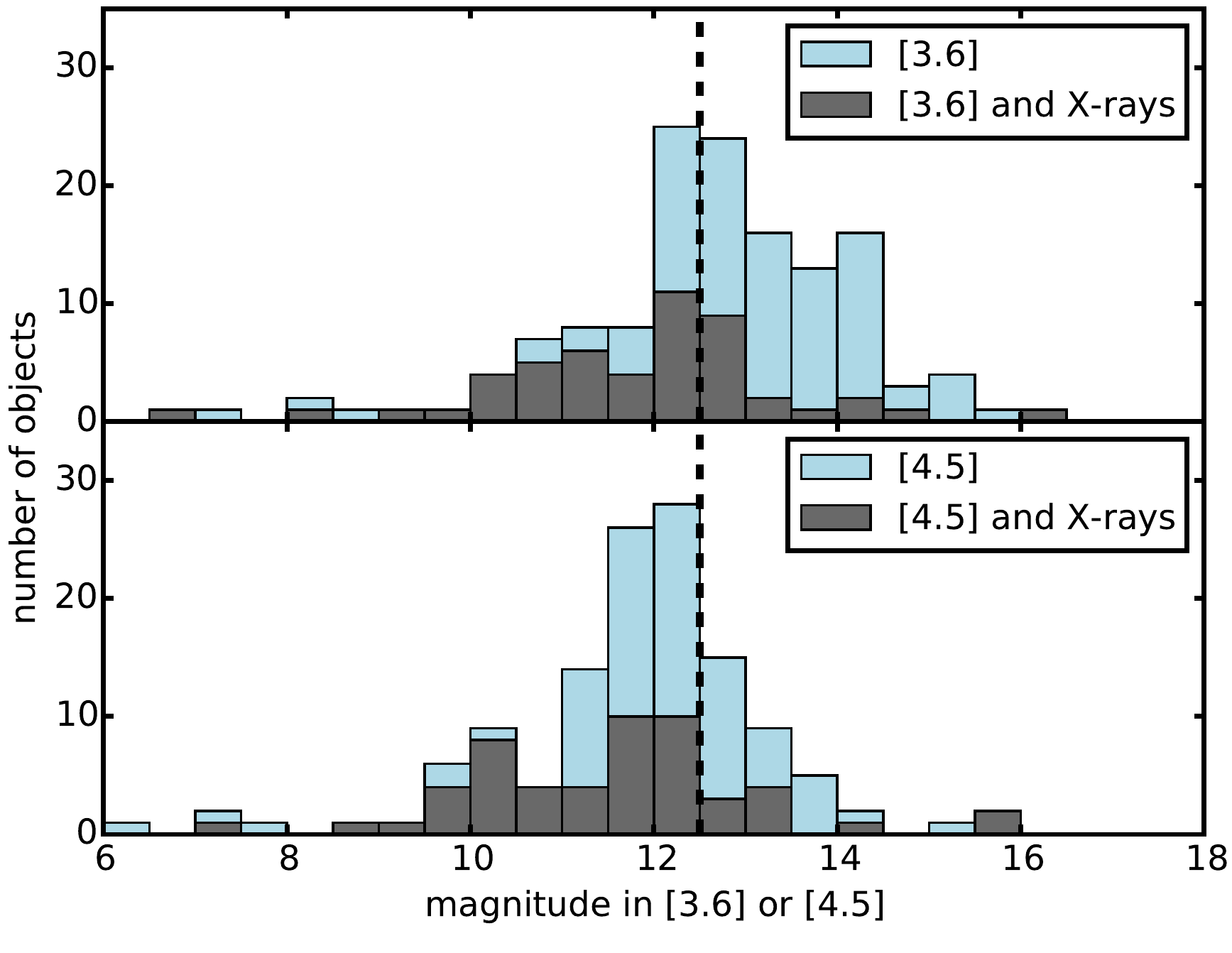}
\caption{Distribution of [3.6] and [4.5] brightness of standard set cluster members (light blue); X-ray detected members (dark grey) tend to be bright in the infrared because they are the least absorbed sources. We therefore use a brightness cutoff at 12.5\,mag (dashed line) for comparisons between X-ray detected and undetected members.}
\label{Xraymag_comparison}
\end{figure}
\begin{figure*}
\includegraphics[width=0.48\textwidth]{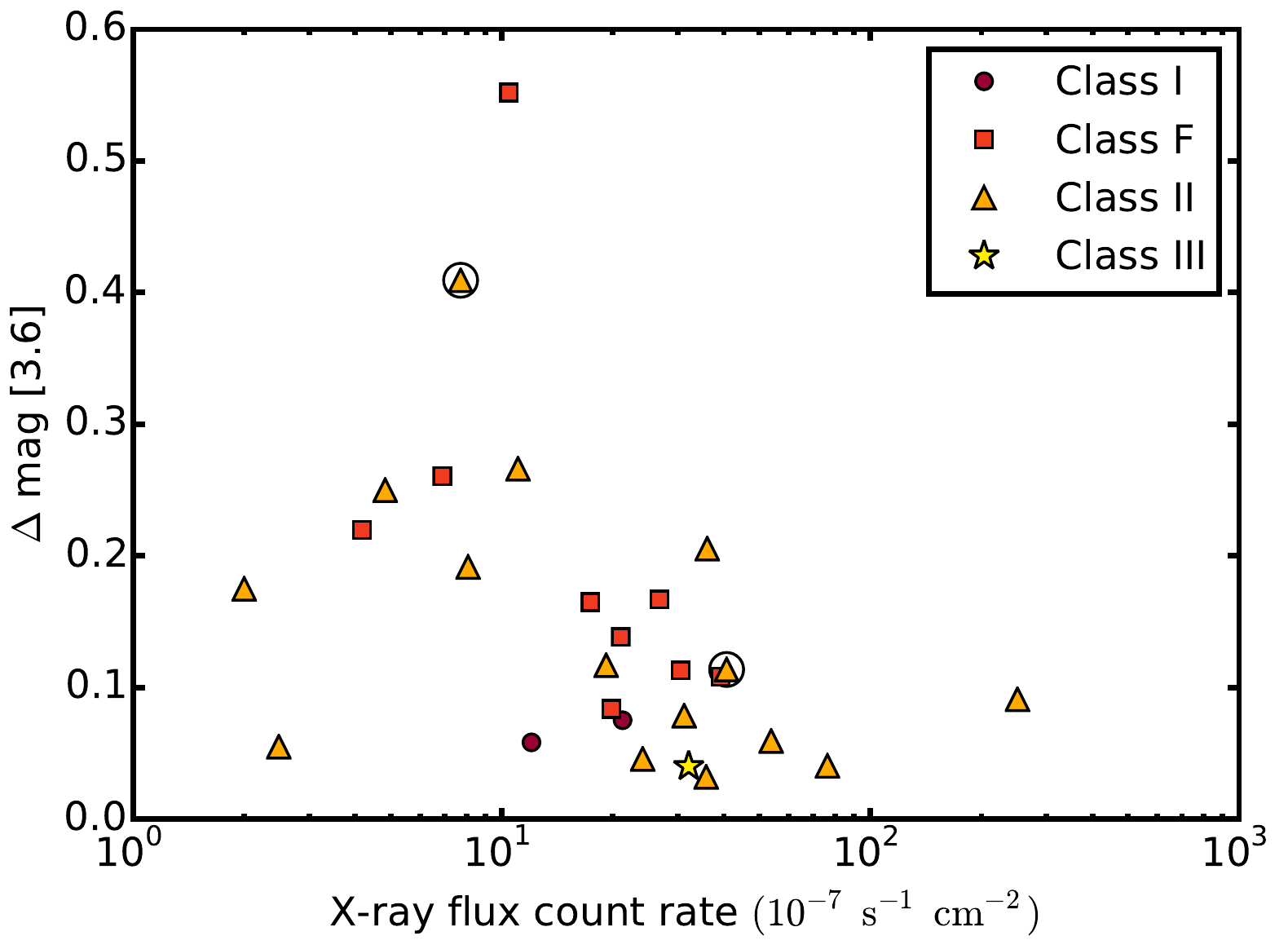}
\includegraphics[width=0.48\textwidth]{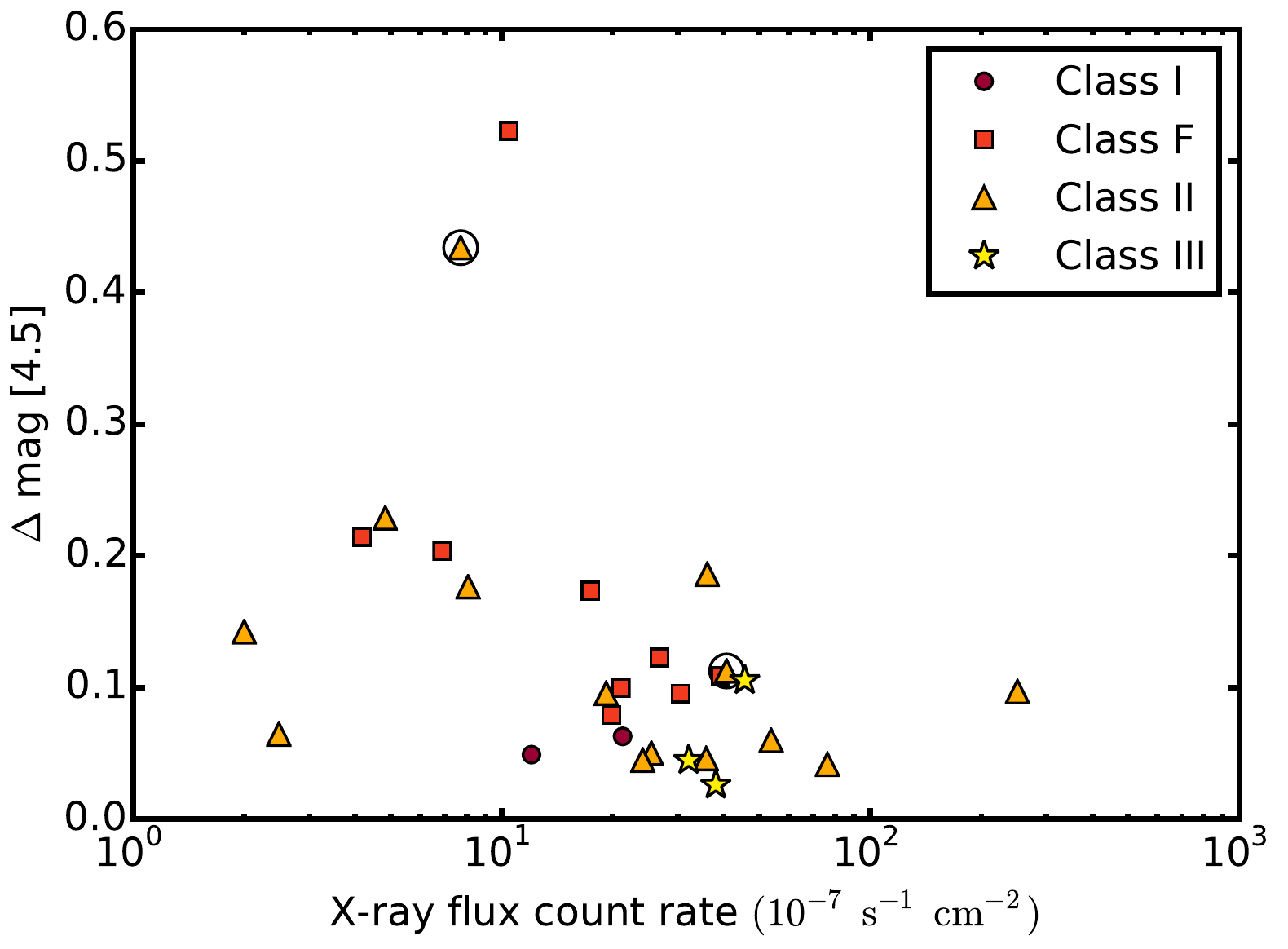}
\caption{The variability amplitudes and the X-ray flux count rate are anticorrelated. The two YSOs that display a blueing CMD slope are marked with additional circles.}
\label{Xrayfluxrate_vs_delta}
\end{figure*}

\subsection{X-rays and infrared variability}
\label{sect:xraysandinfraredvariability}

X-rays from young stars originate from magnetic activity of the stellar corona and from hot spots caused by accretion onto the star. Strong X-ray flares are thought to be able to ionize and alter parts of the circumstellar disk; also, accretion from the disk onto the star has been found to produce an excess of soft X-ray emission through accretion shocks (for a review, see \citealt{Guenther2013review} and references therein). Therefore one may expect IR variability and the observed X-ray properties of the star to be correlated in some way. Recently, \cite{Flaherty2014} analyzed simultaneous X-ray and mid-IR light curves for members of the young stellar cluster IC 348 and found that X-ray variability and IR variability were not detectably correlated, which they interpreted as a sign that X-ray heating does not strongly influence the planet-forming part of the disk on time scales of a few days.

For our cluster IRAS\,20050+2720, we do not have simultaneous X-ray and mid-IR light curves, and therefore resort to testing for general trends of mid-IR properties with X-ray brightness. Specifically, we will test if the X-ray detected and X-ray undetected cluster members show differences in their variability fractions, variability amplitudes, variability time scales, and slopes in the color magnitude diagram. We will also test if these quantities show a trend with X-ray flux count rate within the X-ray detected sample.

We have detected a total of 67 sources in X-rays among the standard set
of members (which is a total of 181 sources) using \textit{Chandra} 
(see section~\ref{sect:chandradata}), with X-ray properties as given 
in \cite{Guenther2012}. Ignoring the objects with light curve 
artifacts, we have 58 X-ray detected, clean light curve objects 
among 155 standard set members with clean light curves. We want to 
compare the IR variability properties of the X-ray detected sample 
to the X-ray undetected sample. However, we have to be careful 
to consider systematic differences in IR brightness: The X-ray 
detected sources tend to be the least absorbed sources and are 
therefore on average IR-brighter than the X-ray undetected sources; see Fig.~\ref{Xraymag_comparison}. 
This means that we are able to detect IR variability with smaller 
amplitudes, and are therefore biased to find a larger variability 
fraction among those bright sources. To counteract this bias, 
we have compared the [3.6] and [4.5] apparent brightness distribution of 
the X-ray detected and undetected samples, and found that cutting 
off sources with magnitude $>12.5$ in the [3.6] and [4.5] band 
yields a similar apparent brightness distribution for the two samples.

First, we investigated if the X-ray detected YSOs are more variable than the ones that are not detected in X-rays. We list the variability fractions per SED class for X-ray detected and undetected sources in Table~\ref{tab:xrayvar}. We find that within statistical error margins the fractions of IR-variable sources are the same among X-ray detected and undetected YSOs. 

The quantity where we do find a significant trend with X-ray flux count rate is the variability amplitude in both IR bands. The distributions of variability amplitudes are statistically indistinguishable for the X-ray detected and undetected samples; however, in the X-ray detected sample, the variability amplitudes are strongly anticorrelated with the X-ray flux. We show this trend in Figure~\ref{Xrayfluxrate_vs_delta}. A rank correlation test with Spearman's $\rho$ yields an anticorrelation with $p$-values of 0.008 in [4.5] and 0.004 in [3.6], i.e.\ a very small chance that two uncorrelated samples would happen to show this or a more extreme anticorrelation. This is not driven by any systematic IR brightness trends within the X-ray detected sample, because both the X-ray flux count rate and the variability amplitude are uncorrelated with the median [3.6] and [4.5] brightness. Even when excluding the two sources with the largest infrared variability, we still find an anticorrelation with $p$-values of 0.037 in [4.5] and 0.013 in [3.6]. 

The trend of sources with high X-ray flux count rates to have lower-amplitude mid-IR light curve changes therefore seems to be real. Note that in Figure~\ref{Xrayfluxrate_vs_delta} the sources with low-amplitude mid-IR variability populate both the low and high X-ray flux count rate parts of the plot, whereas the high-amplitude variables are only detected with low X-ray fluxes. This is not a direct effect of accretors being highly variable and X-ray bright: we only have two sources with blueing CMD slopes in the sample, marked with circles in Figure~\ref{Xrayfluxrate_vs_delta}. A possible interpretation of the trend is that we are seeing an effect of the disk inclination -- X-ray absorption is caused by heavier elements in the gas phase of the disk, such as oxygen, nitrogen, and carbon. The mid-IR absorption, on the other hand, is mainly driven by dust particles. It is therefore possible that the sources with large detected X-ray fluxes are the sources we observe mostly face-on, so that the X-ray flux is not strongly absorbed. At the same time, extinction processes in the disk will not show up in the mid-IR light curves as well, because there is very little dust along the line of sight. Unfortunately, the large distance of IRAS~20050+2720 prevents a spectral X-ray analysis for many of the sources in this sample; one would expect to see absorption effects by gas from the disk on the soft part of the X-ray spectrum. An analysis of other, more nearby clusters has the potential to show if this X-ray brightness vs.\ mid-IR variability trend can be used as a selector for disk inclination.

\begin{table*}
\begin{center}
\caption{\label{xyso_table} X-ray identified YSO candidates with \textit{Spitzer} light curves in IRAS 20050+2720.}
\begin{tabular}{ccccccc}
\hline\hline
IAU name of source & SED class & LC artifacts? & median [3.6] & median [4.5] & variable? & CMD slope type \\
\hline
SSTYSV J200706.00+272901.7 & I & yes & - & - & - & - \\
SSTYSV J200710.32+272853.8 & I & no & 16.02 & 15.09 & no & - \\
SSTYSV J200708.20+272839.5 & I & no & 13.65 & 12.83 & no & - \\
SSTYSV J200705.96+272910.1 & F & no & - & 11.98 & no & - \\
SSTYSV J200707.90+272901.9 & F & yes & - & - & - & - \\
SSTYSV J200707.31+272859.9 & II & no & 11.53 & - & no & - \\
SSTYSV J200707.74+272852.5 & II & yes & - & - & - & - \\
SSTYSV J200706.04+272856.6 & II & no & 9.62 & - & no & - \\
SSTYSV J200705.11+272919.5 & III & no & 12.54 & 12.36 & Stetson, $\chi^2$ & reddening \\
SSTYSV J200706.90+272812.0 & III & no & 13.78 & 13.38 & periodic & reddening \\
\hline
\end{tabular}
\end{center}
\end{table*}

We furthermore test if other properties of the X-ray detected sample differ from the X-ray undetected sample, such as the color-magnitude slopes, the variability amplitudes or the time scales of variability. 
We do not find any significant differences in CMD slope distributions among the X-ray detected and undetected sources, nor a trend of the slope angles with the X-ray flux count rate in the X-ray detected sample alone. With respect to the coherence times, i.e.\ the time scales for variability, we find a slight trend of X-ray sources to have longer coherence times (using the magnitude-limited samples for comparison). However, an Anderson-Darling test between the two samples shows that two random samples drawn from the same distribution have a moderate chance to yield the same or a more extreme difference in coherence times; we find  p-values of  0.063 (0.019 for all sources with disks) in [4.5] and 0.18 (0.13 for all sources with disks) in [3.6]. Note however that Wolk et al. 2014 (submitted) find this trend with higher significance for the YSOVAR cluster GDD1215; they are able to test for this difference on a larger sample of sources. In IRAS~20050+2720, no significant correlations between the magnitude of the X-ray flux count rate and the coherence time are present in [3.6] or [4.5] within the X-ray detected sample.

\subsubsection{X-ray identified Young Stellar Objects}
\label{sect:xrayidentifiedyoungstellarobjects}

\label{xysos}
IRAS~20050+2720 contains a number of sources near the center of the cluster that have not been classified by \cite{Gutermuth2009} because they have not been detected in enough bands to allow a reliable de-reddening. However, \cite{Guenther2012} have detected these sources in X-rays and have listed them are likely cluster members which are obscured by high extinction of $A_V > 10$\,mag and therefore only detected in a small number of infrared and optical bands. We have collected mid-IR light curves for 10 out of the thus identified 18 sources. We list their identities, our SED classes (which do not use a de-reddening scheme), and the median of the [3.6] and [4.5] \textit{Spitzer} light curves in Table~\ref{xyso_table}. 

Two out of those sources are also listed in our standard set of members, because they are compatible with an SED class of type III and, by definition of being an XYSO (``X-ray identified young stellar object''), detected in X-rays. However, we list them here again together with the other XYSOs.

We find that three out of the 10 XYSOs with mid-IR 
light curves display light curve artifacts; out of 
the remaining seven, two are found to be variable. 
The variability fraction among those sources without 
artifacts is therefore $0.29^{+0.17 }_{-0.13 }$, 
which is slightly lower (at ca. $2\sigma$ level) 
than the variability fraction among the standard 
set of members of $0.68^{+0.04}_{-0.04 }$. The lower 
variability fraction is not a bias induced by faint 
sources and therefore hard-to-detect variability, 
as all but one of the sources are relatively bright, 
see Table~\ref{xyso_table}; this may be an indication 
that not all of the X-ray identified sources are actually YSOs. 
Among the sources with clean light curves in two bands and 
large enough variability amplitudes to fit the CMD slopes 
reliably, we find that they display reddening slopes 
throughout.

\section{Conclusion}
\label{sect:conclusion}

\label{conclusion}

We have presented an analysis of mid-infrared light curves of 181 young stellar objects in the young cluster IRAS 20050+2720. Our main findings are:

\begin{itemize}

\item{The variability fraction of sources with disks is high
with 66\%, 80\%, and 73\% for SED classes I, F, and II. Disk-less 
class III sources are detected to be variable in 50\% of the cases. 
While the overwhelming majority of the disk-bearing sources are 
detected with irregular variability patterns, i.e.\ $\chi^2$ or 
Stetson variability, the class III sources are mainly detected to 
be periodically variable, consistent with the interpretation that 
photospheric cool spots are the main driver of variability 
for these stars. }

\item{The detected amplitudes of variability are typically in 
the range of 0.14~mag for disk-bearing sources, with 
some sources displaying amplitudes of up to 0.55~mag in 
[3.6] and [4.5]. Disk-free sources show low-amplitude 
variability around 0.08~mag.}

\item{The time scales for variability tend to be longer for disk-bearing sources, with a wide distribution out to 14 days for periodic sources. For non-periodic sources, we find through an analysis of coherence times that variability time scales reach out to 30 days in our dataset. For the disk-free sources, we find a strong clustering around shorter time scales of 3-5 days (periodic and nonperiodic), consistent with the scenario of stellar spin-up after the magnetic star-disk coupling interrupted by the dissolution of the circumstellar disk.}

\item{We analyzed periodic light curves for additional processes on top of the detected periodicity. We find that class I and F sources always display additional processes in our sample. Class II can display clean periodicity, but also strong additional variability signatures depending on the source. We generally find for disk-bearing sources that all objects with periods larger than 8 days display a large amount of extra variability on top of the periodicity, whereas (disk-free) class III sources display clean periodicity. }

\item{We use color-magnitude diagrams (CMDs) to distinguish between spots, extinction, and accretion as the dominant drivers of variability of a given source. We find that 10\% of our sources with both [3.6] and [4.5] light curves show CMD slopes that turn bluer when fainter, consistent with accretion signatures during the time of the observations; however, other explanations like scattered light or unresolved binaries with brightness changes of one of the components are possible. We also find that sources with blueing slopes show variability on significantly longer time scales than sources compatible with variability by extinction or spot modulation.}

\item{We tested for different mid-IR variability properties among the cluster members that have been detected in X-rays. We find that the variability fraction is indistinguishable between the X-ray detected and undetected members, after controlling for different mid-IR brightness distributions of the sample. We find a weak trend of X-ray detected sources to display variability on longer time scales than their undetected counterparts. Among the X-ray detected members we find that their X-ray flux count rate and amplitude of mid-IR variability are anticorrelated, surprisingly. This may be an effect of disk inclination, causing X-ray absorption by gas in the YSO-disk system. The sources that are the most variable in the mid-IR may be the ones we are observing more edge-on.}

\item{All light curves will be made available through the IRSA database. The pYSOVAR code \citep{pysovar} used for the analysis of the light curves and color-magnitude diagrams is available through \url{https://github.com/YSOVAR}.}
\end{itemize}

\acknowledgments

The authors thank Joshua Bloom and Mike Skrutskie for access to the PAIRITEL telescope, and Elaine Winston for helpful comments on the manuscript.

This work is based on observations made with the Spitzer Space Telescope, which is operated by the
Jet Propulsion Laboratory, California Institute of Technology under a contract with NASA. This research made use of Astropy, a
community-developed core Python package for Astronomy \citep{Astropy}. This research
has made use of the SIMBAD database and the VizieR catalogue access tool \citep{VizieR}, both
operated at CDS, Strasbourg, France and of data products from the Two Micron All Sky Survey, which is a
joint project of the University of Massachusetts and the Infrared Processing and Analysis Center/California
Institute of Technology, funded by NASA and the National Science Foundation. 
K.P.'s work was supported in part by the Jet Propulsion Laboratory (JPL) funded by NASA through the Sagan Fellowship Program executed by the NASA Exoplanet Science Institute.
H.M.G.\ acknowledges Spitzer grant 1490851. P.P.\ acknowledges the JPL Research
and Technology Development and Exoplanet Exploration programs. S.J.W.\ was supported by
NASA contract NAS8-03060.

%\clearpage

\bibliographystyle{aa}
\bibliography{katjasbib.bib}

\end{document}